\documentclass[twocolumn,aps,groupedaddress]{revtex4}

\usepackage{graphicx}
\usepackage{dcolumn}
\usepackage{bm}
\usepackage{amsmath,amssymb,amsfonts}
\usepackage{color}
\usepackage{ulem}
\usepackage{braket}

\begin{document}

\title{Supermetal}

\author{Hiroki Isobe}
\author{Liang Fu}
\affiliation{Department of Physics, Massachusetts Institute of Technology, Cambridge, Massachusetts 02139, USA}

\begin{abstract}
We study the effect of electron interaction in an electronic system with a high-order Van Hove singularity, where the density of states shows a power-law divergence.  Owing to scale invariance, we perform a renormalization group (RG) analysis to find a nontrivial metallic behavior where various divergent susceptibilities coexist but no long-range order appears.  We term such a metallic state as a \textit{supermetal}.  Our RG analysis reveals noninteracting and interacting fixed points, which draws an analogy to the $\phi^4$ theory.  We further present a finite anomalous dimension at the interacting fixed point by a controlled RG analysis, thus establishing an interacting supermetal as a non-Fermi liquid.
\end{abstract}

\maketitle

\section{Introduction}

A Bloch electron in a crystal is described by the energy dispersion $E_{\bm{k}}$ that relates the energy with its wave vector $\bm{k}$.  For metals, the energy dispersion determines the density of states (DOS) at the Fermi level, which to a large extent governs various thermodynamic properties such as charge compressibility, spin susceptibility, and specific heat.
Van Hove's seminal work \cite{VanHove} revealed that the DOS exhibits non-analyticity at an extremum or a saddle point of the energy dispersion, where $\nabla_{\bm{k}} E_{\bm{k}}=0$.  Importantly, Van Hove singularities (VHS) are guaranteed to exist in every energy band by the continuity and the periodicity of $E_{\bm{k}}$ over the Brillouin zone.
The behavior of the DOS at a VHS depends on whether it is at an energy extremum or a saddle point, and also on the dimensionality of the system.  For example, at a saddle point in two dimensions with $E_{\bm{k}} = k_x^2-k_y^2$, the DOS diverges logarithmically. As the chemical potential crosses the VHS, the topology of Fermi surface changes from electron to hole type, known as an electronic topological transition.

Recently, we have extended the notion of VHS to high-order saddle points, where, besides $\nabla_{\bm{k}}E_{\bm{k}}=0$, the Hessian matrix $D_{ij} =\partial_{k_i} \partial_{k_j} E_{\bm{k}}$ satisfies $\det D(\bm{k})=0$ \cite{high-order}.
These high-order saddle points occur where two Fermi surfaces touch {\it tangentially}, or at the common intersection of three or more Fermi surfaces \cite{Fu2011, multicritical1}. An example of the former is $E_{\bm{k}}= k_x^2 - k_y^4$, and of the latter is $E_{\bm{k}}=k_x^3 - 3 k_x k_y^2$.
Generally speaking, high-order saddle points can be realized by tuning the energy dispersion with one or more control parameters.
At high-order saddle points in two dimensions, the DOS shows a power-law divergence $D(E) \propto |E|^{-\epsilon}$, much stronger than a logarithmic one at ordinary VHS \cite{high-order,multicritical1}.

The existence of high-order VHS has recently been identified in a variety of materials including twisted bilayer graphene near a magic angle, trilayer graphene-hexagonal boron nitride heterostructure \cite{high-order}, and Sr$_3$Ru$_2$O$_7$ \cite{multicritical2}. In particular, a power-law divergent DOS of high-order VHS with exponent $-1/4$ was found in scanning tunneling spectroscopy measurements \cite{Pasupathy} on magic-angle twisted bilayer graphene \cite{high-order}.

In the presence of electron-electron interaction, a large DOS near the Fermi level may have important consequences. On the one hand, it may trigger Stoner instability to ferromagnetism.  On the other hand, a large DOS may result in strong screening of repulsive interaction, so that a Fermi liquid description remains valid at low energy.
For the case of a single conventional VHS with a logarithmically divergent DOS at the Fermi energy, previous works \cite{Dzyaloshinskii,Schulz,Lederer,Gonzalez3,Furukawa,Irkhin,Kampf,LeHur,Raghu,Nandkishore,Gonzalez,Katsnelson,Kallin,Kapustin,Isobe} have shown that repulsive interaction decreases at low energies, likely leading to a marginal Fermi liquid \cite{marginal,Pattnaik,Gopalan,Dzyaloshinskii2,Menashe}.

In this work, we study interacting electron systems with a high-order saddle point near the Fermi level.
Assuming that electron interaction is weak, dominant contributions to low-energy thermodynamic properties of the system come from those states in the vicinity of the saddle point, from which the DOS divergence originates. This allows us to formulate a {\it continuum} field theory of interacting fermions by taking the leading-order energy dispersion relation $E_{\bm{k}}$ near the saddle point and extending the range of momentum to infinity.

In this field theory, when the high-order VHS is right at the Fermi level, the Fermi surface in $\bm{k}$-space becomes {\it scale-invariant}. As the VHS approaches the Fermi level, charge and spin susceptibilities exhibit power-law divergence,  reminiscent of critical phenomena.
Motivated by these observations, we develop a renormalization group (RG) theory for interacting fermions near high-order VHS, which parallels Wilson--Fisher RG approach to the $\phi^4$ theory \cite{Wilson-Fisher,Wilson}.
By introducing a small parameter $\epsilon$ associated with the DOS divergence, we present a {\it controlled} RG analysis and find that short-range repulsive interaction is relevant at the noninteracting fixed point and drives the system into a nontrivial $T=0$ interacting fixed point. The former is the analog of the Gaussian fixed point in Fermi systems, and the latter the analog of the Wilson--Fisher fixed point.

The metallic state at the interacting fixed point exhibits scale-invariance in space/time and power-law divergent charge and spin susceptibility, but finite pairing susceptibility. In other words, this is a metal on the verge of charge separation and ferromagnetism.
We call such a critical state of metal with various divergent susceptibilities but without any long-range order, a \textit{supermetal}.
This terminology is motivated by a comparison with a metal and a semimetal.  All three are conductors without a band gap at the Fermi level, but differ in the DOS.  A semimetal has a vanishing DOS, a metal has a finite DOS, and a supermetal has a divergent DOS.

We further show by a two-loop RG calculation for a high-order saddle point that the fermion field acquires a finite anomalous dimension. Hence the interacting supermetal we found is a non-Fermi liquid, as opposed to a  marginal  Fermi liquid for the case of a conventional VHS.
The singular DOS of supermetal $D(E) \propto |E|^{-\epsilon}$  plays a pivotal role by making a non-Fermi liquid possible under \textit{weak} repulsive interaction. The DOS exponent $\epsilon$ naturally serves as a small parameter that allows a controlled analysis via perturbative RG calculation.

The outline of the paper is as follows:
In Sec.~\ref{sec:model}, we introduce a tight-binding model with a high-order VHS and calculate the power-law divergent DOS, whose exponent is determined from the scaling property of energy dispersion near the high-order saddle point.
We show that a high-order VHS appears generically when the energy dispersion around a saddle point is modified by changing just a single hopping parameter.

In Sec.~\ref{sec:mean-field}, we present a mean-field analysis of interacting electrons with  a high-order saddle point near Fermi level.  We find that in the presence of repulsive contact interaction, as the chemical potential approaches the Van Hove energy, a first-order transition to a ferromagnetic metal occurs, displaying a  discontinuous change in spin polarization and charge density.

In Sec.~\ref{sec:energy-shell}, we perform the energy-shell RG analysis step by step.  We first define the energy shell as a region of momentum space.  Then, the tree-level and one-loop RG equations for the chemical potential and interaction strength are derived in sequence, which resembles the case of the $\phi^4$ theory.  We identify the noninteracting fixed point and the nontrivial interacting fixed point, which is the analog of the Wilson--Fisher fixed point in Fermi system.  We next consider other relevant perturbations to the system, including Zeeman and pairing fields as well as additional symmetry-allowed terms in the energy dispersion.  A discussion about a higher-loop RG analysis follows, while an actual two-loop calculation appears in a later section.

In Sec.~\ref{sec:analysis}, we combine the results from the mean-field and RG analyses to propose a phase diagram of interacting electrons near a high-order VHS in the parameter space of chemical potential, interaction strength, and detuning of single-particle energy dispersion from the high-order VHS.
We show that a supermetal appears on a line in the phase diagram, which can be reached by tuning two parameters.
We then perform the scaling analysis for thermodynamic quantities and correlation functions.  The generic formalism is first presented, followed by the one-loop result for various exponents of divergent susceptibilities.
In addition, we discuss the Ward identity, which results from charge conservation and gives relations among the field renormalization and scaling exponents.

In Sec.~\ref{sec:field_theory}, we introduce another RG scheme, the field theory approach with a soft UV energy cutoff, which is confirmed to satisfy the Ward identity.
Compared to the energy-shell RG analysis, it has the advantage in calculating higher-order perturbative corrections.  The one-loop calculation reproduces the energy-shell RG analysis in Sec.~\ref{sec:energy-shell}.  Furthermore, the two-loop calculation shows the finite anomalous dimension of the fermion field at a high-order saddle point.  This result directly establishes the non-Fermi liquid nature of an interacting supermetal.

In Sec.~\ref{sec:lifetime}, we evaluate the quasiparticle lifetime at finite temperature due to electron interaction.  From a perturbative calculation, we find an unusual temperature dependence in the quasiparticle lifetime, which also implies the non-Fermi liquid behavior.

In Sec.~\ref{sec:discussions}, we summarize the results and discuss their significance in the broad context of Van Hove physics, RG approaches to Fermi systems, and non-Fermi liquids. We compare interacting supermetal with other non-Fermi liquid systems, such as one-dimensional systems \cite{Tomonaga,Luttinger,Haldane}, quantum critical metals \cite{NFL1,NFL1a,review2,NFL2,Hertz,Moriya,Millis,Pankov,Rech,Senthil2,Mross,Lee1,Fitzpatrick1,Fitzpatrick2,NFL3,Berg,Abrikosov1,Abrikosov2,Gonzalez2,DasSarma,Son,Moon,Savary,Isobe2,Cho}, and doped Mott insulators \cite{review1,review_LNW,Kivelson1,Fradkin,Emery,orthogonal,Fisher}.
We also discuss experimental signatures of a supermetal.

\section{Model}
\label{sec:model}

\subsection{An example of high-order VHS in two dimensions}
\label{sec:tight-binding}

We consider a tight-binding model on an anisotropic square lattice
\begin{align}
\hat{H} = - \sum_j \left( t_x c_{j+\hat{x}}^\dagger c_j + t_y c_{j+\hat{y}}^\dagger c_j + t'_y c_{j+2\hat{y}}^\dagger c_j \right) + \text{H.c.} \label{Htb}
\end{align}
$t_x$ and $t_y$ are the nearest-neighbor hopping amplitudes along the $x$ and $y$ directions, respectively, and $t'_y$ is the second-nearest neighbor hopping along the $y$ direction.  The energy dispersion is obtained as
\begin{equation}
\label{eq:tb_dispersion_full}
E_{\bm{k}} = -2t_x \cos(k_xa) -2t_y \cos(k_ya) -2t'_y \cos(2k_ya),
\end{equation}
with the lattice constant $a$.

\begin{figure}
\centering
\includegraphics[width=\hsize]{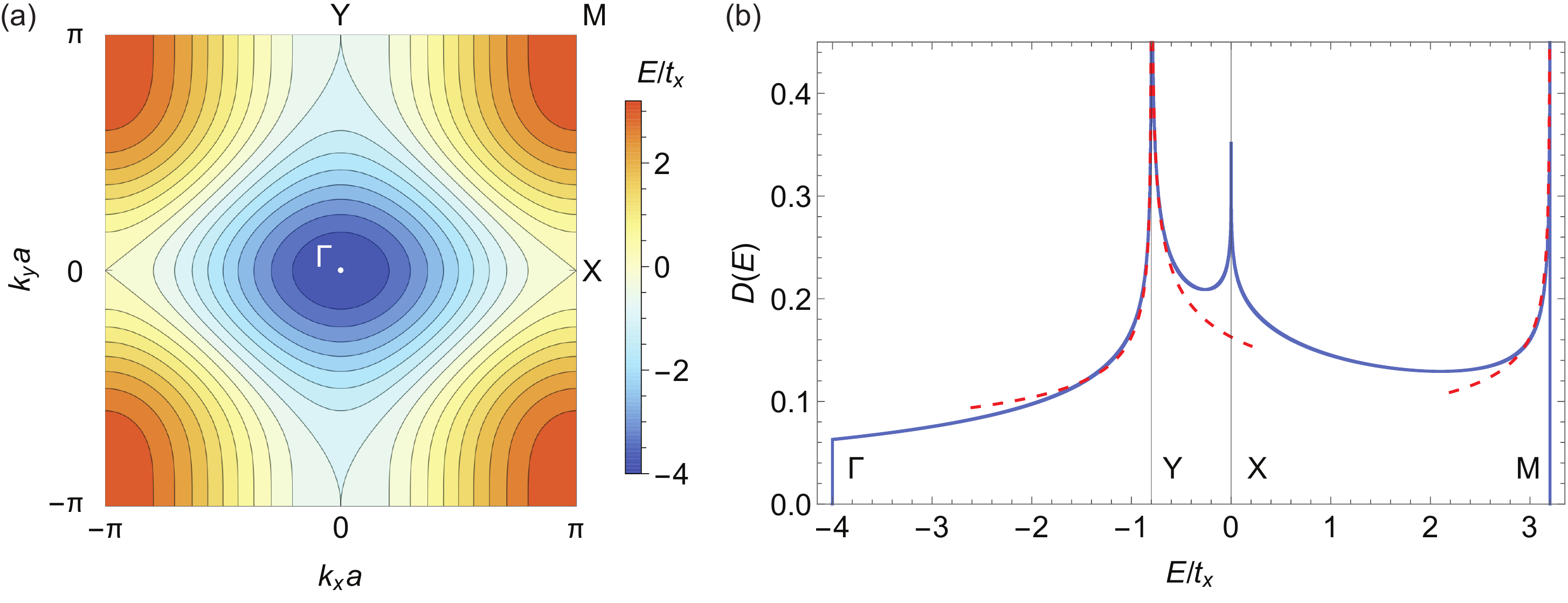}
\caption{
Lattice model for a high-order VHS.
(a) Energy contour plot with $t_y/t_x=0.8$ and $t'_y/t_x=0.2$. We find the energy minimum at $\Gamma$, the maximum at $M$, and the two saddle points at $X$ and $Y$. $Y$ and $M$ are high-order VHS points.  At $Y$, we see that the two Fermi surfaces touch tangentially while they cross linearly at $X$.
(b) DOS for the energy dispersion in (a).  The four VHS points give rise to analytic singularities in the DOS, where the corresponding points are labeled in the figure.
The two peaks at $Y$ and $M$ correspond to high-order VHS, fitted by the analytic formula for the continuum theory Eq.~\eqref{eq:DOS}.
}
\label{fig:tb}
\end{figure}

For $|t_y| \geq 4|t'_y|$, there are four VHS points in the Brillouin zone at the high symmetry points: $\Gamma=(0,0)$, $X=(\pi/a,0)$, $Y=(0,\pi/a)$, and $M=(\pi/a,\pi/a)$.
With $t_x$, $t_y$, $t'_y>0$, the energy minimum and maximum are located at $\Gamma$ and $M$ points, respectively, and $X$ and $Y$ points are the saddle points [Fig.~\ref{fig:tb}(a)].
Near $Y$ point, the energy dispersions takes the form
\begin{equation}
\label{x2y4}
E_{\bm{k}} = k_x^2 - k_y^4 - \lambda k_y^2,
\end{equation}
where $k_x$ and $k_y$ are rescaled to eliminate the coefficients of $k_x^2$ and $k_y^4$.

\begin{figure}
\centering
\includegraphics[width=\hsize]{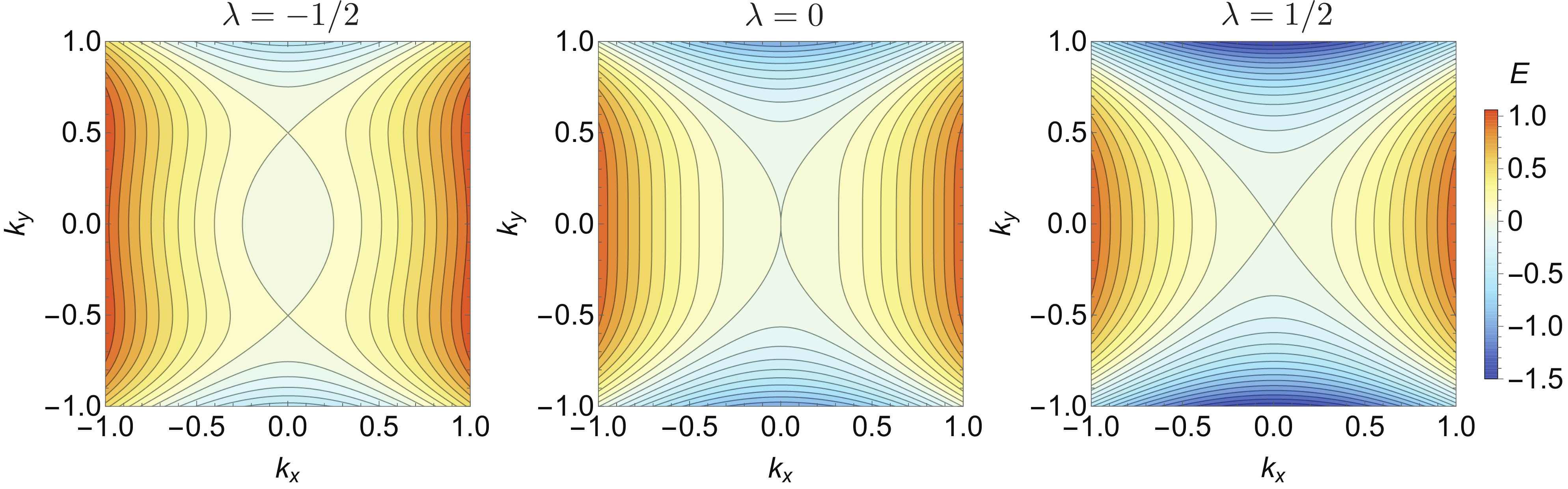}
\caption{
Energy contours for the dispersion Eq.~\eqref{x2y4}.  The parameter $\lambda$ describes a perturbation around a high-order saddle point.  The sign of $\lambda$ controls the topology of the Fermi surface; there is one VHS point at $(k_x, k_y) =(0,0)$ for $\lambda \geq 0$, and it splits into two located at $(0, \pm \sqrt{|\lambda|/2})$ for $\lambda<0$.
}
\label{fig:perturbation}
\end{figure}

The evolution of the Fermi surface by changing $\lambda$ is shown in Fig.~\ref{fig:perturbation}.  For $\lambda>0$ $(t_y>4t'_y)$, there is an ordinary saddle point with a logarithmically divergent DOS, where the Fermi surfaces cross at a point at $\mu=0$.  At $\lambda=0$ $(t_y=4t'_y)$, the two Fermi surfaces touch tangentially to realize a high-order saddle point.  For $\lambda<0$ $(t_y<4t'_y)$, the singular point splits into two saddle points and one minimum.  Those saddle points are located at $(k_x,k_y)=(0,\pm\sqrt{|\lambda|/2})$ with the energy $\lambda^2/4$.
We can see that the high-order VHS is realized around the conventional VHS point(s) by controlling the single parameter $\lambda$ in the energy dispersion \cite{high-order}.
We add that, at $t_y = 4t'_y$, the energy dispersion near $M$ point becomes $k_x^2 + k_y^4$, describing a high-order energy extremum.

The specific tight-binding model (\ref{Htb}) illustrates a general feature of Bloch electron's energy dispersion: the existence of saddle points is mathematically guaranteed \cite{VanHove}, and tuning a single parameter can turn an ordinary saddle point into a high-order one \cite{high-order}.

A VHS manifests itself as an analytic singularity in the DOS
\begin{equation}
\label{eq:DOS_formula}
D(E) = \int_{\bm{k}} \delta(E-E_{\bm{k}}),
\end{equation}
where $\int_{\bm{k}}=\int\frac{d^dk}{(2\pi)^d}$ stands for the momentum integration in $d$ dimensions.  The DOS for the present model $(d=2)$ is depicted in Fig.~\ref{fig:tb}(b).
We find four singularities in the DOS and each of them is tied to the individual VHS of the model.  The band bottom at $\Gamma$ gives rise to a discontinuity in the DOS and the saddle point at $X$ shows a logarithmic divergence in the DOS.  Those two are conventional VHS, known since the original work of Van Hove~\cite{VanHove}.
Here we focus on the high-order VHS at $Y$ and $M$. They exhibit distinct behavior: the DOS has a power-law divergence as $|E|^{-1/4}$ instead of a logarithm.  In addition, the divergence at $Y$ is stronger on the electron side by the factor $\sqrt{2}$ than on the hole side.  Such an asymmetry is not seen for a conventional VHS with a logarithmic divergence at $X$.
The two Fermi surfaces touch tangentially at $Y$ at the Van Hove energy.  When the chemical potential $\mu$ crosses the Van Hove energy, the Fermi surface topology changes from being closed to open in the $k_y$ direction.

In Fig.~\ref{fig:tb}(b), the DOS peaks at the two high-order VHS in our tight-binding model are fitted by the analytical expressions of the DOS calculated from the continuum models in their vicinities. The calculation will be shown in the next subsection.
We can see a close fit within a finite energy range. Since the divergent DOS and hence susceptibilities originate from the vicinity of the high-order VHS, the continuum model is expected to capture universal features at low energy.
Using the continuum model has the advantage of removing non-universal aspects associated with high-energy regions away from the high-order VHS in the tight-binding model. We will show that infrared (IR) scaling properties are not indeed affected by the UV cutoff in the continuum model.

Before proceeding, we briefly mention the Fermi surfaces in strained Sr$_2$RuO$_4$ \cite{ruthenate1,ruthenate2,ruthenate3}.  It has a quasi-two-dimensional electronic structure with a layered perovskite structure.  Under uniaxial pressure, a Lifshitz transition occurs on the Brillouin zone boundary \cite{ruthenate2}.  At the transition point, there is one VHS in the Brillouin zone at the Fermi energy.  The Fermi surface of the band of interest resembles the one obtained from Eq.~\eqref{eq:tb_dispersion_full}.

\subsection{Generalization}
\label{sec:generalization}

From now on, we study a continuum model of fermions with a high-order energy dispersion.
For the purpose of a controlled RG analysis later, here we consider the generalized energy dispersion in the $d$-dimensional $\bm{k}$-space
\begin{equation}
\label{eq:model}
E_{\bm{k}} = A_+ k_+^{n_+}- A_- k_-^{n_-}.
\end{equation}
The momentum is denoted by
\begin{equation}
\bm{k} = (\bm{k}_+, \bm{k}_-),
\end{equation}
where $\bm{k}_\pm$ are $d_\pm$-dimensional vectors with $d_++d_-=d$, and $k_\pm = |\bm{k}_\pm|$.
Analyticity of the energy dispersion requires $n_\pm$ to be positive integers.
We consider the case of even $n_\pm$, so that $E_{\bm{k}} = E_{-\bm{k}}$ satisfies time-reversal symmetry.
When at least one of $n_\pm$ is greater than two, this energy dispersion has a high-order VHS at $k=0$, which is defined as a point where the Hessian matrix $D_{ij} =\partial_{k_i} \partial_{k_j} E_{\bm{k}}$ fulfills $\det D_{ij}=0$.

The energy dispersion Eq.~\eqref{eq:model} follows the scaling relation
\begin{equation}
\label{eq:scaling_E}
E_{\bm{k}} = b E_{\bm{k}'} \text{ with } \bm{k}'=(\bm{k}_+/b^{1/n_+},\bm{k}_-/b^{1/n_-}).
\end{equation}
It then follows from Eqs.~\eqref{eq:DOS_formula} and \eqref{eq:scaling_E} that the DOS satisfies
\begin{gather}
\label{eq:DOS}
D(E) =
\begin{cases}
D_+ E^{-\epsilon} & (E>0) \\
D_- (-E)^{-\epsilon} & (E<0),
\end{cases}
\end{gather}
where the DOS singularity exponent $\epsilon$ is
\begin{equation}
\epsilon = 1-\frac{d_+}{n_+}-\frac{d_-}{n_-}.
\end{equation}
Throughout this work, we consider the case $\epsilon>0$.
For example, the high-order VHS introduced in the preceding section corresponds to the case of $d_+ = d_-=1$, $n_+=2$, $n_-=4$, so that $\epsilon = 1/4$.

We calculate the prefactors $D_\pm$ for the dispersion Eq.~\eqref{eq:model} explicitly and find
\begin{subequations}
\label{eq:DOS_prefactor}
\begin{gather}
D_s = D_0 \sin\left(\frac{\pi d_s}{n_s}\right) \quad (s=\pm),
\end{gather}
with the common factor
\begin{equation}
D_0 = \frac{4\Gamma(\epsilon)}{\pi(4\pi)^{d/2}} \prod_{s=\pm} \frac{\Gamma\left(d_s/n_s\right)}{n_s A_s^{d_s/n_s} \Gamma\left(d_s/2\right)}.
\end{equation}
\end{subequations}
We note that in calculating the DOS, the $d$-dimensional momentum integral over $\bm k\in (-\infty, \infty)^d$ is convergent for all $E \neq 0$.
Also, note that $D_+ \neq D_-$ for $d_+/n_+ \neq d_-/n_-$.  It describes the asymmetry in the DOS above and below $E=0$. This is a feature of the high-order saddle points defined by Eq.~\eqref{eq:model}, distinct from conventional saddle points in two dimensions, where the logarithmically divergent DOS peak is symmetric.

We also find it useful to consider another generalization
\begin{equation}
\label{eq:model_2}
E_{\bm{k}} = A_+ k_+^2 - A_- (k_-^2)^2,
\end{equation}
with $d_+ = 1$ and $d_- = 2-4\epsilon$. The original problem in two dimensions corresponds to $\epsilon=1/4$, while the generalized problem is defined in $3-4\epsilon$ dimensions, in a similar spirit as Wilson--Fisher theory in $4-\epsilon$ dimension. Now, the DOS $D(E)$ has a power-law divergence at $E=0$ for $\epsilon>0$ with the same form as Eq.~\eqref{eq:DOS}, but the coefficients are replaced with
\begin{subequations}
\label{eq:DOS_prefactor_2}
\begin{gather}
D_+ = \frac{1}{(2\pi)^{d/2}} \sqrt{\frac{2}{\pi}} \frac{\Gamma(\epsilon)}{\Gamma(1-\epsilon)} \frac{1}{A_+^{1/4} A_-^{d_-/4}}, \\
D_- = D_+ \cos (\pi\epsilon).
\end{gather}
\end{subequations}

The nontrivial interacting fixed point to be shown later is controlled by the smallness of $\epsilon$.
For the model defined by Eq.~\eqref{eq:model}, the exponent can be any rational number between $0<\epsilon<1$.
By choosing positive integers $n_\pm$ and $d_\pm$ judiciously, we can make $\epsilon$ arbitrarily small in high-dimensional crystals, while keeping the energy-momentum dispersion an analytic function.

We now introduce our model of interacting electrons near a high-order VHS:
\begin{align}
\label{eq:model_hamiltonian}
\hat{H} &= \int d^dr \left[ c_{\bm{r}\sigma}^\dagger (E_{-i\partial_{\bm{r}}} - \mu) c_{\bm{r}\sigma} + g\hat{n}_{\bm{r}\uparrow} \hat{n}_{\bm{r}\downarrow} \right]
\end{align}
with the density operator $\hat{n}_{\bm{r}\sigma} = c_{\bm{r}\sigma}^\dagger c_{\bm{r}\sigma}$.
$g$ denotes the coupling constant for the contact interaction between electrons with opposite spins and the summation over the spin index $\sigma=\uparrow (+)$, $\downarrow (-)$ is implicit.
The corresponding action is given by
\begin{align}
\label{eq:action}
S =&\int_0^{1/T} d\tau \int d^dr
\Big[ \bar{\psi}_{\sigma} (\partial_\tau + E_{-i\partial_{\bm{r}}} -\mu) \psi_{\sigma}
+ g \bar{\psi}_{\uparrow} \bar{\psi}_{\downarrow} \psi_{\downarrow} \psi_{\uparrow} \Big]
\end{align}
with the fermionic field $\psi_\sigma$.
We set $k_B=\hbar=1$ throughout the paper.
Here we formulate the model at temperature $T$.  Temperature $T$ is regarded as the system size $L_\beta \equiv 1/T$ in the imaginary time direction.
Later, we shall consider the effects of other interactions and external fields.

From the action, we define the noninteracting Green's function
\begin{equation}
G_0(\bm{k},\omega_n)= \frac{1}{i\omega_n - E_{\bm{k}}},
\end{equation}
with the fermionic Matsubara frequency $\omega_n=(2n+1)\pi T$ ($n$: integer).
The partition function $\mathcal{Z}$ is expressed as
\begin{equation}
\mathcal{Z} = \int D\bar{\psi}D\psi e^{-S}. \label{Z}
\end{equation}
We are interested in thermodynamic quantities such as specific heat. These are obtained from the free energy density
\begin{equation}
\label{eq:free_energy}
F=-\frac{T}{V}\ln \mathcal{Z},
\end{equation}
where $V$ is the volume of the system.

\section{Mean-field analysis}
\label{sec:mean-field}

\begin{figure*}
\centering
\includegraphics[width=\hsize]{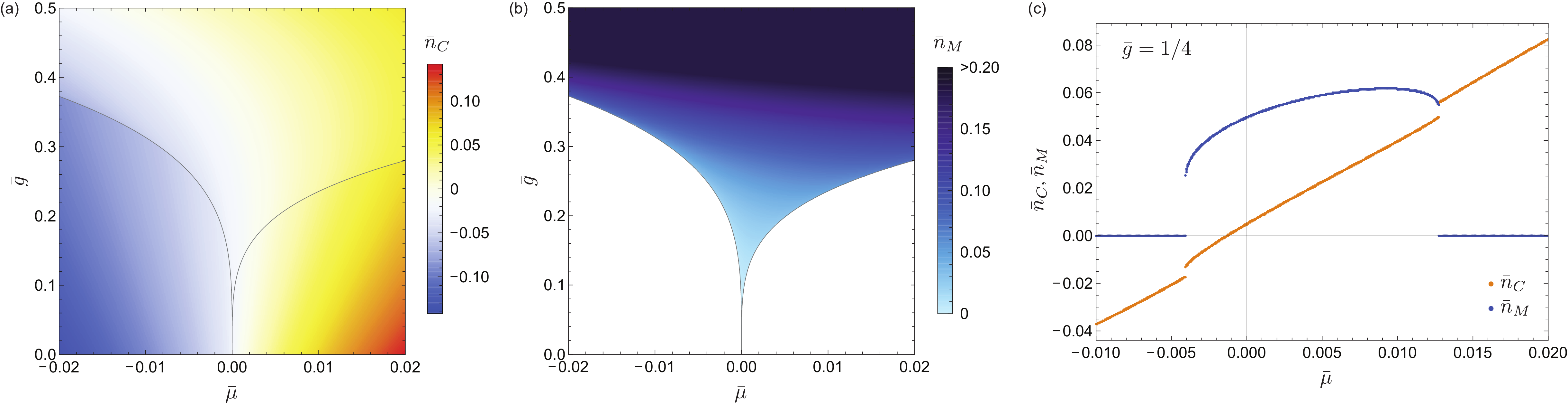}
\caption{
Mean-field results of the order parameters for (a) charge $\bar{n}_C$ and (b) magnetization $\bar{n}_M$ by changing $\bar{\mu}$ and $\bar{g}$.  We set $\epsilon=1/4$ and $\bar{D}_r=1/\sqrt{2}$.  We find discontinuities approximately along $\bar{g}_c(\bar{\mu}) \propto D^{-1}(\bar{\mu})$ and ferromagnetic states when the interaction is stronger than $\bar{g}_c$.  (c) The line cut with a fixed $\bar{g}$ shows that discontinuities in $\bar{n}_C$ and $\bar{n}_M$ occur at the same $\bar{\mu}$.
}
\label{fig:MF-phase}
\end{figure*}

We first consider the effect of interaction in Eq.~\eqref{eq:model_hamiltonian} at $T=0$ with a mean-field approximation.
We assume repulsive interaction $(g>0)$ and minimize the energy expectation value $\braket{\Psi_0|\hat{H}|\Psi_0}$ with the variational wave function given by
\begin{equation}
\label{eq:psi_var}
\ket{\Psi_0 (E_\uparrow,E_\downarrow)} = \prod_\sigma \prod_{\bm{k}\in W(E_\sigma)} c_{\bm{k}\sigma}^\dagger \ket{0}.
\end{equation}
This wave function has two independent variational parameters $E_\uparrow$ and $E_\downarrow$, corresponding to Fermi energies for spin-up and spin-down electrons, respectively.  $W(E_\sigma)$ denotes the region in the momentum space where the energy $E_{\bm{k}}$ is below the variational parameter $E_\sigma$: $W(E_\sigma) = \{\bm{k} | E_{\bm{k}}\leq E_\sigma \}$.  We note that for $g<0$ the system becomes unstable against pairing and hence the variational wave function Eq.~\eqref{eq:psi_var} is inapplicable.

The variational wave function $\ket{\Psi_0}$ gives the exact ground state at $g=0$ by choosing the two variational parameters $E_\uparrow = E_\downarrow = \mu$.
For $g \neq 0$, the energy expectation value becomes
\begin{align}
&\quad \braket{\Psi_0|\hat{H}|\Psi_0} \nonumber\\
&= \sum_\sigma \left[ \int_{\bm{k}\in W(E_\sigma)} E_{\bm{k}} -\mu n(E_\sigma) \right] + g n(E_\uparrow) n(E_\downarrow),
\end{align}
where the electron density for spin $\sigma$ at an energy $E_\sigma$ is given by
\begin{equation}
n(E_\sigma) = \int_{\bm{k}\in W(E_\sigma)} 1
= \int_{-\Lambda}^{E_\sigma} dE D(E).
\end{equation}
We introduce a lower bound in the energy integral, i.e., the UV cutoff $\Lambda(>0)$, which corresponds to the inverse of the microscopic lattice scale.
Since $D(E) > 0$, the electron density $n(E_\sigma)$ is a monotonic function of $E_\sigma$.  The one-to-one correspondence allows an inverse function of $n(E_\sigma)$; we define $n_\sigma \equiv n(E_\sigma) - n(0)$ to write $E_\sigma$ as a function of $n_\sigma$:
\begin{align}
E_\sigma(n_\sigma) =
\begin{cases}
\left( \dfrac{1-\epsilon}{D_+} n_\sigma \right)^{\frac{1}{1-\epsilon}} & (n_\sigma \geq 0) \\
-\left( \dfrac{1-\epsilon}{D_-} (-n_\sigma) \right)^{\frac{1}{1-\epsilon}} & (n_\sigma < 0).
\end{cases}
\end{align}

Now we can write the energy expectation value $\braket{\Psi_0|\hat{H}|\Psi_0}$ as a function of $n_\sigma$:
\begin{align}
\braket{\Psi_0|\hat{H}|\Psi_0}
&= \sum_\sigma \int_{-\Lambda}^{E_\sigma(n_\sigma)} dE D(E) E -\mu(n_\uparrow+n_\downarrow) \nonumber\\
&\quad + g [n_\uparrow+n(0)][n_\downarrow+n(0)] \nonumber\\
&= \sum_\sigma \left[ \Phi(n_\sigma) - \tilde{\mu} n_\sigma \right] + gn_\uparrow n_\downarrow + \text{const.},
\end{align}
where we introduce
\begin{gather}
\Phi(n_\sigma) = \int_0^{E_\sigma(n_\sigma)} dE D(E) E, \\
\tilde{\mu} = \mu - gn(0).
\end{gather}
It is convenient to express the energy expectation value with the dimensionless quantities defined by
\begin{gather}
\bar{g} = gD(\Lambda), \quad \bar{n}_\sigma = \frac{n_\sigma}{\Lambda D(\Lambda)}, \quad \bar{\mu} = \frac{\tilde{\mu}}{\Lambda}, \quad \bar{D}_r = \frac{D_-}{D_+}.
\end{gather}
Then, we obtain
\begin{align}
\braket{\Psi_0|\hat{H}|\Psi_0}
= \Lambda^{2-\epsilon} D_+ \bar{\mathcal{E}}(\bar{n}_\uparrow, \bar{n}_\downarrow) + \text{const.},
\end{align}
where the dimensionless function
\begin{align}
\bar{\mathcal{E}}(\bar{n}_\uparrow, \bar{n}_\downarrow) = \sum_\sigma \left[ \bar{\Phi}(\bar{n}_\sigma) -\bar{\mu}\bar{n}_\sigma \right] + \bar{g} \bar{n}_\uparrow \bar{n}_\downarrow
\end{align}
is to be minimized by varying $\bar{n}_\uparrow$ and $\bar{n}_\downarrow$.  The function $\bar{\Phi}(\bar{n}_\sigma)$ is given by
\begin{align}
\bar{\Phi}(\bar{n}_\sigma) =
\begin{cases}
\dfrac{(1-\epsilon)^{\frac{2-\epsilon}{1-\epsilon}}}{2-\epsilon} \bar{n}_\sigma^{\frac{2-\epsilon}{1-\epsilon}} & (\bar{n}_\sigma \geq 0) \\
\dfrac{(1-\epsilon)^{\frac{2-\epsilon}{1-\epsilon}}}{2-\epsilon} \bar{D}_r^{-\frac{1}{1-\epsilon}} (-\bar{n}_\sigma)^{\frac{2-\epsilon}{1-\epsilon}} & (\bar{n}_\sigma < 0).
\end{cases}
\end{align}

The electron densities $\bar{n}_\uparrow$ and $\bar{n}_\downarrow$ are order parameters in the mean-field analysis.  Instead of $\bar{n}_\uparrow$ and $\bar{n}_\downarrow$, we use
\begin{gather}
\bar{n}_C = n_\uparrow + n_\downarrow, \quad
\bar{n}_M = n_\uparrow - n_\downarrow,  	
\end{gather}
where they corresponds to the order parameters for charge and magnetization, respectively.  The values of $\bar{n}_C$ and $\bar{n}_M$ are obtained by minimizing the function $\bar{\mathcal{E}}(\bar{n}_\uparrow,\bar{n}_\downarrow)$ with the chemical potential $\bar{\mu}$ and the coupling constant $\bar{g}$ given.
The numerical result for $\epsilon=1/4$ and $\bar{D}_r=1/\sqrt{2}$ is shown in Fig.~\ref{fig:MF-phase}.  We find discontinuities in $\bar{n}_C$ and $\bar{n}_M$ at the same $\bar{\mu}$ and $\bar{g}$, which characterizes a first-order phase transition and defines a critical value of the coupling constant $\bar{g}_c(\bar{\mu})$.  Finite magnetization $\bar{n}_M$ above $\bar{g}_c$ characterizes a ferromagnetic state with the spin-rotational symmetry broken.
The phase boundary in the numerical result well obeys $\bar{g}_c(\bar{\mu}) \propto D^{-1}(\bar{\mu})$, as expected from Stoner criterion for ferromagnetism.

\section{Energy-shell RG analysis}
\label{sec:energy-shell}

The mean-field analysis in the previous section leads to a first-order transition to ferromagnetism with discontinuous changes of the charge density and magnetization. The ferromagnetic region shrinks and the discontinuity at the transition weakens as the interaction decreases. Nonetheless, in the mean-field theory, this transition occurs with infinitesimal repulsive interaction at $\bar{\mu}=0$ because of the divergent DOS at VHS. However, this is an artifact of the mean-field analysis that neglects long-wavelength fluctuations, which becomes increasingly important as the first-order transition becomes weaker. In this section, we perform an energy-shell RG analysis to study the role of these fluctuations near high-order VHS ($|\bar{\mu}|\ll 1$) with weak repulsion ($\bar{g} \ll 1$).

\subsection{Formalism at zero temperature}

Here, we adopt the Wilsonian approach to the RG equations for the action Eq.~\eqref{eq:action}.
For clarity, we consider first the action at $T=0$, where we will find fixed points.  Then, the Matsubara frequency becomes continuous $\omega_n \to \omega$, and the action is written with frequency $\omega$ and momentum $\bm{k}$ as
\begin{align}
S &= \int\frac{d\omega}{2\pi} \int_{\bm{k}} \bigg[ \bar{\psi}_\sigma(k) (-i\omega + E_{\bm{k}} -\mu) \psi_\sigma(k) \nonumber\\
&\quad + g \bigg( \prod_{j=1}^{4} \int\frac{d\omega_j}{2\pi} \int_{\bm{k}_j} \bigg) (2\pi)^{d+1} \delta (k_1+k_2-k_3-k_4) \nonumber\\
&\quad\ \times
\bar{\psi}_{\uparrow}(k_1) \bar{\psi}_{\downarrow}(k_2) \psi_{\downarrow}(k_3) \psi_{\uparrow}(k_4).
\end{align}
We introduce the shorthand notation $k = (\bm{k},\omega)$.

We impose a UV energy cutoff $\Lambda$ on this action to remove unphysical UV divergences that appear in electron density of the ground state, etc.
We note that the UV cutoff here is imposed on energy, but not on momentum directly.  The region in $\bm{k}$-space with $|E_{\bm{k}}| \leq \Lambda$ still extends to infinity.
Importantly, this UV cutoff does not affect universal scaling properties of IR fixed points in the analysis of high-order VHS, as we shall show.  The UV cutoff merely appears in the prefactors of IR scaling functions.

We use two different energy cutoff schemes in this paper: an energy shell with a hard cutoff and a soft energy cutoff.
The former scheme allows the Wilsonian RG approach, which offers a rather simple analysis and understanding.  The latter requires a field theoretical analysis, which is apparently complicated, but high-order perturbative corrections become more tractable.

\begin{figure}
\centering
\includegraphics[width=\hsize]{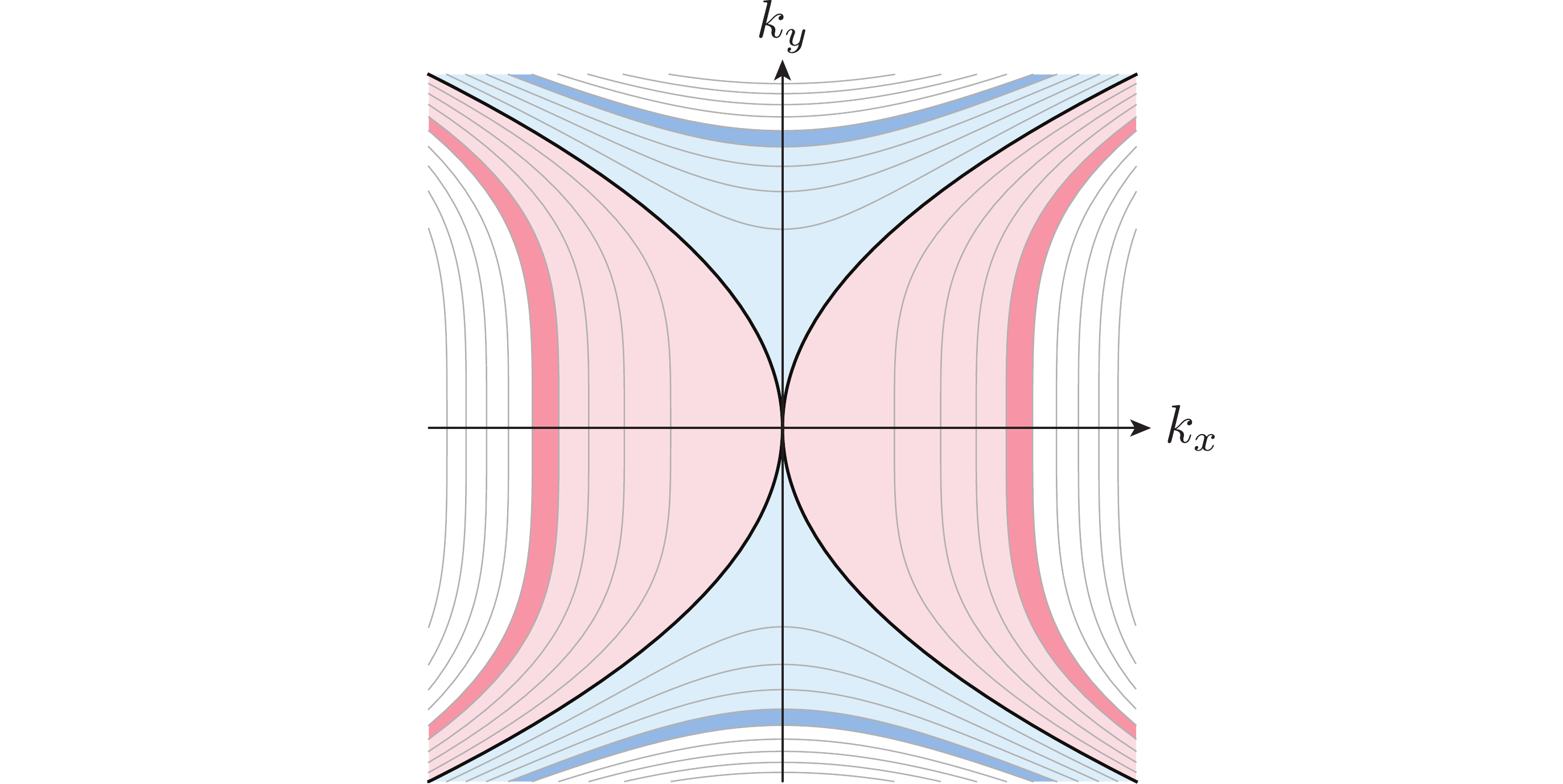}
\caption{
Energy contour plot for $E_{\bm{k}} = k_x^2 - k_y^4$.  The thick line is the Fermi surface at the Van Hove energy, which is scale-invariant.  The colored region has the energy inside the cutoff $\Lambda$, where the red (blue) area corresponds to $E>0$ ($E<0$).  At every RG step of the energy-shell RG scheme, high-energy modes within the energy shell shown in darker colors are integrated out.  In the field theory approach, all states below the cutoff $\Lambda$ are integrated over at once.
}
\label{fig:contour}
\end{figure}

This section focuses on the energy-shell RG scheme, which imposes a constraint on momentum integrals.  By converting the momentum integral to an energy integral with the help of the DOS, we write the momentum integral with the cutoff $\Lambda$ as
\begin{align}
\int_{\bm{k}}^\Lambda \mathcal{F}(E_{\bm{k}}) = \int_{-\Lambda}^{\Lambda} dE D(E) \mathcal{F}(E),
\end{align}
for an arbitrary function $\mathcal{F}$.
We denote the action with the energy cutoff $\Lambda$ as $S_\Lambda$, obtained by replacing the momentum integral $\int_{\bm{k}}$ by $\int_{\bm{k}}^\Lambda$.
The UV energy cutoff designates an unbounded region in $\bm{k}$-space, reflecting the extended Fermi surface with scale invariance (Fig.~\ref{fig:contour}).
Note that frequency integrals still range from $-\infty$ to $+\infty$.
In a high-order VHS, divergences of momentum integrals arise from a singularity at $k=0$ but not $k\to\infty$.
We will show that this simplifies the energy-shell RG analysis, which includes only the UV energy cutoff $\Lambda$.
This is in contrast to a conventional VHS with a logarithmic divergence of the DOS \cite{Kallin,Kapustin}; it additionally requires a UV momentum cutoff.  A further discussion can be found in Sec.~\ref{sec:discussions}.

We now sketch how an RG transformation works with the energy-shell RG scheme.  To access the IR behavior, we progressively eliminate UV modes and focus more on remaining modes.
In the energy-shell RG scheme, we first split the energy range into two parts; one corresponds to lower energies $E_{\bm{k}} \in [-\Lambda/b, \Lambda/b]$ and the other to higher energies $E_{\bm{k}} \in[-\Lambda,-\Lambda/b) \lor (\Lambda/b,\Lambda]$ $(b>1)$.  Accordingly, the fermion field $\psi$ is decomposed as
\begin{align}
\psi_\sigma(k) = \psi_\sigma^<(k) + \psi_\sigma^>(k),
\end{align}
where $\psi_\sigma^<$ represents the low-energy modes and $\psi_\sigma^>$ the high-energy modes.  We write a momentum integral in the same way:
\begin{equation}
\int_{\bm{k}}^\Lambda = \int_{\bm{k}}^< + \int_{\bm{k}}^>.
\end{equation}
Due to this division, the action is decomposed into the three parts as
\begin{equation}
S_\Lambda[\psi] = S^<[\psi^<] + S^>[\psi^>] + S^{<>}[\psi^<,\psi^>].
\end{equation}
The first term $S^<[\psi^<]$ consists only of the low-energy modes $\psi^>$ and the second term $S^>[\psi^>]$ of the high-energy modes $\psi^>$.  The last term $S^{<>}[\psi^<,\psi^>]$ describes the coupling of the low- and high-energy modes, which arises when the interaction is finite $(g \neq 0)$.
To obtain the effective action without the high-energy modes, we need to integrate out the high-energy modes:
\begin{align}
\label{eq:effective}
&\quad S_{\Lambda/b}[\psi^<] \nonumber\\
&=
S^<[\psi^<] - \ln \left( \int D\bar{\psi}^> D\psi^> e^{-S^>[\psi^>] -S^{<>}[\psi^<,\psi^>]} \right) \nonumber\\
&=
S^<[\psi^<] - \ln \left( \int D\bar{\psi}^> D\psi^> e^{-S^{<>}[\psi^<,\psi^>]} \right) + \text{const}.
\end{align}

Now the high-energy modes are eliminated and the new action has the smaller cutoff $\Lambda/b$.  One may be tempted to compare $S_\Lambda[\psi]$ and $S_{\Lambda/b}[\psi^<]$ to look into low-energy properties.  However, it is like ``comparing apples to oranges'' \cite{Shankar} as the two actions are defined in different domains.  For a fair comparison, we should make a change of variables ($\bm{k}$, $\omega$, and $\psi$) to restore the  cutoff $\Lambda$.  This procedure, called rescaling, completes the RG step.
It results in the change of parameters in the model, which is described by RG equations.

The RG equations describe the flow of the parameters under a scale transformation.
When the parameters do not change under a scale transformation, the system reaches an RG fixed point and exhibits scale-invariant properties.
Away from a fixed point, the parameters flow.  If the flow converges to a fixed point in its vicinity, then the fixed point is called a stable fixed point.  If the parameters flow away from a fixed point, then it is an unstable fixed point.
The RG equations also tell us how various susceptibilities and correlation lengths diverge as the critical point is approached, and the scaling properties of correlation functions  at the critical point.

\subsection{Tree-level analysis}

The mixing term $S^{<>}$ can be calculated by expanding the logarithm in powers of the coupling constant $g$.  We first consider the zeroth-order contribution in $g$.
Since the remaining terms are described by tree diagrams without loops, the approximation is referred to as the tree-level analysis.

At tree-level, the effective action with the cutoff $\Lambda/b$ becomes $S_{\Lambda/b}[\psi^<] = S^<[\psi^<]$.  To compare with $S_\Lambda[\psi]$, we need to change the variables to put the cutoff $\Lambda/b$ back to $\Lambda$.  Now we change the variables so that the energy satisfies the relation
\begin{subequations}
\begin{equation}
\label{eq:rescaling_energy}
E_{\bm{k}'} = bE_{\bm{k}}.
\end{equation}
For the energy dispersion given by Eq.~\eqref{eq:model}, this immediately leads to rescaling of the momentum
\begin{equation}
\label{eq:rescaling_momentum}
k_+' = b^{1/n_+} k_+, \quad
k_-' = b^{1/n_-} k_-,
\end{equation}
while the coefficients do not change:
\begin{gather}
A'_+ = A_+, \quad A'_-=A_-.
\end{gather}
\end{subequations}
To retain the form of the action, we also need to rescale the field $\psi$, frequency $\omega$, chemical potential $\mu$, and coupling constant $g$ to be
\begin{subequations}
\begin{gather}
\psi' = b^{-(3-\epsilon)/2} \psi^<, \\
\label{eq:omega_tree}
\omega' = b \omega, \\
\label{eq:mu_tree}
\mu' = b \mu, \\
\label{eq:g_tree}
g' = b^{\epsilon} g.
\end{gather}
\end{subequations}

When we look at the parameters of the model, the chemical potential $\mu$ and the coupling constant $g$ change after an RG step, whereas the coefficients of the energy dispersion $A_\pm$ do not.  The flow of an parameter under an infinitesimal scale transformation $(b\to 1)$ is described by a differential equation, namely the RG equation.  For $\mu$ and $g$, the RG equations are obtained from Eqs.~\eqref{eq:mu_tree} and \eqref{eq:g_tree}:
\begin{gather}
\label{eq:RG_tree}
\frac{d\mu}{dl} = \mu, \quad
\frac{dg}{dl} = \epsilon g.
\end{gather}
with $l = \ln b$.

In the present case, we find the noninteracting fixed point at $\mu=g=0$ in Eq.~\eqref{eq:RG_tree}, where the partition function takes a functional form of the Gaussian integral. If the parameters are away from the fixed point, they grow as $l$ increases i.e., in low energies, and flow away from the fixed point.  Therefore, the fixed point at $\mu=g=0$ is unstable and both $\mu$ and $g$ are relevant perturbations to the unstable fixed point.

So far we have only considered the contact interaction.  However, electron-electron interactions can take a more complicated form.  Other types of interactions will be generated under RG  even if not present initially, and thus their effects should be considered as well. In general, a finite-range interaction can be expanded in powers of spatial derivatives, with contact interaction being the lowest order term.  The next leading term $g_- (\bar{\psi}_\uparrow \partial_{r_-} \bar{\psi}_\downarrow) (\psi_\downarrow \partial_{r_-} \psi_\uparrow)$ contains two spatial derivatives, and has a different scaling relation: $g_-' = b^{\epsilon-2/n_-} g_-$, which has a much smaller exponent than $\epsilon$ for the contact interaction.
As an example, for the energy dispersion (\ref{x2y4}) in two dimensions, we have  $\epsilon=1/4$ and $n_-=4$, so that $g_-$ is irrelevant.
It is therefore legitimate to retain only the contact interaction in RG analysis.

\subsection{One-loop analysis}
\label{sec:energy-shell_one-loop}

In the presence of interaction, elimination of the high-energy modes gives rise to corrections in the effective action through the mixing of low- and high-energy modes in $S^{<>}[\psi^<,\psi^>]$.  When depicted diagrammatically, $S^{<>}[\psi^<,\psi^>]$ involves diagrams with loops, corresponding to integrations of the high-energy modes.  We here consider perturbative corrections to one-loop order.

The effective action Eq.~\eqref{eq:effective} can be calculated perturbatively with respect to the coupling constant $g$ when it is small.  We also treat the chemical potential $\mu$ as a perturbation as we are interested in critical phenomena where there is no characteristic scale in the system.
Including perturbative corrections, we write down the action in the form
\begin{align}
\label{eq:action_one-loop}
&\quad S_{\Lambda/b}[\psi^<] \nonumber\\
&= \int\frac{d\omega}{2\pi} \int_{\bm{k}}^< \bar{\psi}^<_\sigma(k) (-i\omega + E_{\bm{k}} -\mu +\Sigma) \psi^<_\sigma(k) \nonumber\\
&\quad + (g+\delta g) \bigg( \prod_{j} \int\frac{d\omega_j}{2\pi} \int_{\bm{k}_j}^< \bigg) \nonumber\\
&\quad\ \times (2\pi)^{d+1} \delta (k_1+k_2-k_3-k_4) \nonumber\\
&\quad\ \times
\bar{\psi}^<_{\uparrow}(k_1) \bar{\psi}^<_{\downarrow}(k_2) \psi^<_{\downarrow}(k_3) \psi^<_{\uparrow}(k_4) \nonumber\\
&\quad+ \cdots,
\end{align}
where $\delta g$ is a correction to the coupling constant and $(\cdots)$ consists of interactions with derivatives that may be generated after integrating out the high-energy modes.  As we have discussed above, finite-range interactions are irrelevant, so that we can safely neglect them.

\begin{figure}
\centering
\includegraphics[width=\hsize]{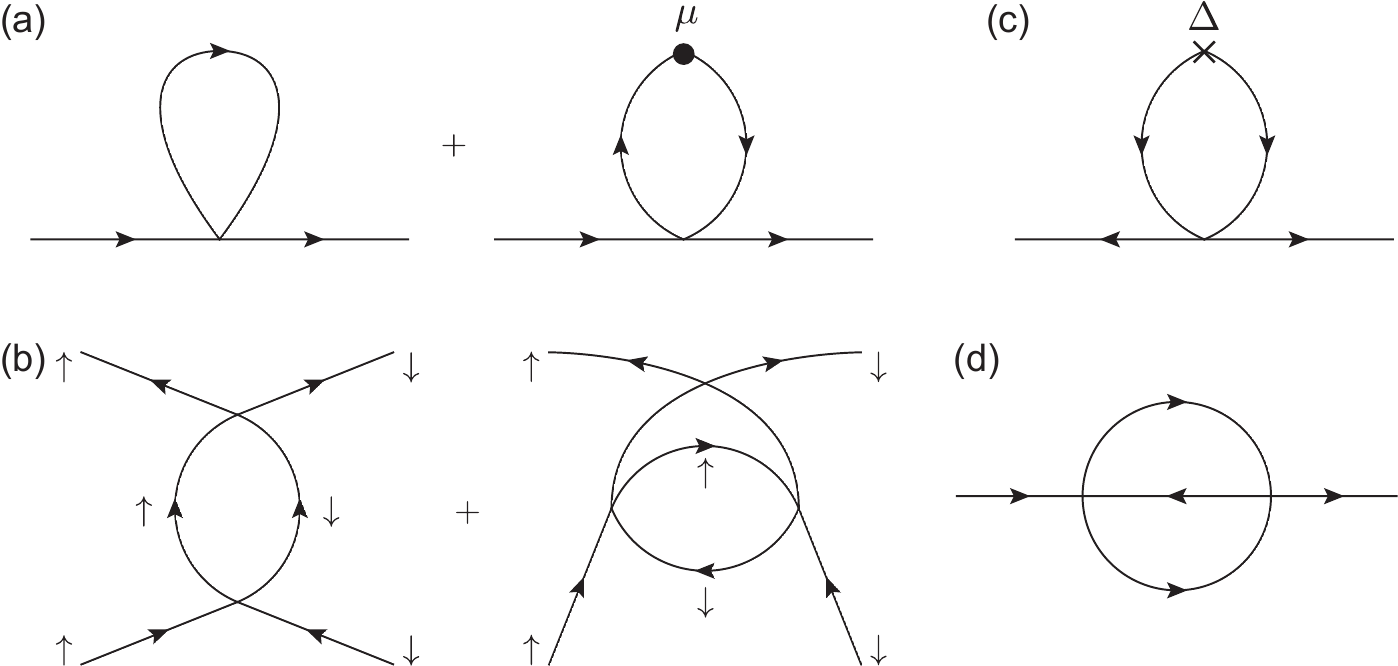}
\caption{
Diagrammatic representation of perturbative corrections.
The solid lines with arrows are the noninteracting electron propagators $G_0$. Each vertex corresponds to the contact interaction with the coupling constant $g$.
(a) Self-energy $\Sigma$ to one-loop order.  The first term represents the Hartree term and the second shows the one-loop correction linear in the chemical potential $\mu$.
(b) Correction to the coupling constant $\delta g$.  There are particle-particle (left) and particle-hole (right) contributions.
(c) One-loop correction to the pairing field $\Delta$.
(d) Two-loop correction to the self-energy, which gives rise to the finite field renormalization, and thus to the anomalous dimension.
}
\label{fig:loop}
\end{figure}

Perturbative corrections to the lowest order, namely to one-loop order, are diagrammatically depicted in Fig.~\ref{fig:loop}(a) and (b), corresponding to $\Sigma$ and $\delta g$, respectively.
We find that the one-loop corrections to the self-energy $\Sigma$ and the coupling constant $\delta g$ can be written as
\begin{subequations}
\label{eq:one-loop_sigma_g}
\begin{gather}
\label{eq:sigma_t0}
\Sigma = g\Sigma_\text{H} - g\mu \Pi_\text{ph}, \\
\label{eq:delta_g_t0}
\delta g = -g^2 (\Pi_\text{pp}+\Pi_\text{ph}).
\end{gather}
\end{subequations}
We emphasize that the all loop corrections should be evaluated at zero external frequency and momentum.
The one-loop corrections are obtained to $O(l)$ as
\begin{subequations}
\label{eq:one-loop_T0}
\begin{gather}
\Sigma_\text{H} = \int\frac{d\omega}{2\pi} \int_{\bm{k}}^> G_0(\bm{k},\omega) \simeq -l c_\text{H} \Lambda D(\Lambda) , \\
\label{eq:one-loop_T0_pp}
\Pi_{\text{pp}} = \int\frac{d\omega}{2\pi} \int_{\bm{k}}^> G_0(\bm{k},\omega) G_0(-\bm{k},-\omega) \simeq l c_\text{pp} D(\Lambda) , \\
\Pi_\text{ph} = \int\frac{d\omega}{2\pi} \int_{\bm{k}}^> G_0(\bm{k},\omega) G_0(\bm{k},\omega) =0,
\end{gather}
\end{subequations}
where $D(\Lambda)$ is the DOS at the cutoff energy and the dimensionless constants $c_\text{H}$ and $c_\text{pp}$ are
\begin{subequations}
\begin{gather}
\label{eq:c_H_0}
c_\text{H} =  \frac{1}{2} \left( 1-\frac{D_-}{D_+} \right), \\
\label{eq:c_pp_0}
c_\text{pp} = \frac{1}{2} \left( 1+\frac{D_-}{D_+} \right).
\end{gather}
\end{subequations}
We can see that the particle-hole contribution vanishes identically after the frequency integration, i.e., at $T=0$ there is no particle-hole screening coming from states near the cutoff energy $\Lambda$.
On the other hand, the particle-particle loop has a finite contribution.  The Hartree contribution $\Sigma_\text{H}$ can be finite only when the DOS is asymmetric on the electron and hole side $(D_+ \neq D_-)$, leading to a finite $c_\text{H}$ at most of order $\epsilon$.

There is no frequency or momentum dependence in the self-energy to one-loop order, so that the self-energy only renormalizes the chemical potential $\mu$.  The field renormalization or renormalization of the energy dispersion does not appear at one-loop order.  They appear at two-loop order from the diagram shown in Fig.~\ref{fig:loop}(d), which will be examined with the field theory approach in Sec.~\ref{sec:field_theory}.

With the one-loop corrections obtained, the new parameters $\mu'$ and $g'$ after rescaling are
\begin{subequations}
\begin{gather}
\mu' = b (\mu-\Sigma) \simeq b [\mu + lc_\text{H}\Lambda gD(\Lambda)], \\
g' = b^\epsilon (g + \delta g) \simeq b^\epsilon g [ g- l c_\text{pp} g^2 D(\Lambda) ],
\end{gather}
\end{subequations}
which lead to the RG equations for the chemical potential $\mu$ and the coupling constant $g$.
It is convenient to define the dimensionless chemical potential $\bar{\mu}$ and coupling constant $\bar{g}$ as
\begin{gather}
\label{eq:dimensionless_mu_g}
\bar{\mu} = \frac{\mu}{\Lambda}, \quad
\bar{g} = g D(\Lambda).
\end{gather}
Then, we obtain RG equations for $\bar{\mu}$ and $\bar{g}$ as
\begin{subequations}
\label{eq:RG_mu_g_0}
\begin{gather}
\label{eq:RG_mu_0}
\frac{d\bar{\mu}}{dl} = \bar{\mu} + c_\text{H} \bar{g}, \\
\label{eq:RG_g_0}
\frac{d\bar{g}}{dl} = \epsilon \bar{g} - c_\text{pp} \bar{g}^2.
\end{gather}
\end{subequations}

\begin{figure}
\centering
\includegraphics[width=0.75\hsize]{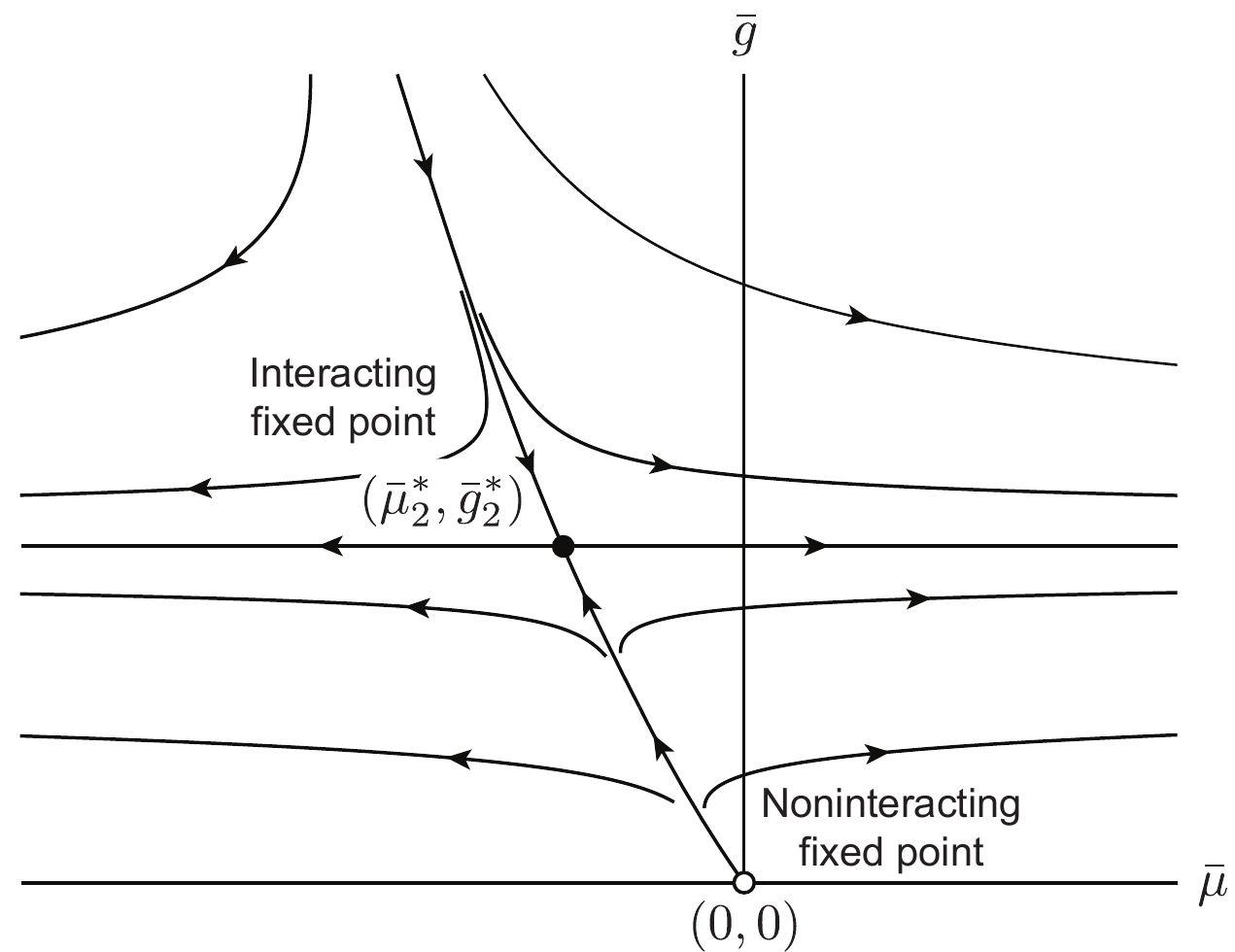}
\caption{
RG flow of the chemical potential $\bar{\mu}$ and the coupling constant $\bar{g}$.  There are two fixed points: the fixed point with $\bar{g}_1^*=0$ corresponds to the noninteracting fixed point and the other with $\bar{g}_2^*=\epsilon/c_\text{pp}$ to the nontrivial interacting fixed point.  The interacting fixed point is stable along a line that connects the two fixed points, whereas the noninteracting fixed point is unstable on the plane.  $\bar{g}_2^*$ is a positive number of order $\epsilon$, i.e., the interacting fixed point has weak repulsive interaction with its strength controlled by the DOS singularity exponent $\epsilon$.
We assume $D_+ > D_-$ for the RG flow, which makes the DOS larger on the electron side.  In such a case, the interacting fixed point is shifted from zero to be $\bar{\mu}=O(\epsilon^2)<0$.
Note that the coupling constant $\bar{g}$ monotonically grows when the interacting is attractive $(\bar{g}<0)$.
}
\label{fig:RGflow_2}
\end{figure}

Since we are interested in the low-energy behavior, we consider the RG flow by increasing $l$.  The RG flow is shown in Fig.~\ref{fig:RGflow_2}.
From the RG equations (\ref{eq:RG_mu_g_0}) for the coupling constant $\bar{g}$ and the chemical potential $\bar{\mu}$, we find two fixed points
\begin{gather}
\label{eq:fixed_points_1}
\bar{g}_1^* = 0, \quad \bar{\mu}_1^*=0  \\
\label{eq:fixed_points_2}
\bar{g}_2^* = \frac{\epsilon}{c_\text{pp}}, \quad \bar{\mu}_2^*=-\frac{\epsilon c_\text{H}}{c_\text{pp}}.
\end{gather}
$\bar{g}_1^*=\bar{\mu}_1^*=0$ corresponds to the noninteracting fixed point.
The new fixed point at $\bar{g}_2^*>0$, $\bar{\mu}_2^*<0$ is the nontrivial interacting fixed point with finite repulsive interaction, whose strength is of order $\epsilon$.  The smallness of the coupling constant allows a controlled analysis by the DOS singularity exponent $\epsilon$ about the interacting fixed point.

We can find the similarity to the $\phi^4$ theory in the structure of the RG equation \eqref{eq:RG_mu_g_0}: the coefficient of the quadratic term $r \phi^2$ corresponds to the chemical potential $\bar{\mu}$ and the quartic interaction term $\phi^4$ to the coupling constant $\bar{g}$.  From this viewpoint, our theory can be regarded as the fermionic analog of the $\phi^4$ theory. Like the Wilson--Fisher fixed point, our perturbative RG analysis is analytically controlled thanks to the smallness of the coupling constant on the order of $\epsilon$ at the interacting fixed point. While the $\phi^4$ theory in three dimensions corresponds to $\epsilon=1$ in Wilson--Fisher RG, in our theory for high-order VHS in two dimensions $\epsilon$ takes the value of $1/4$, given by the DOS exponent.

In the $\phi^4$ theory, the RG flow of $r$ describes the phase transition between ordered and disordered states: the RG flow to $r \gg 0$ corresponds to the disordered state and $r \ll 0$ to the ordered state, where the field $\phi$ has a finite expectation value associated with spontaneous symmetry breaking.
The parameter $r$ is analogous to the chemical potential $\mu$ in the present fermionic model, where $\mu \gg 0$ yields the electron Fermi surface and $\mu \ll 0$ the hole Fermi surface.  The sign change of $\mu$ thus describes a topological transition between electron and hole Fermi liquids, which involves a change of Fermi surface topology without symmetry breaking.

Note that at the interacting fixed point $\bar{\mu}^*_2$ is nonzero when there is a finite contribution from the Hartree term $c_\text{H} \neq 0$ due to the asymmetry of DOS at $E>0$ and $E<0$: $D(\pm |E|) = D_\pm |E|^{-\epsilon}$ with $D_- \neq D_+$.  This means that in the presence of repulsive interaction, the chemical potential at which scale-invariant Fermi surface appears is shifted from the noninteracting case, similar to  the deviation of $r$ at Wilson--Fisher fixed point from the mean-field value. For small $\epsilon$, it follows from the expressions for $D_\pm$ that $\bar{\mu}^*_2$ is at most of order $\epsilon^2$.

\subsection{Relevant perturbations}
\label{sec:relevant_perturbations}

We have identified the two fixed points: the noninteracting and interacting fixed points.  With the chemical potential tuned at the fixed points, the noninteracting fixed point at $\bar{g}_1^*$ is an unstable fixed point and the interacting fixed point at $\bar{g}_2^*$ is a stable fixed point.  The chemical potential is a relevant perturbation around both fixed points.  We have included the chemical potential even in the analysis of the simplest case above as it can be generated by interaction due to  the absence of particle-hole symmetry in the single-particle DOS.

In addition to the chemical potential, we consider other relevant perturbations to the fixed points, including the magnetic field $h$ and the $s$-wave pairing field $\Delta$.
Those relevant perturbations add the following terms to the action at criticality:
\begin{gather}
-\mu \bar{\psi} \psi, \quad
h (\bar{\psi_\uparrow} \psi_\uparrow - \bar{\psi_\downarrow} \psi_\downarrow ), \quad
\Delta \bar{\psi}_\uparrow \bar{\psi}_\downarrow + \Delta^* \psi_\downarrow \psi_\uparrow.
\end{gather}
Finite temperature is also a relevant perturbation.  Its effect is taken account of via Matsubara frequencies; see Appendix~\ref{sec:finite-temperature}.
We further consider other relevant perturbations.
For an energy dispersion $E_{-i\partial_{\bm{r}}} = -\partial_x^2 - \partial_y^4$, i.e., $E_{\bm{k}} = k_x^2 -k_y^4$, the fermion bilinear terms with derivatives $\partial_{x}$, $\partial_{y}$, $\partial_{y}^2$, $\partial_{y}^3$, $\partial_{x} \partial_{y}$ are also relevant perturbations.

Perturbations to the system are subject to symmetry constraints:
Particle conservation forbids the pairing term, spin-rotational symmetry nonzero $h$, and reflection symmetry odd-derivative terms in $x$ or $y$.
With all three symmetries present, only two terms $\mu\bar{\psi}\psi$ and $\bar{\psi}\partial_{y}^2\psi$ are allowed as perturbations to the system with $E_{\bm{k}}=k_x^2-k_y^4$; see Fig.~\ref{fig:perturbation}.  This means that we need to tune two parameters to reach the critical metallic state governed by the interacting RG fixed point shown earlier.

In our RG analysis so far, the starting point is the single-particle dispersion at the high-order VHS, where the term $\bar{\psi}\partial_{y}^2\psi$  is absent. To one-loop order, this term is not generated from the interaction since the self-energy $\Sigma$ is independent of momentum. However, it may be generated at higher-loop order. As we shall show later, this means that in the presence of interacting, the critical state is reached when the $\bar{\psi}\partial_{y}^2\psi$ term is present in the single-particle dispersion and its coefficient is tuned to a particular value.

The pairing field can be introduced by proximity to an external superconductor, or it can be regarded as a test field for studying $s$-wave pairing susceptibility.  Likewise, the magnetic field $h$ can be externally introduced or regarded as a test field for the spin susceptibility.
In this viewpoint, the chemical potential is conjugate to the particle number, and hence it is related to the charge compressibility.

Corrections to the perturbations $h$ and $\Delta$ are calculated similarly as those for $\mu$ and $g$ at $T=0$.  To consider a correction to the pairing field $\Delta$, we include the particle-particle loop diagram, where the one-loop diagram is shown in Fig.~\ref{fig:loop}(c).
We include the corrections to write the magnetic field $h + \delta h$ and the pairing field $\Delta + \delta\Delta$.

Integrating out the high-energy modes is followed by rescaling.
The parameters of the model should be rescaled at tree level as $h' = bh$ and $\Delta' = b\Delta$.  Those parameters are relevant and thus their values increase as we proceed with RG steps.
When the perturbative corrections are included, the new parameters after an RG step are
\begin{gather}
h' = b (h + \delta h), \quad
\Delta' = b (\Delta + \delta\Delta).
\end{gather}

To one-loop order, the correction terms are expressed as
\begin{gather}
\delta h = 0, \quad \delta\Delta = -g\Pi_\text{pp}.
\end{gather}
The one-loop correction $\Pi_\text{pp}$ is obtained in Eq.~\eqref{eq:one-loop_T0_pp}.  Then, the parameters change as
\begin{gather}
h' = b h, \quad
\Delta' \simeq b [\Delta - lc_\text{pp}gD(\Lambda)].
\end{gather}
With the dimensionless quantities
\begin{gather}
\bar{h}=\frac{h}{\Lambda}, \quad
\bar{\Delta}=\frac{\Delta}{\Lambda},
\end{gather}
we reach the RG equations
\begin{subequations}
\label{eq:RG_h_delta}
\begin{gather}
\label{eq:RG_h}
\frac{d\bar{h}}{dl} = \bar{h}, \\
\label{eq:RG_Delta}
\frac{d\bar{\Delta}}{dl} = ( 1- c_\text{pp}\bar{g} ) \bar{\Delta}.
\end{gather}
\end{subequations}
We confirm that the perturbations $h$ and $\Delta$ are relevant around the two fixed points, given in Eqs.~\eqref{eq:fixed_points_1} and \eqref{eq:fixed_points_2}.  Finite temperature is also a relevant perturbation, which scales in the same manner as energy and frequency.  All low-energy fixed points are found at $T=0$, and thus we focus on zero temperature in the main part.  The one-loop RG equations at finite temperature are presented in Appendix~\ref{sec:finite-temperature}.
The physical consequences, i.e., scaling properties of thermodynamic quantities, are discussed in the next section.

\subsection{Structure of higher-order RG}
\label{sec:RG_Wilson_discussion}

So far, we have made the energy-shell RG analysis to one-loop order.  We now illustrate how it works in the case with higher-order corrections.  Again,  for clarity we consider here the minimal case at $T=0$ without symmetry-breaking fields.  Inclusion of other relevant contributions such as $T$, $h$, and $\Delta$ is straightforward.

Higher-order perturbative corrections give rise to the frequency and momentum dependence in the self-energy $\Sigma$ in Eq.~\eqref{eq:action_one-loop}, while the one-loop corrections are independent of frequency or momentum and depend only on the DOS as we have seen.  We expand the self-energy with respect to the frequency and momentum to find corrections to the field, energy dispersion, and chemical potential.

As we shall show later in Sec.~\ref{sec:field_theory}, the momentum dependence may give corrections to the energy dispersion.  In that case, one has to be wary of the generation of relevant corrections in the single-particle energy dispersion even when they are initially absent.  We represent such a term as $\lambda \tilde{k}^n$, where the coefficient $\lambda$ transforms under Eq.~\eqref{eq:scaling_E} as $\lambda' = b^a \lambda$ $(a>0)$.  For instance, for the case of $E_{\bm{k}}=k_x^2-k_y^4$, this term corresponds to $\lambda k_y^2$ $(a=1/2)$, which is the only relevant perturbation to the energy dispersion.

In order to keep track of such relevant term(s), we include $\lambda \tilde{k}^n$ in the energy dispersion:
\begin{equation}
\label{eq:model+perturbation}
E_{\bm{k}} = A_+ k_+^{n_+} - A_- k_-^{n_-} + \lambda \tilde{k}^n.
\end{equation}
At least one such relevant perturbation term exists for a high-order VHS, and if present, turns a high-order saddle point in the noninteracting single-particle dispersion into ordinary one. For the generalized dispersion  Eq.~\eqref{eq:model_2}, there is only one relevant perturbation to find $E_{\bm{k}} = A_+ k_+^2 - A_- (k_-^2)^2 + \lambda k_-^2$ because the original dispersion is rotationally invariant in the $k_-$ submanifold. For other types of dispersion, it is possible to have multiple relevant terms.  An extension to a case with multiple relevant terms is straightforward.

Then, the expansion of the self-energy is generally given by
\begin{align}
\label{eq:sigma_expansion}
\Sigma &= \Sigma_0 + (i\omega) \Sigma_\omega + \delta A_+ k_+^{n_+} - \delta A_- k_-^{n_-} + \delta\lambda \tilde{k}^n -\delta\mu \nonumber\\
&\quad + \text{(high-order terms)},
\end{align}
where irrelevant high-order terms are safely neglected.
After integrating out the high-energy modes within the energy shell, we obtain the effective action
\begin{widetext}
\begin{align}
S_{\Lambda/b}[\psi^<]
&= \int\frac{d\omega}{2\pi} \int_{\bm{k}}^< \bar{\psi}^<_\sigma(k)
\{ -i\omega (1-\Sigma_\omega) + (A_+ +\delta A_+) k_+^{n_+} -(A_- + \delta A_-) k_-^{n_-} + (\lambda + \delta\lambda) \tilde{k}^n -(\mu + \delta \mu) \} \psi^<_\sigma(k) \nonumber\\
&\quad + (g+\delta g) \bigg( \prod_{j} \int\frac{d\omega_j}{2\pi} \int_{\bm{k}_j}^< \bigg)
(2\pi)^{d+1} \delta (k_1+k_2-k_3-k_4)
\bar{\psi}^<_{\uparrow}(k_1) \bar{\psi}^<_{\downarrow}(k_2) \psi^<_{\downarrow}(k_3) \psi^<_{\uparrow}(k_4).
\end{align}
\end{widetext}

The next step in the energy-shell RG analysis is to rescale the momentum and restore the energy cutoff $\Lambda/b$ to $\Lambda$; see Eqs.~\eqref{eq:rescaling_energy} and \eqref{eq:rescaling_momentum}.
However, the effective action $S_{\Lambda/b}$ still evidently has a different form from $S_\Lambda$.  To recover the form of the action, we rescale the other quantities as follows:
\begin{subequations}
\begin{gather}
\label{eq:scaling_omega}
\omega' = b\omega , \\
\label{eq:scaling_A}
A'_\pm = (1-\Sigma_\omega)^{-1} (A_\pm + \delta A_\pm) \equiv b^{\gamma_{A_\pm}} A_\pm, \\
\label{eq:scaling_lambda}
\lambda' = b^a (1-\Sigma_\omega)^{-1} (\lambda+\delta\lambda) , \\
\label{eq:scaling_mu}
\mu' = b (1-\Sigma_\omega)^{-1} (\mu+\delta\mu) \equiv b^{\gamma_\mu} \mu, \\
\label{eq:scaling_g}
g' = b^\epsilon (1-\Sigma_\omega)^2 (g+\delta g), \\
\psi' = b^{-(3-\epsilon)/2} (1-\Sigma_\omega)^{-1/2} \psi^< \equiv b^{-(3-\epsilon)/2} b^{\gamma_\psi/2} \psi^<.
\end{gather}
\end{subequations}
Here we introduce the scaling exponents $\gamma_{A_\pm}$, $\gamma_\mu$, and $\gamma_\psi$.
Note that there is an ambiguity in defining $\omega'$ and $\psi'$ as the factor $(1-\Sigma_\omega)$ can be imposed on either $\omega'$ or $\psi'$.  We choose to scale $\omega$ linearly in $b$ and hence the factor $(1-\Sigma_\omega)$ contributes to the field renormalization.

For $\gamma_{A_\pm} \neq 0$, if we continue to rescale momentum according to Eq.~\eqref{eq:rescaling_momentum} and the coefficients $A_\pm$ according to Eq.~\eqref{eq:scaling_A}, the cutoff energy $\Lambda/b$ is not mapped to $\Lambda$.  To remedy this issue, we rescale momentum as
\begin{equation}
\label{eq:scaling_k_tilde}
k'_\pm = b^{1/\tilde{n}_\pm} k_\pm \text{ with } \tilde{n}_\pm = \frac{n_\pm}{1+\gamma_{A_\pm}},
\end{equation}
so that $E_{\bm{k}'} = bE_{\bm{k}}$ is satisfied.
In this way, the coefficients $A_\pm$ do not change under rescaling.

At a fixed point, the parameters in the action are determined to satisfy scale invariance; i.e., they do not vary under rescaling [Eqs.~\eqref{eq:scaling_lambda}--\eqref{eq:scaling_g}].  To reach a fixed point, initial values of the relevant perturbations $\mu$ and $\lambda$ should be tuned so that they cease to flow when the coupling constant $g$ reaches the fixed point value.

Rescaling of the magnetic field $h$ and the pairing field $\Delta$ can be considered similarly.  Including the field renormalization, we obtain
\begin{subequations}
\begin{gather}
\label{eq:scaling_h}
h' = b (h+\delta h) (1-\Sigma_\omega)^{-1} \equiv b^{\gamma_h} h, \\
\label{eq:scaling_Delta}
\Delta' = b (\Delta + \delta\Delta) (1-\Sigma_\omega)^{-1} \equiv b^{\gamma_\Delta} \Delta,
\end{gather}
\end{subequations}
where we define the exponents $\gamma_h$ and $\gamma_\Delta$.
We shall show later that the Ward identity requires $\gamma_\mu = \gamma_h = 1$.

\section{Analysis}
\label{sec:analysis}

In this section, we combine the results obtained from the mean-field and RG analyses to present a phase diagram of interacting electrons near a high-order VHS.
We then describe scaling properties for thermodynamic quantities and correlation functions near the supermetal critical point.
In addition, we discuss the Ward identity, which results from charge conservation and gives relations among scaling exponents of electronic specific heat, magnetic susceptibility, and charge compressibility.

\subsection{Phase diagram}

\begin{figure*}
\centering
\includegraphics[width=.85\hsize]{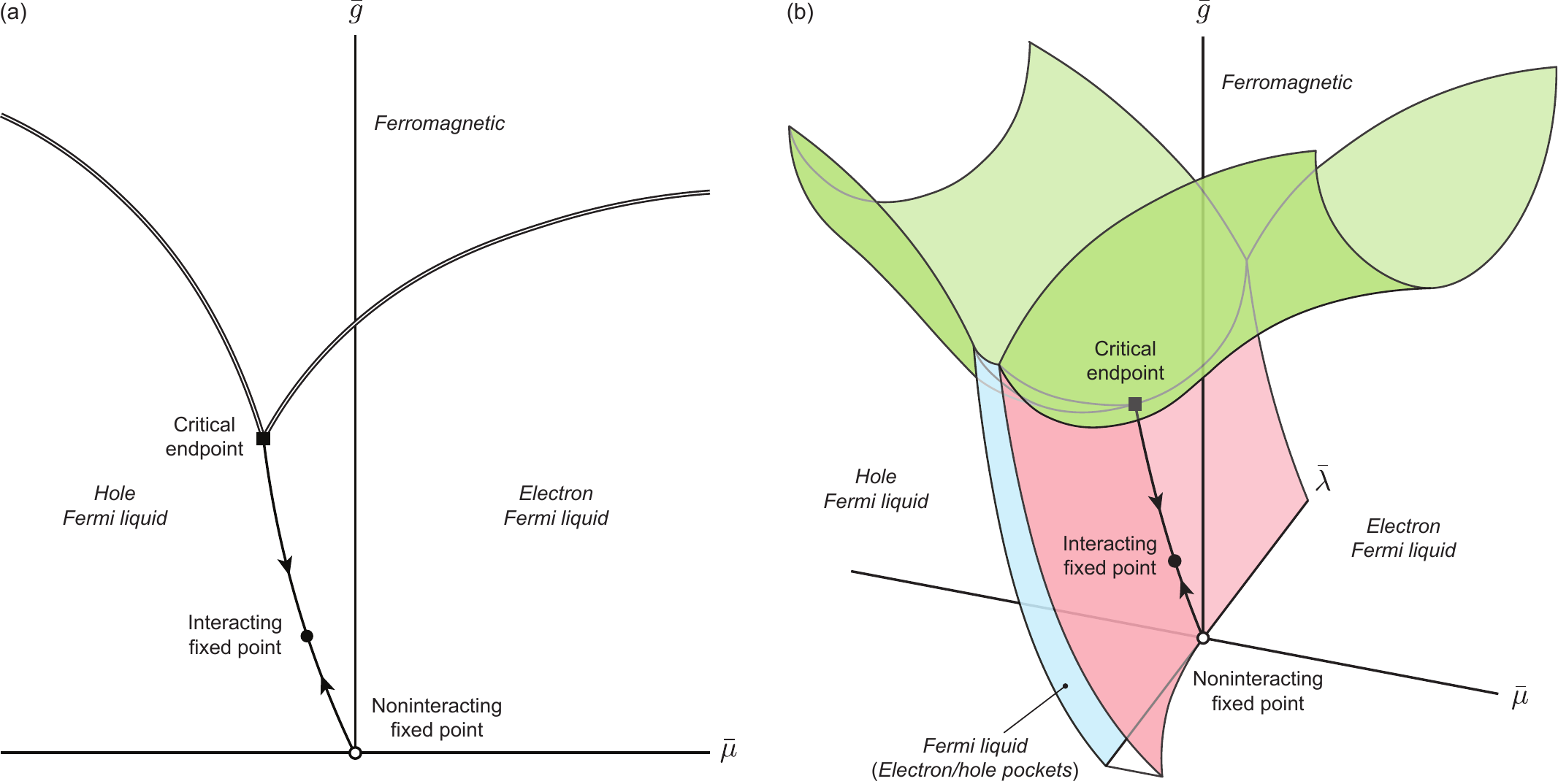}
\caption{
Schematic phase diagrams.  There are two tuning parameters to realize a high-order VHS: the dimensionless chemical potential $\bar{\mu}$ and the perturbation to the high-order VHS $\bar{\lambda}$.
The two-dimensional diagram (a) shows a cut of a generic three-dimensional diagram (b) to include a scale-invariant line.
The line with arrows depicts the RG flow and the system is scale invariant along the line, where susceptibilities shows power-law divergence.  The interacting and noninteracting fixed points lie on the line.  They are multicritical points in this phase space.
Small $\bar{g}$ does not cause symmetry breaking or a long-range order, so that electron and hole Fermi liquid states are separated by planes [red and blue in (b), respectively], which include the scale-invariant line.  The parameter $\bar{\lambda}$ can split the VHS, which yields an electron or hole pocket in the course of a transition from the electron Fermi liquid to the hole Fermi liquid.
Lifshitz transitions occur on the planes that separate the Fermi liquid states.
A first-order phase transition occurs for large $\bar{g}$ [above the double line in (a) and the green plane in (b)], to trigger a broken-symmetry state with ferromagnetism.  The scale-invariant line terminates with the first-order transition at a critical endpoint.
}
\label{fig:phase_diagram-conj}
\end{figure*}

As we have discussed in Sec.~\ref{sec:model}, realization of a high-order VHS requires tuning of the energy dispersion in addition to the chemical potential.  There is at least one parameter for a relevant perturbation in the energy dispersion to be tuned; see Sec.~\ref{sec:relevant_perturbations}.  Therefore, to present a global phase diagram near high-order VHS, we need three axes for the coupling constant $g$, the chemical potential $\mu$, and a tuning parameter $\lambda$ of the energy dispersion.
All those three are relevant perturbations at noninteracting fixed point; we use dimensionless parameters defined by $\bar{g}=g/\Lambda^\epsilon$, $\bar{\mu}=\mu/\Lambda$, and $\bar{\lambda}=\lambda/\Lambda^a$.

We now incorporate the results from the mean-field analysis (Fig.~\ref{fig:MF-phase}) and the RG analysis (Fig.~\ref{fig:RGflow_2}).
The mean-field analysis is expected to be qualitatively correct for relatively large $\bar{g}$, while the RG analysis is valid for small $\bar{\mu}$ and $\bar{g}$.
Based on these considerations, we propose a global phase diagram of interacting electrons near high-order VHS shown in Fig.~\ref{fig:phase_diagram-conj}.

For large $\bar{g}$, mean-field calculation reveals that an itinerant ferromagnetic metal exists over a wide range of chemical potential, and the ferromagnetic transition is first order. Due to the finite correlation length, we expect these results to be  qualitatively correct and continue to hold in the presence of a small $\lambda$.

On the other hand, for small $\bar{g}$,  there is no spontaneous symmetry breaking or long-range order.  When the DOS is not divergent, the system is either a electron or a hole Fermi liquid depending on the sign of $\bar{\mu}$. These two Fermi liquid states are indistinguishable by symmetry but differ in the Fermi surface topology.
A transition between electron and hole Fermi liquids, i.e., a Lifshitz transition, occurs as the chemical potential crosses the VHS. This transition occurs on a surface in the three-dimensional phase diagram.

Our RG analysis reveals that by tuning both $\mu$ and $\lambda$, a multicritical line on the Lifshitz transition surface can be reached, where the system displays various divergent susceptibilities and scale-invariant Fermi surface.
We coin a term, \textit{supermetal}, to describe such an unusual metallic state.
At the end of this multicritical line $g=\mu=\lambda=0$, the noninteracting supermetal exhibits divergent charge, spin and pairing susceptibilities determined by the power-law divergent DOS.
In contrast, for $g\neq 0$, the interacting supermetal displays universal critical properties governed by the nontrivial interacting fixed point, located at $\bar{g}^* = \epsilon/c_\text{pp}$, $\bar{\mu}^*=0$, $\bar{\lambda}^*=0$ to first order in $\epsilon$.
As we shall show in next subsection,
at this fixed point, while the charge compressibility and spin susceptibility diverge, the $s$-wave pairing susceptibility remains finite. We shall also show later by a two-loop RG calculation that the electron Green's function has the scaling form $G(\omega)\propto 1/|\omega|^{1-\eta}$ with $\eta>0$, thus establishing the non-Fermi liquid nature of an interacting supermetal.

The interacting fixed point is stable along the multicritical line and unstable in two other directions. One of the unstable direction (roughly speaking $\bar{\lambda}$) lies within the Lifshitz transition surface, while the other direction (roughly speaking $\bar{\mu}$) drives the system into electron or hole Fermi liquid.
Note that a finite $\bar{\lambda}$ converts a high-order VHS to a conventional VHS ($\bar{\lambda}>0$) or splits it to two conventional VHS points ($\bar{\lambda}<0$). For the latter case for Eq.~\eqref{x2y4}, over a finite range of the chemical potential (Fig.~\ref{fig:perturbation}) there is an extra small pocket around $k=0$ in addition to  large Fermi surfaces.

Since the relevant perturbations $\mu$ and $\lambda$ introduce an intrinsic momentum scale to the system, the resulting Fermi liquids may be unstable to superconductivity at very low temperature via the Kohn--Luttinger mechanism associated with non-analyticity of susceptibility at momentum $2k_F$  \cite{Kohn-Luttinger}. This scenario is neglected in the phase diagram (Fig.~\ref{fig:phase_diagram-conj}).
In contrast, being a quantum critical state of metal at $T=0$, the interacting supermetal is immune from the Kohn--Luttinger mechanism since its Fermi surface is scale-invariant without any intrinsic scale.

Finally, we conjecture how the ferromagnetic transition at large $\bar{g}$ and $\bar{\mu}$ and the Lifshitz transition at small $\bar{g}$ and $\bar{\mu}$ meet together. A plausible scenario is that the multicritical line of supermetal meets  the first-order ferromagnetic transition line at a tricritical point between electron Fermi liquid, hole Fermi liquid and ferromagnetic metal. The nature of this tricritical point is an interesting open question.

\subsection{Scaling analysis}

\subsubsection{Generic case}
\label{sec:scaling_generic}

Scale invariance at the fixed points enables us to extract various scaling relations.
Since the partition function $\mathcal{Z}$ is invariant under the scale transformation, the free energy density $F$, defined in Eq.~\eqref{eq:free_energy}, reflects the scaling of the factor $T/V$:
\begin{equation}
F' = b^{1 + d_+/\tilde{n}_+ + d_-/\tilde{n}_-} F,
\end{equation}
where the volume $V$ scales according to Eq.~\eqref{eq:scaling_k_tilde} and temperature scales the same manner as energy and frequency.
For convenience, we rewrite the exponent as
\begin{align}
\label{eq:epsilon_tilde}
1 + \frac{d_+}{\tilde{n}_+} + \frac{d_-}{\tilde{n}_-} &= 2 - \left( \epsilon - \frac{d_+ \gamma_{A_+}}{n_+} - \frac{d_- \gamma_{A_-}}{n_-} \right) \nonumber\\
&\equiv 2-\tilde{\epsilon}.
\end{align}
By explicitly showing the parameters of $F$, we obtain the scaling relation of the free energy density
\begin{equation}
\label{eq:scaling}
F(\mu,h,\Delta;T) = b^{-2+\tilde{\epsilon}} F(b^{\gamma_\mu} \mu , b^{\gamma_h}h, b^{\gamma_\Delta}\Delta; b T).
\end{equation}
Here, the scaling exponents $\gamma_{A_\pm}$, $\gamma_\mu$, $\gamma_h$, and $\gamma_\Delta$ correspond to the values at a fixed point $\bar{g}^*$.
We later see $\gamma_\mu = \gamma_h =1$, but we keep them in the following scaling analysis.
The coupling constant $g$ itself does not appear in the scaling relation of the free energy density $F$, but the effect is imprinted on $\gamma_\Delta$ and $\tilde{\epsilon}$ as the fixed point properties.
We shall see that $\gamma_{A_\pm}$ are at most of order $\epsilon^2$ at the interacting fixed point and thus $\tilde{\epsilon}$ is also a small positive quantity.

We then consider the critical exponents of the charge compressibility $\kappa$, magnetic susceptibility $\chi$, heat capacity per unit volume $C_V$, and $s$-wave pairing susceptibility $\chi_\text{BCS}$.  From Eq.~\eqref{eq:scaling}, we find
\begin{gather}
\label{eq:scaling_compressibility}
\kappa = -\left(\frac{\partial^2 F}{\partial\mu^2}\right)_T \sim
\begin{cases}
T^{-(\tilde{\epsilon}+2\gamma_\mu-2)} \\
|\mu|^{-(\tilde{\epsilon}+2\gamma_\mu-2)/\gamma_\mu},
\end{cases} \\
\label{eq:scaling_susceptibility}
\chi = -\lim_{h\to0}\left(\frac{\partial^2 F}{\partial h^2}\right)_T \sim
\begin{cases}
T^{-(\tilde{\epsilon}+2\gamma_h-2)} \\
|\mu|^{-(\tilde{\epsilon}+2\gamma_h-2)/\gamma_\mu},
\end{cases} \\
\label{eq:scaling_heat-capacity}
\frac{C_V}{T} = -\left(\frac{\partial^2 F}{\partial T^2}\right)_V \sim
\begin{cases}
T^{-\tilde{\epsilon}} \\
|\mu|^{-\tilde{\epsilon}/\gamma_\mu} \\
|h|^{-\tilde{\epsilon}/\gamma_h} \\
|\Delta|^{-\tilde{\epsilon}/\gamma_\Delta},
\end{cases} \\
\label{eq:scaling_pairing}
\chi_\text{BCS} = \left(\frac{\partial^2 F}{\partial\Delta \partial\Delta^*}\right)_T
\sim
\begin{cases}
T^{-(\tilde{\epsilon}+2\gamma_\Delta-2)} \\
|\Delta|^{-2+(2-\tilde{\epsilon})/\gamma_\Delta}.
\end{cases}
\end{gather}

We also examine the pair correlation function
\begin{align}
C(\bm{r},\tau) &= \langle (\psi_\uparrow \psi_\downarrow) (\bm{r},\tau) (\bar{\psi}_\downarrow \bar{\psi}_\uparrow) (0,0) \rangle \nonumber\\
&\quad - \langle (\psi_\uparrow \psi_\downarrow) (\bm{r},\tau) \rangle \langle (\bar{\psi}_\downarrow \bar{\psi}_\uparrow) (0,0)\rangle,
\end{align}
with $\langle \mathcal{O} \rangle = \int D\bar{\psi} D\psi \mathcal{O} e^{-S}/\mathcal{Z}$.  From the comparison between $\chi_\text{BCS}$ and $C(\bm{r},\tau)$, we obtain the scaling form
\begin{align}
C(r_+,r_-,\tau) = \nu^{2(2-\tilde{\epsilon}-\gamma_\Delta)} \hat{c}\left( r_+ \nu^{1/\tilde{n}_+}, r_- \nu^{1/\tilde{n}_-}, \tau \nu \right),
\end{align}
where $\nu$ is an arbitrary energy scale and $\hat{c}$ is a scaling function.

The field renormalization with the exponent $\gamma_\psi$ appears in the two-point correlation function $G(\bm{r},\tau)$.  We shall show the derivation later with the field theory approach.  In the critical region, the exponent $\gamma_\psi$ can be replaced with a constant $\eta = \gamma_\psi(\bar{g}^*)$; the scaling form is given by
\begin{subequations}
\label{eq:correlation_scaling}
\begin{align}
\label{eq:correlation_scaling_real}
G(r_+,r_-,\tau) = \nu^{1-\tilde{\epsilon}+\eta} \hat{g}\left( r_+ \nu^{1/\tilde{n}_+}, r_- \nu^{1/\tilde{n}_-}, \tau \nu \right),
\end{align}
or its Fourier transform is
\begin{align}
\label{eq:correlation_scaling_frequency}
G(k_+,k_-,\omega) = \nu^{-(1-\eta)} \hat{g}' \left( \frac{k_+}{\nu^{1/\tilde{n}_+}}, \frac{k_-}{\nu^{1/\tilde{n}_-}}, \frac{\omega}{\nu} \right),
\end{align}
\end{subequations}
where $\hat{g}$ and $\hat{g}'$ are scaling functions.
Particularly, we see the frequency dependence $G(\bm{k}=\bm{0},\omega) \propto 1/|\omega|^{1-\eta}$, which differs from the noninteracting correlation function $G(\bm{k}=\bm{0}, \omega)\propto 1/|\omega|$ with finite $\eta$.  $\eta$ corresponds to the anomalous dimension and specifies the non-Fermi liquid behavior.

\subsubsection{One-loop results}

To one-loop order, we find from the RG equations \eqref{eq:RG_mu_0} and \eqref{eq:RG_h_delta} the exponents at the fixed points
\begin{gather}
\label{eq:exponents_one-loop_1}
\gamma_\mu=1, \quad
\gamma_h=1, \\
\begin{aligned}[b]
\gamma_\Delta &= 1-c_\text{pp}(0)\bar{g}_j^* \\
&=
\begin{cases}
1 & \text{(Noninteracting fixed point)} \\
1-\epsilon & \text{(Interacting fixed point)}.
\end{cases}
\end{aligned}
\end{gather}
with $\tilde{\epsilon} = \epsilon$.  Most exponents in Eqs.~\eqref{eq:scaling_compressibility}--\eqref{eq:scaling_heat-capacity} are the same at the noninteracting and interacting fixed points, which is identical to that of the DOS in the noninteracting state.
The difference is found when the pairing field $\Delta$ is involved.  The exponent for the pairing field $\gamma_\Delta$ renders different exponents for the pairing susceptibility $\chi_\text{BCS}$:
\begin{equation}
\chi_\text{BCS} \sim
\begin{cases}
T^{-\epsilon},\ |\Delta|^{-\epsilon} & \text{(Noninteracting fixed point)} \\
T^{+\epsilon},\ |\Delta|^{+\epsilon} & \text{(Interacting fixed point)}.
\end{cases}
\end{equation}
The $s$-wave pairing susceptibility remains finite at the interacting fixed point whereas it diverges at the noninteracting fixed point.  We also find a difference in the pair correlation function
\begin{align}
&\quad C(r_+,r_-,\tau) \nonumber\\
&= \hat{c} \left( r_+ \nu^{1/\tilde{n}_+}, r_- \nu^{1/\tilde{n}_-}, \tau \nu \right) \nonumber\\
&\quad \times
\begin{cases}
\nu^{-2(1-\epsilon)} & \text{(Noninteracting fixed point)} \\
\nu^{-2} & \text{(Interacting fixed point)}.
\end{cases}
\end{align}
It shows a faster decay at the interacting fixed point, reflecting the suppressed pairing susceptibility.

\subsection{Ward identity}

In Sec.~\ref{sec:scaling_generic}, we mentioned the relations $\gamma_\mu=1$ and $\gamma_h=1$.  They result from the conservation laws for charge and spin.
The Ward identity (more generally the Ward--Takahashi identity) describes a conservation law \cite{Ward,Takahashi}.  The identity is regarded as the quantum analog to Noether's theorem.  We present how the Ward identity works in our present analysis.  The identity should hold even after an RG analysis, and thus it can be used to check the validity of an RG scheme, or specifically a choice of a cutoff.  It also gives relations between the exponents for thermodynamic quantities Eqs.~\eqref{eq:scaling_compressibility}--\eqref{eq:scaling_heat-capacity}.

Now we investigate the structure of the self-energy and vertex corrections.
To be concrete, we look into the expansion of the self-energy Eq.~\eqref{eq:sigma_expansion} to find a relation between $\Sigma_\omega$ and $\delta\mu$.
The Ward identity concludes
\begin{equation}
\label{eq:Ward_Sigma}
\Sigma_\omega = -\frac{\delta\mu}{\mu}
\end{equation}
at $T=0$.
The identity is based on charge conservation or the U(1) gauge invariance; the action and correlation functions are invariant under the transformations $\psi \mapsto e^{i\theta(\bm{r},\tau)} \psi$ and $\bar{\psi} \mapsto \bar{\psi} e^{-i\theta(\bm{r},\tau)}$ with a smooth scalar function $\theta(\bm{r},\tau)$.
In the present model, charge conservation holds for each spin separately, thus leading to
\begin{equation}
\label{eq:Ward_magnetic}
\Sigma_\omega = -\frac{\delta h}{h}.
\end{equation}
Then, Eqs.~\eqref{eq:scaling_mu} and \eqref{eq:scaling_h} yield
\begin{equation}
\label{eq:Ward_exponents}
\gamma_\mu = 1, \quad
\gamma_h = 1.
\end{equation}

The result of the energy-shell RG analysis to one-loop order in Sec.~\ref{sec:energy-shell} satisfies the Ward identity, which means that the conservation laws are correctly taken account of.  We notice that a frequency shell instead of the energy shell violates the Ward identity.

Furthermore, Eq.~\eqref{eq:Ward_exponents} makes some ratios among scaling relations Eqs.~\eqref{eq:scaling_compressibility}--\eqref{eq:scaling_heat-capacity} constant as functions of temperature $T$. One is the Wilson ratio $R_W$ between the electronic specific heat $C_V$ and the magnetic susceptibility $\chi$ and the other is the ratio $R_C$ between the charge compressibility $\kappa$ and $C_V$:
\begin{subequations}
\begin{gather}
R_W = \frac{T\chi(T)}{C_V(T)} = \text{const.}, \\
R_C = \frac{T\kappa(T)}{C_V(T)} = \text{const.}
\end{gather}
\end{subequations}

\begin{figure}
\centering
\includegraphics[width=0.9\hsize]{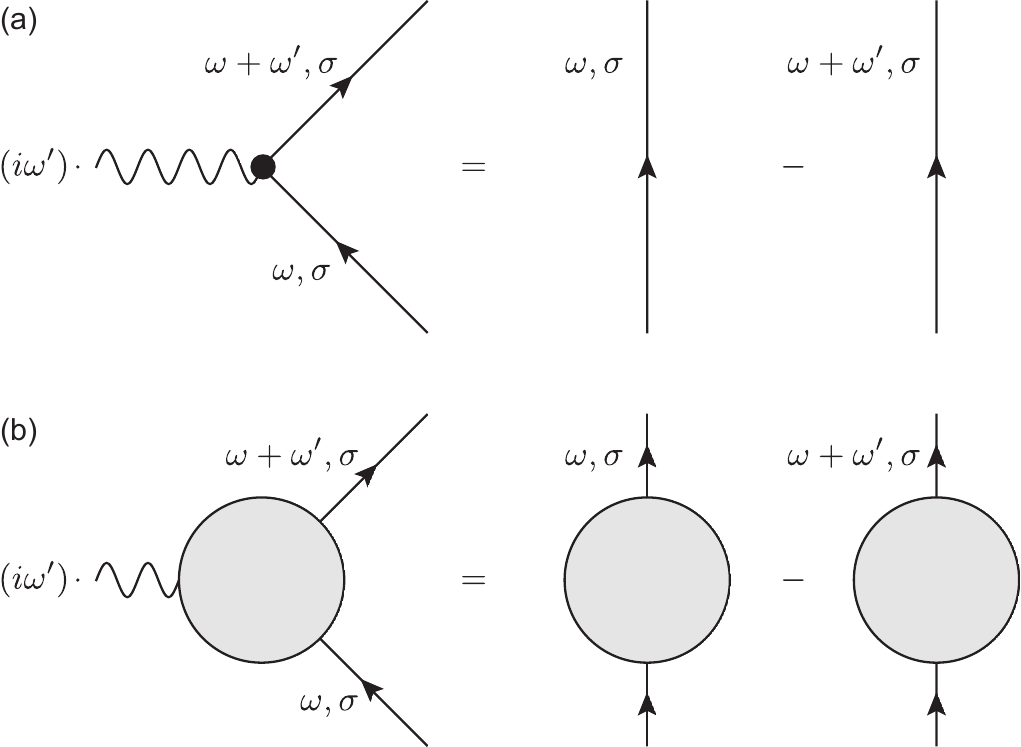}
\caption{
(a) Relation between the bare vertex and the noninteracting Green's function.
(b) Diagrammatic representation of the Ward--Takahashi identity.
}
\label{fig:WT}
\end{figure}

In the following, we sketch the derivation of the Ward identity from the diagrammatic point of view.  A detailed derivation is given in Appendix~\ref{sec:ward-takahashi}.
In the present analysis, the Ward identity relates the frequency derivative of the self-energy and the vertex function corresponding to the coupling term $\alpha_\sigma\bar{\psi}_\sigma \varphi \psi_\sigma$ in the action.  $\alpha_\sigma$ is the spin-dependent coupling constant and the $\varphi$ is a bosonic field.
We write the vertex function as $\Gamma^{(2,\alpha_\sigma)}_\sigma(\omega+\omega',\omega)$, where we focus only on the frequency dependence.  The vertex function modifies the coupling term to be $\alpha_\sigma \Gamma^{(2,\alpha_\sigma)}_\sigma(\omega+\omega',\omega) \bar{\psi}_\sigma(\omega+\omega') \varphi(\omega') \psi_\sigma(\omega)$.
The derivation of the Ward identity makes use of the equality $G_{0}^{-1}(\bm{k},\omega+\omega') - G_{0}^{-1}(\bm{k},\omega) = i\omega'$,
or equivalently
\begin{equation}
\label{eq:Ward_Green-0}
G_0(\bm{k},\omega+\omega') (i\omega') G_0(\bm{k},\omega)
= G_0(\bm{k},\omega) - G_0(\bm{k},\omega+\omega').
\end{equation}
This equation is diagrammatically shown in Fig.~\ref{fig:WT}(a).  It relates the noninteracting vertex function $\Gamma^{(2,\alpha_\sigma)}_\sigma(\omega+\omega',\omega) = 1$ and the noninteracting Green's function $G_0$.  Now we add corrections to the self-energy, depicted in Fig.~\ref{fig:WT}(b) as shaded blobs.
The dressed vertex function is obtained from the dressed self-energy by attaching the external scalar field $\varphi$ to every internal fermion line.
As a result, we find the Ward--Takahashi identity
\begin{align}
&\quad G(\bm{k},\omega+\omega') (i\omega') \Gamma^{(2,\alpha_\sigma)}_\sigma(\omega+\omega',\omega) G(\bm{k},\omega) \nonumber\\
&= G(\bm{k},\omega) - G(\bm{k},\omega+\omega').
\end{align}
The full Green's function $G(\bm{k},\omega)$ is given by
\begin{equation}
G(\bm{k},\omega) = \frac{1}{i\omega - E_{\bm{k}} - \Sigma(\bm{k},\omega)},
\end{equation}
with the full self-energy $\Sigma$.
Taking the zero frequency limit $\omega' \to 0$, we obtain the Ward identity
\begin{equation}
\label{eq:Ward_vertex-2}
\Gamma^{(2,\alpha_\sigma)}_\sigma(\omega,\omega) = 1 - \frac{\partial\Sigma(\omega)}{\partial(i\omega)}.
\end{equation}

The vertex function $\Gamma^{(2,\alpha_\sigma)}_\sigma(\omega,\omega)$ gives the quantum correction to the coupling $\alpha_\sigma$ to be $\alpha_\sigma + \delta \alpha_\sigma = \alpha_\sigma \Gamma^{(2,\alpha_\sigma)}_\sigma(\omega,\omega)$.  $\alpha_\sigma$ represents the chemical potential with $\alpha_\uparrow = \alpha_\downarrow = -\mu$ and the magnetic field with $\alpha_\uparrow = - \alpha_\downarrow = h$.
Since the right-hand side of Eq.~\eqref{eq:Ward_vertex-2} is independent of spin $\sigma$, we confirm Eqs.~\eqref{eq:Ward_Sigma} and \eqref{eq:Ward_magnetic}.

\section{Field theory approach}
\label{sec:field_theory}

This section focuses on the RG analysis from the field theory approach.  To begin with, we briefly argue the two RG schemes: the energy-shell RG analysis and the field theory approach.
We then confirm that the two methods give the same result at one-loop order.  We also perform a two-loop analysis of the self-energy (two-point function) from the field theory approach to show the anomalous dimension and the correction to the energy dispersion.

\subsection{RG schemes}
\label{sec:schemes}

An objective of RG analyses is to track the flow of parameters in a theory under a scale transformation.  Here, we illustrate two different RG schemes: the Wilsonian approach, including the preceding energy-shell RG analysis, and the field theory approach.  The common feature is to divide the integration manifold (frequency and momentum in the present case) into two parts and integrate out modes belonging to one of them.  The two schemes differ in intervals of integrations.  The first scheme involves an integration within a hard shell.  In the energy-shell RG analysis, fluctuations inside the thin energy shell $E\in[-\Lambda,-\Lambda/b) \lor (\Lambda/b,\Lambda]$ are eliminated. This mode elimination followed by rescaling enables us to keep track of the change of parameters under a scale transformation.  On the other hand, in the field theory approach, we integrate out all low-energy fluctuations below the cutoff $\Lambda$.  Then, we deduce the RG flow of parameters by comparing results at different cutoffs $\Lambda$ and $\Lambda'$.

The two schemes have advantages in different aspects.
In the Wilsonian approach, the frequency-momentum space is progressively integrated over, so the interpretation of the RG procedure is rather simple.  The inclusion of low-energy modes results in a theory at low energies with different parameters.
In spite of its simple interpretation, higher-loop calculations are not easy with the Wilsonian approach.  In a one-loop calculation, we have only one shell to be concerned about.  However, higher-loop diagrams consist of many internal lines (virtual states), so that we have to take care of shells for each of them.
On the other hand, the field theory approach does not require such error-prone steps as it deals with all modes below the cutoff at once.  This makes higher-loop calculations more tractable.  Although not as intuitive as the Wilsonian approach, the field theory approach leads to the same results about critical phenomena.  More descriptions about the comparison between the two schemes can be found in e.g. Ref.~\cite{Shankar}.  A brief review of the field theory approach is given in Appendix~\ref{sec:field-theory_review}.

\subsection{Soft cutoff}

In the field theory approach, we calculate the connected $N$-point correlation function $G^{(N)}$ or the one-particle irreducible $N$-point function $\Gamma^{(N)}$.  If we face a UV divergence in calculating them, we need to cure the divergence to obtain physically meaningful results.  There are several ways to do so; we here choose to employ the UV energy cutoff $\Lambda$ to make a comparison to the preceding energy-shell RG analysis.

The functions $G^{(N)}$ and $\Gamma^{(N)}$ can be obtained perturbatively with the noninteracting Green's function $G_0$.  We introduce the UV energy cutoff by suppressing the high-energy contributions in $G_0$.
We define the noninteracting Green's function with the energy cutoff $G_{0\Lambda}(\bm{k},\omega_n)$ as
\begin{align}
G_{0\Lambda}(\bm{k},\omega_n) &= G_0(\bm{k},\omega_n) K_\Lambda(E_{\bm{k}}) \nonumber\\
&= \frac{K_\Lambda(E_{\bm{k}})}{i\omega_n - E_{\bm{k}}},
\end{align}
with the UV energy cutoff factor
\begin{equation}
K_\Lambda(E) = \frac{\Lambda^2}{\Lambda^2+E^2}.
\end{equation}
Note that the cutoff factor smoothly varies from 0 to 1 and thus works as a soft energy cutoff.  This is in contrast to the energy-shell RG analysis, where the interval of an energy integration is cut off abruptly at $\Lambda/b$ and $\Lambda$.

We can interpret the modified Green's function as a Green's function with an energy-dependent quasiparticle weight $K_\Lambda(E)$.
The weight fades away in the high-energy limit $E\to\pm\infty$ to eliminate UV divergences, while $K_\Lambda(E) \to 1$ for energies much lower than the cutoff $\Lambda$.
One may be tempted to see the modified Green's function in a different way.  For example, it can be rewritten as
\begin{align}
G_{0\Lambda}(\bm{k},\omega_n)
&= \frac{1}{i\omega_n-E_{\bm{k}}} - \frac{1}{i\omega_n-E_{\bm{k}}} \frac{E_{\bm{k}}^2}{\Lambda^2+E_{\bm{k}}^2}.
\end{align}
It may be viewed as a variation of the Pauli--Villars regularization, where the additional term cures a UV divergence but vanishes in the limit $\Lambda\to\infty$.  However, we cannot think of it as a propagator with a large mass term since we cannot add a mass term for the electronic energy dispersion which is continuous and unbounded.

It should be noted that the cutoff factor $K_\Lambda(E)$ does not depend on frequency.  It potentially causes a violation of the Ward identity, which would result in wrong conclusions.  For example, if one chooses a cutoff factor of the form $\Lambda^2/(\Lambda^2+E^2+\omega_n^2)$, it invalidates the Ward identity.
The absence of the frequency in the cutoff factor ensures the Ward identity.

\subsection{Formalities}

\subsubsection{Structure of the RG analysis}

To derive RG equations and see scaling properties, we calculate the one-particle irreducible $N$-point function $\Gamma^{(N)}_\Lambda$ with the cutoff $\Lambda$ and examine its cutoff dependence.  The cutoff dependence is seen by comparing two $N$-point functions at different cutoffs; see Eq.~\eqref{eq:Gamma_scale}.  Specifically, we compare $\Gamma^{(N)}_\Lambda$ to one at a reference point $\Gamma^{(N)}_R$.  The energy scale at the reference point is referred to as the renormalization scale.  The procedure of fixing the model to the reference is equivalent to setting the initial parameters in the Wilsonian approach.

We first analyze the case with $T=h=\Delta=0$.  We impose the renormalization conditions
\begin{gather}
\label{eq:condition_2}
\Gamma_R^{(2)}(k) = -i\omega_n + E_{\bm{k},0} -\mu_0, \\
\label{eq:condition_4}
\Gamma_R^{(4)}(k_1,k_2;k_3,k_4) = g_0,
\end{gather}
where the condition for $\Gamma^{(4)}$ should be considered at $k_1+k_2 = k_1+k_3 = k_1 +k_4 = 0$.
The subscript $0$ denotes quantities at the renormalization scale.
The interaction dresses the two-point and four-point functions and they acquire cutoff-dependent corrections.
We here use the energy dispersion Eq.~\eqref{eq:model+perturbation}, which includes a relevant perturbation to a high-order VHS, since such a term could be generated under the RG analysis at two-loop order or higher; see the discussion in Sec.~\ref{sec:RG_Wilson_discussion}.
Then, the two-point and four-point functions at the cutoff $\Lambda$ can be expressed as
\begin{gather}
\label{eq:Gamma_2_Lambda}
\begin{aligned}[b]
\Gamma_\Lambda^{(2)} =& -i\omega_n Z_\psi^{-1} + Z_{A_+}^{-1} A_+ k_+^{n_+} - Z_{A_-}^{-1} A_- k^{n_-} \\
&+ Z_\lambda^{-1} \tilde{k}^n - Z_\mu^{-1} \mu,
\end{aligned}\\
\label{eq:Gamma_4_Lambda}
\Gamma_\Lambda^{(4)} = Z_g^{-1} g,
\end{gather}
where the corrections $Z_\psi$, $Z_{A_\pm}$, $Z_\lambda$, $Z_\mu$, and $Z_g$ are calculated perturbatively.
The $N$-point functions at the renormalization scale and the cutoff $\Lambda$ are related by
\begin{gather}
\Gamma_R^{(N)} = Z_\psi^{N/2} \Gamma_\Lambda^{(N)}.
\end{gather}
We note the structure of the RG analysis is general, so that an analysis of other energy dispersions such as Eq.~\eqref{eq:model_2} is straightforward.

The last equation leads to the RG equations.  Since the left-hand side does not depend on the cutoff $\Lambda$, we obtain the differential equation
\begin{align}
\Lambda \frac{d}{d\Lambda} \Gamma_R^{(N)} = 0,
\end{align}
leading to the Callan--Symanzik equation \cite{Callan,Symanzik1,Symanzik2}.  We obtain the Callan--Symanzik for the one-particle irreducible $N$-point function
\begin{widetext}
\begin{gather}
\label{eq:CS}
\left[ \Lambda\frac{\partial}{\partial\Lambda} -\beta(\bar{g})\frac{\partial}{\partial\bar{g}} - \beta_\mu(\bar{g},\bar{\mu}) \frac{\partial}{\partial\bar{\mu}} - \beta_{A_+}(\bar{g},A_\pm) \frac{\partial}{\partial A_+} - \beta_{A_-}(\bar{g},A_\pm) \frac{\partial}{\partial A_-} - \beta_\lambda(\bar{g},\bar{\lambda}) \frac{\partial}{\partial\lambda} - \frac{N}{2} \gamma_\psi(\bar{g}) \right] \Gamma_\Lambda^{(N)} = 0.
\end{gather}
\end{widetext}
The beta functions and $\gamma_\psi$ are defined by
\begin{subequations}
\begin{gather}
\label{eq:beta_g}
\beta(\bar{g}) = -\left( \Lambda\frac{\partial\bar{g}}{\partial\Lambda} \right)_{\bar{g}_0,\bar{\mu}_0,A_{\pm,0},\lambda_0}, \\
\label{eq:beta_mu}
\beta_\mu(\bar{g},\bar{\mu}) = -\left( \Lambda\frac{\partial\bar{\mu}}{\partial\Lambda} \right)_{\bar{g}_0,\bar{\mu}_0,A_{\pm,0},\lambda_0}, \\
\label{eq:beta_A}
\beta_{A_\pm}(\bar{g},A_\pm) = -\left( \Lambda\frac{\partial A_\pm}{\partial\Lambda} \right)_{\bar{g}_0,\bar{\mu}_0,A_{\pm,0},\lambda_0}, \\
\label{eq:beta_lambda}
\beta_\lambda(\bar{g},\bar{\lambda}) = -\left( \Lambda\frac{\partial\bar{\lambda}}{\partial\Lambda} \right)_{\bar{g}_0,\bar{\mu}_0,A_{\pm,0},\lambda_0}, \\
\label{eq:gamma_def}
\gamma_\psi = -\left( \Lambda\frac{\partial}{\partial\Lambda} \ln Z_\psi \right)_{\bar{g}_0,\bar{\mu}_0,A_{\pm,0},\lambda_0}.
\end{gather}
\end{subequations}
Since the renormalized values are given by
\begin{subequations}
\begin{gather}
\bar{g} = Z_g Z_\psi^{-2} \bar{g}_0, \\
\bar{\mu} = Z_\mu Z_\psi^{-1} \bar{\mu}_0, \\
A_\pm = Z_{A_\pm} Z_\psi^{-1} A_{\pm,0}, \\
\bar{\lambda} = Z_\lambda Z_\psi^{-1} \lambda_0,
\end{gather}
\end{subequations}
we can rewrite the beta functions as
\begin{subequations}
\begin{gather}
\beta = \bar{g} \left( \epsilon -\Lambda\frac{\partial}{\partial\Lambda} \ln Z_g -2\gamma_\psi \right) , \\
\label{eq:beta_mu_mod}
\beta_\mu = \bar{\mu} \left( 1-\Lambda\frac{\partial}{\partial\Lambda} \ln Z_\mu -\gamma_\psi \right) , \\
\label{eq:beta_A_mod}
\beta_{A_\pm} = A_\pm \left( -\Lambda\frac{\partial}{\partial\Lambda} \ln Z_{A_\pm} -\gamma_\psi \right) , \\
\label{eq:beta_lambda_mod}
\beta_\lambda = \bar{\lambda} \left( a -\Lambda\frac{\partial}{\partial\Lambda} \ln Z_\lambda -\gamma_\psi \right).
\end{gather}
\end{subequations}
Those equations show that the field renormalization gives additional effects to the beta functions and hence the scaling properties.

\subsubsection{Solutions}

The Callan--Symanzik equation can be solved by the method of characteristics; see Appendix~\ref{sec:field-theory_review}.  The beta functions describe the RG flows of the parameters:
\begin{subequations}
\begin{gather}
\label{eq:RG_g_beta}
\frac{d\bar{g}}{dl} = \beta(\bar{g}), \\
\label{eq:RG_mu_beta}
\frac{d\bar{\mu}}{dl} = \beta_{\mu} (\bar{g}, \bar{\mu}), \\
\label{eq:RG_A_beta}
\frac{dA_\pm}{dl} = \beta_{A_\pm} (\bar{g},A_\pm), \\
\label{eq:RG_lambda_beta}
\frac{d\bar{\lambda}}{dl} = \beta_\lambda (\bar{g},\bar{\lambda}).
\end{gather}
\end{subequations}
$l=\ln{\Lambda_0/\Lambda}$ denotes the RG scale, measured relative to the renormalization scale $\Lambda_0$.
Those RG equations are to be compared with those obtained by the energy-shell RG analysis in Sec.~\ref{sec:energy-shell}.
In general, they are coupled differential equations and zeros of the beta functions determine fixed points.

We can write the beta functions $\beta_\mu$, $\beta_{A_\pm}$, and $\beta_\lambda$ around a fixed point with $\bar{g}^*$ as
\begin{subequations}
\begin{gather}
\beta_\mu(\bar{g}^*,\bar{\mu}) = \gamma_\mu(\bar{g}^*) \bar{\mu}, \\
\label{eq:beta_A_gamma}
\beta_{A_\pm}(\bar{g}^*,A_\pm) = \gamma_{A_\pm}(\bar{g}^*) A_{\pm}, \\
\label{eq:beta_lambda_gamma}
\beta_\lambda(\bar{g}^*,\bar{\lambda}) = \gamma_\lambda(\bar{g}^*) \bar{\lambda}.
\end{gather}
\end{subequations}
$\gamma_\mu(\bar{g}^*)$, $\gamma_{A_\pm}(\bar{g}^*)$, and $\gamma_\lambda(\bar{g}^*)$ give the exponents in the scaling region.  Recall that $\gamma_\mu=1$ is required by the Ward identity, regardless of $\bar{g}$.
From the beta functions around the fixed point, we observe the scaling properties
\begin{gather}
\bar{\mu}(l) \sim \bar{\mu}_0 e^{l}, \quad
\label{eq:scaling_A_2}
A_\pm(l) \sim A_{\pm,0} e^{\gamma_{A_\pm}(\bar{g}^*)l}, \quad
\bar{\lambda}(l) \sim \bar{\lambda}_0 e^{\gamma_\lambda(\bar{g}^*)l}.
\end{gather}
Since the energy dispersion does not receive correction at one-loop order, we have $\gamma_{A_\pm}(\bar{g}^*) = O(\epsilon^2)$ and $\gamma_\lambda(\bar{g}^*) = a + O(\epsilon^2)$.

The shift of the chemical potential and the generation of the relevant perturbation to the energy dispersion are also seen from the beta functions.
When $\beta_\mu(\bar{g},0) \neq 0$, the chemical potential is displaced from zero under the RG analysis, while it does not alter the scaling behavior of $\bar{\mu}$.  Similarly, a finite relevant perturbation $\bar{\lambda}$ is generated if $\beta_\lambda(\bar{g},0) \neq 0$, even when it is initially absent.

The function $\gamma_\psi$ is ascribed to the anomalous dimension $\eta$ when it is computed at a fixed point.
To see this, we solve the Callan--Symanzik equation \eqref{eq:CS}; see Appendix~\ref{sec:field-theory_review} for details.
The solution of the two-point function is given by
\begin{align}
\label{eq:two-point_solution}
&\quad
\Gamma^{(2)}_\Lambda \left(e^{-l/n_\pm}k_{\pm,0}, e^{-l}\omega_0; \bar{g}(0), \bar{\mu}(0), A_\pm(0), \bar{\lambda}(0) \right) \nonumber\\
&= e^{-l} \Gamma_\Lambda^{(2)} \left( k_{\pm,0}, \omega_0; \bar{g}(l), \bar{\mu}(l), A_\pm(l), \bar{\lambda}(l) \right) \nonumber\\
&\quad\times \exp \left[ \int_0^l dl' \gamma_\psi(\bar{g}(l')) \right].
\end{align}
We now examine the behavior in the critical region as a function of $\omega$, $k_+$, and $k_-$.  We assume the two-point function is a function of $A_+ k_+^{n_+}$, $A_- k_-^{n_-}$, $\omega$ in the scaling region.
Since those three quantities, $\Lambda$, and $\Gamma^{(2)}_\Lambda$ have the dimension of energy, the two-point function can be written as
\begin{align}
\Gamma^{(2)}_\Lambda \left( k_+, k_-, \omega; A_+, A_- \right) = \Lambda \hat{\Gamma}^{(2)}_\Lambda \left( \frac{A_+ k_+^{n_+}}{\Lambda}, \frac{A_- k_-^{n_-}}{\Lambda}, \frac{\omega}{\Lambda} \right),
\end{align}
where $\hat{\Gamma}^{(2)}_\Lambda$ is a dimensionless scaling function.
Here, we do not need to assume homogeneity for $\hat{\Gamma}^{(2)}_\Lambda$ but determine the exponents for $A_+ k_+^{n_+}/\Lambda$, $A_- k_-^{n_-}/\Lambda$, and $\omega/\Lambda$, separately.
In Eq.~\eqref{eq:two-point_solution}, $l$ is an arbitrary quantity; to inspect the scaling behavior in terms of $\omega$, we set $l=\ln (\omega_0/\omega)$ and $k_+=k_-=0$.
The momentum dependence is considered in the same manner with $l=\ln (k_{\pm,0}/k_\pm)^{n_\pm}$ and Eq.~\eqref{eq:scaling_A_2}.
We then find
\begin{align}
&\quad \Gamma^{(2)}_\Lambda (k_+,k_-,\omega) \nonumber\\
&\propto
\begin{cases}
\left(k_+^{n_+/[1+\gamma_{A_+}(\bar{g}^*)]}\right)^{1-\eta} & (k_-=\omega=0) \\
\left(k_-^{n_-/[1+\gamma_{A_-}(\bar{g}^*)]}\right)^{1-\eta} & (k_+=\omega=0) \\
\omega^{1-\eta} & (k_+=k_-=0),
\end{cases}
\end{align}
where $\gamma_\psi(\bar{g}(l))=\eta$ is used.
It confirms the scaling relation of the two-point correlation function Eq.~\eqref{eq:correlation_scaling} along with the relation $G = [\Gamma^{(2)}]^{-1}$.

\begin{figure*}
\centering
\includegraphics[width=\hsize]{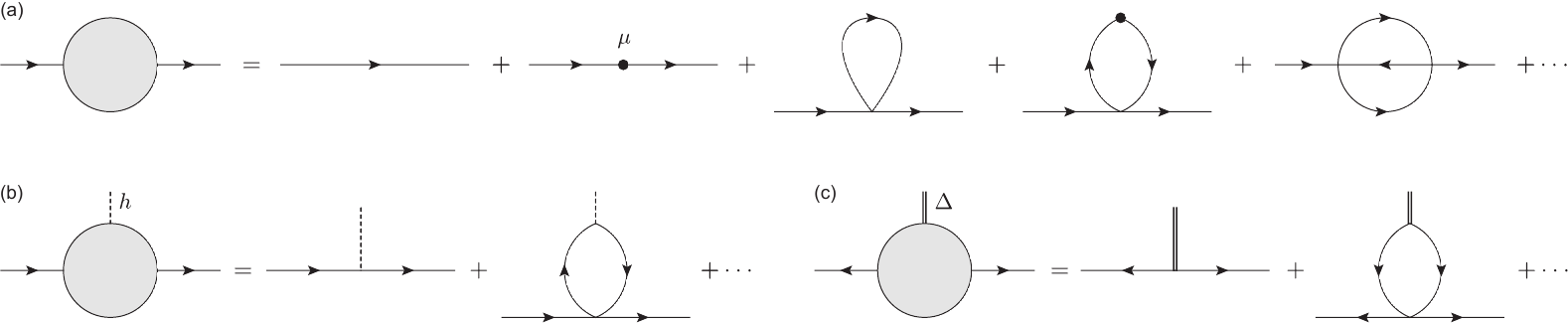}
\caption{
Perturbative corrections to connected correlation functions.  An directed solid line represents the free electron propagator, or the noninteracting Green's function.  Shaded blobs involve all possible irreducible diagrams.  $(\cdots)$ includes higher-order corrections.
(a) Two-point correlation function $G^{(2)}=[\Gamma^{(2)}]^{-1}$.  A dot denotes the perturbation with respect to the chemical potential $\mu$, corresponding an insertion of $-\mu\bar{\psi}\psi$.
(b), (c) Vertex functions for the magnetic field $h$ and the pairing field $\Delta$, respectively.
}
\label{fig:loop_two-point}
\end{figure*}

\begin{figure*}
\centering
\includegraphics[width=0.95\hsize]{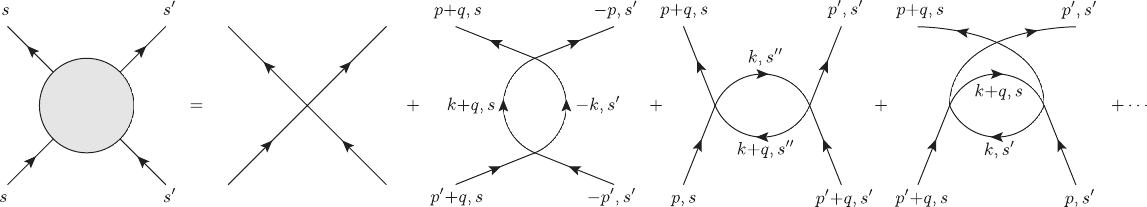}
\caption{
Four-point correlation function.  Here, $p, k, q$ denote both frequency and momentum, and $s$ includes a spin $\sigma$ and valleys/orbitals, if exist.  The four-point function represents the coupling constant for the contact interaction when it is evaluated with $q=0$.  There are three one-loop diagrams, which are regarded as the BCS, density-density, and exchange channels (from left to right).  Among the three, the density-density contribution does not exist in the present analysis because of the Pauli exclusion principle for the contact interaction; there is no way to appropriately assign the spin $\sigma''(\neq \sigma, \sigma')$ under the condition $\sigma \neq \sigma'$.  It should be taken account of when the interaction has finite range or there is an additional valley/orbital degree of freedom.
}
\label{fig:stu}
\end{figure*}

\subsection{One-loop calculations $\bm{(h=\Delta=0)}$}
\label{sec:field_RG_one-loop}

We calculate the two-point and four-point functions to obtain the beta functions and $\gamma_\psi(\bar{g})$.  This is accomplished by evaluating the perturbative corrections to the two-point and four-point functions (Figs.~\ref{fig:loop_two-point} and \ref{fig:stu}).
As the corrections to the coupling constant $g$, there are three possible one-loop diagrams shown in Fig.~\ref{fig:stu}.  To determine the perturbative correction $\delta g$, all diagrams should be evaluated with zero momentum transfer $q=0$, which is required by the renormalization condition Eq.~\eqref{eq:condition_4}.  The three one-loop diagrams in Fig.~\ref{fig:stu}(a) correspond to the BCS, density-density, and exchange channels (from left to right).  Out of the three, the density-density channel does not contribute because of the Pauli exclusion principle for the contact interaction.  This contribution is allowed when we assume the density-density interaction in finite range $\bar{\psi}_\sigma\bar{\psi}_{\sigma'} \psi_{\sigma'} \psi_\sigma$ with arbitrary spins $\sigma$, $\sigma'$ or when there is an additional valley of orbital degree of freedom.   (For reference, we note that the three channels are referred to as the BCS, ZS (zero sound), and ZS$'$ in Ref.~\cite{Shankar}; or $s$-, $t$-, and $u$-channels with the Mandelstam variables.)

To one-loop order, the two-point and four-point functions give corrections to the chemical potential and the coupling constant, but not to field or the energy dispersion as we have seen in the energy-shell RG analysis.  One-loop diagrams can be represented by $\Sigma_\text{H}$, $\Pi_\text{pp}$, and $\Pi_\text{ph}$ like Eq.~\eqref{eq:one-loop_sigma_g}.  Then, the two-point and four-point functions become
\begin{gather}
\label{eq:Gamma_2_oneloop}
\Gamma^{(2)}_\Lambda = -i\omega_n + E_{\bm{k}} -\mu +g\Sigma_\text{H} -g\mu\Pi_\text{ph}, \\
\label{eq:Gamma_4_oneloop}
\Gamma^{(4)}_\Lambda = g - g^2 (\Pi_\text{pp}+\Pi_\text{ph}),
\end{gather}
which lead to
\begin{subequations}
\begin{gather}
Z_\psi^{-1} = 1, \quad Z_{A_\pm}^{-1} = 1, \quad Z_\lambda^{-1} = 1, \\
Z_\mu^{-1} = 1 - \frac{g}{\mu}\Sigma_\text{H} + g\Pi_\text{ph}, \\
Z_g^{-1} = 1 - g(\Pi_\text{pp}+\Pi_\text{ph}).
\end{gather}
\end{subequations}

Here we calculate the perturbative corrections with the soft cutoff $K_\Lambda$.
The actual calculations for the beta functions require the $\Lambda$-derivatives instead of the corrections themselves.  We thus obtain the one-loop corrections as follows:
\begin{subequations}
\begin{align}
&\quad \Lambda\frac{\partial}{\partial\Lambda} \Sigma_\text{H} \nonumber\\
&= T\sum_{\omega_n} \int_{\bm{k}} G_0(\bm{k},\omega_n) \Lambda\frac{\partial}{\partial\Lambda} K_\Lambda(E_{\bm{k}}) \nonumber\\
&= T\sum_{\omega_n>0} \int dE D(E) \frac{-2E}{\omega_n^2 + E^2} \frac{2\Lambda^2 E^2}{(\Lambda^2 + E^2)^2} \nonumber\\
&= -\Lambda^2 \int dE D(E) \tanh\left(\frac{E}{2T}\right) \frac{E^2}{(\Lambda^2 + E^2)^2} \nonumber\\
&= -[D(\Lambda)-D(-\Lambda)] \Lambda \int_0^\infty dx \frac{x^{2-\epsilon}}{(1+x^2)^2} \tanh\left(\frac{\Lambda}{2T}x\right) \nonumber\\
&\equiv -\Lambda D(\Lambda) \tilde{c}_\text{H}(\bar{T}) ,
\end{align}
\begin{align}
&\quad \Lambda\frac{\partial}{\partial\Lambda} \Pi_\text{pp} \nonumber\\
&= T\sum_{\omega_n} \int_{\bm{k}} G_0(\bm{k},\omega_n) G_0(-\bm{k},-\omega_n) \Lambda\frac{\partial}{\partial\Lambda} K^2_\Lambda(E_{\bm{k}}) \nonumber\\
&= T\sum_{\omega_n > 0} \int dE D(E) \frac{2}{\omega_n^2 + E^2} \frac{4\Lambda^4 E^2}{(\Lambda^2+E^2)^3} \nonumber\\
&= 2\Lambda^4 \int dE D(E) \frac{E}{(\Lambda^2+E^2)^3} \tanh\left(\frac{E}{2T}\right) \nonumber\\
&= 2[D(\Lambda)+D(-\Lambda)] \int_0^\infty dx \frac{x^{1-\epsilon}}{(1+x^2)^3} \tanh\left(\frac{\Lambda}{2T}x\right) \nonumber\\
&\equiv D(\Lambda) \tilde{c}_\text{pp}(\bar{T}),
\end{align}
\begin{align}
&\quad \Lambda\frac{\partial}{\partial\Lambda} \Pi_\text{ph} \nonumber\\
&= T\sum_{\omega_n} \int_{\bm{k}} G_0^2(\bm{k},\omega_n) \Lambda\frac{\partial}{\partial\Lambda} K^2_\Lambda(E_{\bm{k}}) \nonumber\\
&= T\sum_{\omega_n > 0} \int dE D(E) \frac{-2(\omega_n^2-E^2)}{(\omega_n^2+E^2)^2} \frac{4\Lambda^4 E^2}{(\Lambda^2+E^2)^3} \nonumber\\
&= -\frac{\Lambda^4}{T} \int dE D(E) \frac{E^2}{(\Lambda^2+E^2)^3} \frac{1}{\cosh^2\left(\dfrac{E}{2T}\right)} \nonumber\\
&= -[D(\Lambda)+D(-\Lambda)] \frac{\Lambda}{T} \int_0^\infty dx \frac{x^{2-\epsilon}}{(1+x^2)^3} \frac{1}{\cosh^2\left(\dfrac{\Lambda}{2T}x\right)} \nonumber\\
&\equiv -D(\Lambda) \tilde{c}_\text{ph}(\bar{T}).
\end{align}
\end{subequations}

As a result, we obtain the beta functions Eqs.~\eqref{eq:beta_g} and \eqref{eq:beta_mu}
\begin{subequations}
\begin{gather}
\beta(\bar{g}) = \epsilon\bar{g} - \bar{g}^2 \left[ \tilde{c}_\text{pp}(\bar{T})-\tilde{c}_\text{ph}(\bar{T}) \right], \\
\label{eq:beta_mu_field-theory}
\beta_\mu(\bar{g},\bar{\mu}) = \left[ 1-\tilde{c}_\text{ph}(\bar{T}) \bar{g} \right] \bar{\mu} + \tilde{c}_\text{H}(\bar{T}) \bar{g}.
\end{gather}
\end{subequations}
Note that the tree-level scaling terms appear from the definitions of the dimensionless parameters $\bar{g}_0=g_0 D(\Lambda)$ and $\bar{\mu}_0 = \mu/\Lambda$.
The beta functions are to be compared with the result from the energy-shell RG analysis Eq.~\eqref{eq:RG_mu_g_0}.
To confirm, we first evaluate the coefficients $\tilde{c}_\text{H}$, $\tilde{c}_\text{pp}$, $\tilde{c}_\text{ph}$ at $T=0$:
\begin{subequations}
\begin{gather}
\tilde{c}_\text{H}(0) = \left( 1-\frac{D_-}{D_+} \right) \frac{\pi}{4} (1-\epsilon) \frac{1}{\cos\left(\dfrac{\pi\epsilon}{2}\right)}, \\
\tilde{c}_\text{pp}(0) = \left( 1+\frac{D_-}{D_+} \right) \frac{\pi}{4} \left( 1+\frac{\epsilon}{2} \right) \frac{\epsilon}{\sin\left(\dfrac{\pi\epsilon}{2}\right)}, \\
\tilde{c}_\text{ph}(0) = 0.
\end{gather}
\end{subequations}
The zeros of the beta function $\beta(\bar{g})$ give the two fixed points
\begin{equation}
\label{eq:fixed_points_field}
\bar{g}_1^* = 0, \quad \bar{g}_2^* = \frac{\epsilon}{\tilde{c}_\text{pp}(0)} (>0).
\end{equation}
We now find the noninteracting and interacting fixed points from the field theory approach.  Although the value of $\bar{g}_2^*$ differs in the two schemes, resulting exponents for the thermodynamic quantities are not suffered from the difference as the exponents are not directly dependent on the coupling constant $\bar{g}$ at fixed point.
We explicitly confirm this in the next subsection by calculating the beta functions for the magnetic field $h$ and pairing field $\Delta$.

\subsection{RG equations for $\bm{h}$ and $\bm{\Delta}$}
\label{sec:field_RG_h-Delta}

The beta functions for the magnetic field $h$ and pairing field $\Delta$ can be obtained from the corresponding vertex functions $\Gamma^{(2,h)}$ and $\Gamma^{(2,\Delta)}$, respectively.  Perturbative corrections to them are depicted in Figs.~\ref{fig:loop_two-point}(b) and \ref{fig:loop_two-point}(c).  We impose the renormalization conditions
\begin{subequations}
\begin{gather}
\Gamma^{(2,h)}_R = Z_\psi \Gamma^{(2,h)}_\Lambda = h_0, \\
\Gamma^{(2,\Delta)}_R = Z_\psi \Gamma^{(2,\Delta)}_\Lambda = \Delta_0,
\end{gather}
\end{subequations}
where the vertex functions with the cutoff $\Lambda$ are expressed as
\begin{gather}
\Gamma^{(2,h)}_\Lambda = Z_h^{-1} h, \quad
\Gamma^{(2,\Delta)}_\Lambda = Z_\Delta^{-1} \Delta.
\end{gather}

To obtain the beta functions to one-loop order, it is sufficient to consider the Callan--Symanzik equations without corrections to the energy dispersion and the chemical potential:
\begin{subequations}
\begin{gather}
\left[ \Lambda\frac{\partial}{\partial\Lambda} - \beta(\bar{g})\frac{\partial}{\partial \bar{g}} - \beta_h(\bar{g},\bar{h})\frac{\partial}{\partial\bar{h}} - \gamma_\psi(\bar{g}) \right] \Gamma^{(2;h)} = 0, \\
\left[ \Lambda\frac{\partial}{\partial\Lambda} - \beta(\bar{g})\frac{\partial}{\partial \bar{g}} - \beta_\Delta(\bar{g},\bar{\Delta})\frac{\partial}{\partial\bar{\Delta}} - \gamma_\psi(\bar{g}) \right] \Gamma^{(2;\Delta)} = 0,
\end{gather}
\end{subequations}
where the beta functions for the magnetic field and pairing field are defined by
\begin{subequations}
\begin{gather}
\label{eq:beta_h}
\beta_h(\bar{h}) = -\left( \Lambda\frac{\partial\bar{h}}{\partial\Lambda} \right)_{\bar{g}_0}, \\
\label{eq:beta_Delta}
\beta_{\Delta}(\bar{g}) = -\left( \Lambda\frac{\partial\bar{\Delta}}{\partial\Lambda} \right)_{\bar{g}_0}.
\end{gather}
\end{subequations}
Using the relations
\begin{gather}
\bar{h} = Z_h Z_\psi^{-1} \bar{h}_0, \quad
\bar{\Delta} = Z_\Delta Z_\psi^{-1} \bar{\Delta}_0,
\end{gather}
the beta functions become
\begin{subequations}
\begin{gather}
\beta_h(\bar{g},\bar{h})
= \bar{h} \left( 1-\Lambda\frac{\partial}{\partial\Lambda} \ln Z_h -\gamma_\psi \right), \\
\beta_\Delta(\bar{g},\bar{\Delta})
= \bar{\Delta} \left( 1-\Lambda\frac{\partial}{\partial\Lambda} \ln Z_\Delta -\gamma_\psi \right).
\end{gather}
\end{subequations}
They are related to the exponents $\gamma_h$ and $\gamma_\Delta$ when evaluated at a fixed point:
\begin{gather}
\beta_h(\bar{g}^*,\bar{h}) = \gamma_h(\bar{g}^*) \bar{h}, \quad
\beta_\Delta(\bar{g}^*,\bar{\Delta}) = \gamma_\Delta(\bar{g}^*) \bar{\Delta}.
\end{gather}

We calculate the vertex functions for $h$ and $\Delta$ to one-loop order and find
\begin{gather}
\Gamma^{(2,h)}_\Lambda = h + gh\Pi_\text{ph}, \quad
\Gamma^{(2,\Delta)}_\Lambda = \Delta - g\Delta\Pi_\text{pp}.
\end{gather}
The vertex functions lead to the beta functions
\begin{subequations}
\begin{gather}
\beta_h(\bar{g},\bar{h}) = \left[ 1-\tilde{c}_\text{ph}(\bar{T}) \bar{g} \right] \bar{h}, \\
\beta_\Delta(\bar{g},\bar{\Delta}) = \left[ 1-\tilde{c}_\text{pp}(\bar{T})\bar{g} \right] \bar{\Delta}.
\end{gather}
\end{subequations}
Now we confirm by taking $T \to 0$ that the exponent for the pairing field $\Delta$ is the same independent of the RG schemes.  Particularly at the interacting fixed point, we obtain $\beta_\Delta(\bar{g}_2^*) = (1-\epsilon)\bar{\Delta}$.  This is consistent with the result from the energy-shell RG analysis.  The coefficient $\tilde{c}_\text{pp}$, which determines the value of the coupling constant at the interacting fixed point, does not appear to the exponent of the pairing field.

\subsection{Two-loop calculations}
\label{sec:anomalous_dimension}

So far we have calculated the perturbative corrections from the field theory approach to confirm that the two distinct RG schemes conclude the same physical results.
An advantage of the field theory approach is considerable when we deal with higher-order corrections.  In the following, we consider the two-loop corrections to the two-point correlation function at $T=0$ for the anomalous dimension and the correction to the energy dispersion.

The field renormalization is seen from the frequency dependence of the self-energy.  The linear term $\Sigma_\omega$ in Eq.~\eqref{eq:sigma_expansion} is given by
\begin{align}
\Sigma_\omega &= \frac{\partial}{\partial(i\omega)} \Sigma(\bm{k}=\bm{0},\omega) \bigg|_{\omega=0}
\equiv \sum_{j\geq 2} g^j \Sigma_\omega^{(j)}.
\end{align}
We expand $\Sigma$ with respect to the coupling constant $g$.
On the other hand, the zero-frequency part is related to corrections to the chemical potential and the energy dispersion:
\begin{align}
\Sigma_{\bm{k}} \equiv \Sigma(\bm{k},\omega=0) \equiv \sum_{j\geq 1} g^j \Sigma_{\bm{k}}^{(j)}.
\end{align}
The corrections $\delta\mu$, $\delta A_\pm$, and $\delta\lambda$ are obtained as
\begin{subequations}
\begin{gather}
\delta\mu = -\Sigma_{\bm{k}=\bm{0}} \equiv \sum_{j\geq 1} g^j \delta\mu^{(j)}, \\
\delta A_\pm = \pm\frac{\partial \Sigma_{\bm{k}}}{\partial k_\pm^{n_\pm}} \bigg|_{k=0} \equiv \sum_{j \geq 2} g^j \delta A_\pm^{(j)}, \\
\delta\lambda = \frac{\partial \Sigma_{\bm{k}}}{\partial \tilde{k}^n} \bigg|_{k=0} \equiv \sum_{j \geq 2} g^j \delta\lambda^{(j)}.
\end{gather}
\end{subequations}
We have used the fact that the one-loop correction, i.e., the Hartree contribution, does not yield the frequency or momentum dependence.

The renormalization condition Eq.~\eqref{eq:condition_2} reads
\begin{gather}
Z_\psi^{-1} = 1- \Sigma_\omega, \\
Z_\mu^{-1} = 1 + \frac{\delta\mu}{\mu}, \quad
Z_{A_\pm}^{-1} = 1 + \frac{\delta A_\pm}{A_\pm}, \quad
Z_\lambda^{-1} = 1 + \frac{\delta\lambda}{\lambda}.
\end{gather}
The field renormalization $\gamma_\psi$ is expressed from Eq.~\eqref{eq:gamma_def} as
\begin{align}
\label{eq:gamma_two-loop}
\gamma_\psi &= \Lambda\frac{\partial}{\partial\Lambda} \ln \left( 1 -\sum_{j\geq2} g^j \Sigma_\omega^{(j)} \right) \nonumber\\
&= -g^2 \Lambda\frac{\partial}{\partial\Lambda} \Sigma_\omega^{(2)} + O(g^3),
\end{align}
and the beta functions for the chemical potential and the coefficients of the energy dispersions are obtained from Eqs.~\eqref{eq:beta_mu_mod}--\eqref{eq:beta_lambda_mod} as
\begin{widetext}
\begin{subequations}
\begin{gather}
\label{eq:beta_mu_two-loop}
\beta_{\mu}
= -\Lambda\frac{\partial\bar{\mu}}{\partial\Lambda}
= \bar{\mu} \left[ 1 + g \Lambda\frac{\partial}{\partial\Lambda} \frac{\delta\mu^{(1)}}{\mu} + g^2 \Lambda\frac{\partial}{\partial\Lambda} \left( \frac{\delta \mu^{(2)}}{\mu} + \Sigma_\omega^{(2)} \right) + O(g^3) \right] , \\
\label{eq:beta_A_two-loop}
\beta_{A_\pm}
= -\Lambda\frac{\partial A_\pm}{\partial\Lambda}
= A_\pm \left[ g^2 \Lambda\frac{\partial}{\partial\Lambda} \left( \frac{\delta A_\pm^{(2)}}{A_\pm} + \Sigma_\omega^{(2)} \right) + O(g^3) \right] , \\
\label{eq:beta_lambda_two-loop}
\beta_{\lambda}
= -\Lambda\frac{\partial\bar{\lambda}}{\partial\Lambda}
= \bar{\lambda} \left[ a + g^2 \Lambda\frac{\partial}{\partial\Lambda} \left( \frac{\delta \lambda^{(2)}}{\lambda} + \Sigma_\omega^{(2)} \right) + O(g^3) \right] .
\end{gather}
\end{subequations}

We now calculate the two-loop correction to the self-energy $\Sigma^{(2)}$.
For the case of the contact interaction, there is only one connected two-loop diagram, i.e., the sunrise diagram shown in Fig.~\ref{fig:loop}(d) and \ref{fig:loop_two-point}(a) as the rightmost term.
The frequency and momentum dependent contribution appears from this diagram, calculated from
\begin{align}
\label{eq:two-loop_function}
\Sigma^{(2)}(k) &= -\int_{pql} G_{0\Lambda}(p) G_{0\Lambda}(q) G_{0\Lambda}(l) (2\pi)^{d+1} \delta(p+q-l-k).
\end{align}
We use the shorthand notations $p=(\bm{p},\omega_p)$ and $\int_p = \int\frac{d\omega_p}{(2\pi)}\int\frac{d^dp}{(2\pi)^d}$.
Then, we obtain the $\omega$-linear contribution
\begin{subequations}
\label{eq:two-loop}
\begin{align}
\label{eq:two-loop_frequency}
&\quad -\Lambda\frac{\partial}{\partial\Lambda} \Sigma_\omega^{(2)} \nonumber\\
&= \frac{\partial}{\partial(i\omega_k)} \int_{pql} (2\pi)^{d+1} \delta(p+q-l-k) \frac{1}{i\omega_p-E_{\bm{p}}} \frac{1}{i\omega_q-E_{\bm{q}}} \frac{1}{i\omega_l-E_{\bm{l}}}
\Lambda\frac{\partial}{\partial\Lambda} K_\Lambda(E_{\bm{p}}) K_\Lambda(E_{\bm{q}}) K_\Lambda(E_{\bm{l}}) \bigg|_{k=0} \nonumber\\
&= \left( \int_{\bm{p}\bm{q}}^+ \int_{\bm{l}}^- + \int_{\bm{p}\bm{q}}^- \int_{\bm{l}}^+ \right) (2\pi)^{d} \delta(\bm{p}+\bm{q}-\bm{l}) \frac{1}{(E_{\bm{p}}+E_{\bm{q}}-E_{\bm{l}})^2}
\Lambda\frac{\partial}{\partial\Lambda} K_\Lambda(E_{\bm{p}}) K_\Lambda(E_{\bm{q}}) K_\Lambda(E_{\bm{l}}) \nonumber\\
&= 2\Lambda^{-2\epsilon} \left( \int_{\bar{\bm{p}}\bar{\bm{q}}}^+ \int_{\bar{\bm{l}}}^- + \int_{\bar{\bm{p}}\bar{\bm{q}}}^- \int_{\bar{\bm{l}}}^+ \right) \frac{(2\pi)^{d} \delta(\bar{\bm{p}}+\bar{\bm{q}}-\bar{\bm{l}})}{(E_{\bar{\bm{p}}}+E_{\bar{\bm{q}}}-E_{\bar{\bm{l}}})^2}
\frac{3E_{\bar{\bm{p}}}^2E_{\bar{\bm{q}}}^2E_{\bar{\bm{l}}}^2 + 2(E_{\bar{\bm{q}}}^2E_{\bar{\bm{l}}}^2+E_{\bar{\bm{p}}}^2E_{\bar{\bm{l}}}^2+E_{\bar{\bm{p}}}^2E_{\bar{\bm{q}}}^2) + (E_{\bar{\bm{p}}}^2+E_{\bar{\bm{q}}}^2+E_{\bar{\bm{l}}}^2)}{(1+E_{\bar{\bm{p}}}^2)^2 (1+E_{\bar{\bm{q}}}^2)^2 (1+E_{\bar{\bm{l}}}^2)^2} \nonumber\\
&\equiv D^2(\Lambda) C^{(2)},
\end{align}
and the momentum-dependent part
\begin{align}
\label{eq:two-loop_momentum}
&\quad \Lambda\frac{\partial}{\partial\Lambda} \Sigma^{(2)}_{\bm{k}} \nonumber\\
&= -\int_{pql} (2\pi)^{d+1} \delta(p+q-l-k) \frac{1}{i\omega_p-E_{\bm{p}}} \frac{1}{i\omega_q-E_{\bm{q}}} \frac{1}{i\omega_l-E_{\bm{l}}}
\Lambda\frac{\partial}{\partial\Lambda} K_\Lambda(E_{\bm{p}}) K_\Lambda(E_{\bm{q}}) K_\Lambda(E_{\bm{l}}) \bigg|_{\omega_k=0} \nonumber\\
&= -\left( \int_{\bm{p}\bm{q}}^+ \int_{\bm{l}}^- + \int_{\bm{p}\bm{q}}^- \int_{\bm{l}}^+ \right) (2\pi)^{d} \delta(\bm{p}+\bm{q}-\bm{l}-\bm{k}) \frac{1}{E_{\bm{p}}+E_{\bm{q}}-E_{\bm{l}}}
\Lambda\frac{\partial}{\partial\Lambda} K_\Lambda(E_{\bm{p}}) K_\Lambda(E_{\bm{q}}) K_\Lambda(E_{\bm{l}}) \nonumber\\
&= -2\Lambda^{-2\epsilon} \left( \int_{\bar{\bm{p}}\bar{\bm{q}}}^+ \int_{\bar{\bm{l}}}^- + \int_{\bar{\bm{p}}\bar{\bm{q}}}^- \int_{\bar{\bm{l}}}^+ \right) \frac{(2\pi)^{d} \delta(\bar{\bm{p}}+\bar{\bm{q}}-\bar{\bm{l}}-\bar{\bm{k}})}{E_{\bar{\bm{p}}}+E_{\bar{\bm{q}}}-E_{\bar{\bm{l}}}}
\frac{3E_{\bar{\bm{p}}}^2E_{\bar{\bm{q}}}^2E_{\bar{\bm{l}}}^2 + 2(E_{\bar{\bm{q}}}^2E_{\bar{\bm{l}}}^2+E_{\bar{\bm{p}}}^2E_{\bar{\bm{l}}}^2+E_{\bar{\bm{p}}}^2E_{\bar{\bm{q}}}^2) + (E_{\bar{\bm{p}}}^2+E_{\bar{\bm{q}}}^2+E_{\bar{\bm{l}}}^2)}{(1+E_{\bar{\bm{p}}}^2)^2 (1+E_{\bar{\bm{q}}}^2)^2 (1+E_{\bar{\bm{l}}}^2)^2} \nonumber\\
&\equiv D^2(\Lambda) C^{(2)}_{\bm{k}}.
\end{align}
\end{subequations}
\end{widetext}
Here we denote the dimensionless quantities by adding bars; we define $\bar{\omega} = \omega/\Lambda$, $\bar{p}_+ = p_+/\Lambda^{1/n_+}$, and $\bar{p}_- = p_-/\Lambda^{1/n_-}$.  The momentum is scaled by $\Lambda$ so that the energy becomes dimensionless: $E_{\bar{\bm{k}}} = E_{\bm{k}}/\Lambda$.
$\int^{\pm}_{\bm{p}} = \int_{\bm{p}} \Theta (\pm E_{\bm{p}})$ stands for the momentum integral within the positive (negative) energy domain.
The constraints on the momentum integrals emerge after the frequency integrals.  They can be evaluated by identifying the position of poles on the complex plane, leading to the restricted regions of the momentum integrals.

We expect finite results for the two-loop results Eqs.~\eqref{eq:two-loop_frequency} and \eqref{eq:two-loop_momentum} at a saddle point of an energy dispersion because of the constraints on the momentum integrals $\int_{\bm{p}\bm{q}}^\pm \int_{\bm{l}}^\mp$.  The two-loop contributions vanish at a band edge since there is no sign change in the energy dispersion.

Now we scrutinize the frequency-dependent part $\Sigma_\omega^{(2)}$, which is responsible to the field renormalization and hence the anomalous dimension.  As we have discussed, the contribution vanishes at a band edge and thus an anomalous dimension does not arise.  It can be finite only at an energy saddle point.   In addition, it is worth pointing out that the integrand of Eq.~\eqref{eq:two-loop_frequency} is guaranteed to be positive.  Therefore, if there exists a finite volume that satisfies the constraint of the momentum integrals, we find a finite result: $C^{(2)}>0$.
The constraints on the momentum integrals can be rephrased as follows: There exists a momentum $\bm{l}=\bm{p}+\bm{q}$ such that $\operatorname{sgn}(E_{\bm{l}}) = -\operatorname{sgn}(E_{\bm{p}}) = -\operatorname{sgn}(E_{\bm{q}})$.  Such a momentum $\bm{l}$ in general exists near a saddle point because the energy dispersion near a saddle point comprises two or more filled Fermi seas and the area is not convex.
We do not further evaluate the expression of the two-loop correction as its value depends on the explicit form of the energy dispersion.

Equation~\eqref{eq:two-loop_frequency} defines a numerical factor $C^{(2)}$, which is independent of the cutoff $\Lambda$.  From Eq.~\eqref{eq:gamma_two-loop}, we find the field renormalization
\begin{equation}
\gamma_\psi = C^{(2)} \bar{g}^2 + O(\bar{g}^3).
\end{equation}
This quantity gives the anomalous dimension when evaluated at a fixed point.  It can be finite at the interacting fixed point to become
\begin{equation}
\eta = C^{(2)} \bar{g}_2^{*2} + O(\bar{g}_2^{*3}) (>0).
\end{equation}
A finite anomalous dimension concludes a non-Fermi liquid behavior at the interacting fixed point.  This happens at a saddle point of an energy dispersion with a power-law DOS singularity.

The uniform component of Eq.~\eqref{eq:two-loop_momentum} adds a correction to the beta function for the chemical potential Eq.~\eqref{eq:beta_mu_field-theory}, but it does not change the structure of the RG flow for small $\bar{g}$.  Here, we focus on the momentum dependence, which is absent to one-loop order.  Similarly to $C^{(2)}$, it becomes finite only at a saddle point of an energy dispersion but not at a band edge.  The momentum dependence of $C_{\bm{k}}^{(2)}$ leads to the beta functions
\begin{subequations}
\begin{gather}
\label{eq:beta_A_2}
\beta_{A_\pm} = A_\pm \left[ \bar{g}^2 \left( \pm\frac{1}{A_\pm}\left.\frac{\partial C_{\bm{k}}^{(2)}}{\partial \bar{k}_\pm^{n_\pm}}\right|_{\bm{k}=\bm{0}} - C^{(2)} \right) +O(\bar{g}^3) \right], \\
\label{eq:beta_lambda_2}
\beta_\lambda = \bar{\lambda} \left[ a + \bar{g}^2 \left( \frac{1}{\bar{\lambda}}\left.\frac{\partial C_{\bm{k}}^{(2)}}{\partial \tilde{\bar{k}}^n}\right|_{\bm{k}=\bm{0}} - C^{(2)} \right) +O(\bar{g}^3) \right].
\end{gather}
\end{subequations}
From Eqs.~\eqref{eq:beta_A_gamma} and \eqref{eq:beta_lambda_gamma}, we can identify the scaling exponents $\gamma_{A_\pm}$ and $\gamma_\lambda$.  The former affect the exponents of susceptibilities via Eqs.~\eqref{eq:scaling_k_tilde} and \eqref{eq:epsilon_tilde}.  We can see that a finite $C^{(2)}(>0)$ and hence an anomalous dimension has a negative contribution to $\gamma_{A_\pm}$.  When $\gamma_{A_+}$ and $\gamma_{A_-}$ are negative, we have $\tilde{\epsilon}>\epsilon$, leading to stronger divergences with respect to $T$, $\mu$, and $h$; see Eqs.~\eqref{eq:scaling_compressibility}--\eqref{eq:scaling_heat-capacity}.

When $\beta_\lambda(\bar{g},\bar{\lambda}=0)$ is finite, the relevant perturbation to the energy dispersion $\lambda\tilde{k}^n$ is generated under the RG analysis.  It is observed if $\partial C_{\bm{k}}^{(2)}/\partial \tilde{\bar{k}}^n |_{\bm{k}=\bm{0}}$ does not vanish when it is evaluated with $\lambda=0$.  Then, the beta function for $\lambda$ has the form $\beta_\lambda = (a+ c_1 \bar{g}^2)\bar{\lambda} + c_2 \bar{g}^2 + O(\bar{g}^3)$, where $c_1$ and $c_2$ are determined by Eq.~\eqref{eq:beta_lambda_2}.
This is analogous to the shift of the chemical potential when the Hartree term $\Sigma_\text{H}$ is finite, but it occurs at different order in $\bar{g}$.
Generation of $\bar{\lambda}$ curves the scale-invariant line in the phase diagram [Fig.~\ref{fig:phase_diagram-conj}(b)] in the $\bar{\lambda}$ direction at order $\bar{g}^2$, while a change in the $\bar{\mu}$ direction can happen at order $\bar{g}$.
Lastly, we note that the discussion from Eq.~\eqref{eq:two-loop} is general for any energy dispersion with a power-law divergent DOS, including Eq.~\eqref{eq:model_2} with a relevant perturbation $\lambda k_-^2$.

\section{Quasiparticle decay rate}
\label{sec:lifetime}

The preceding RG analyses focused on the real part of the self-energy or equivalently the two-point function.  They give rise to the corrections to the action, which are captured through the RG equations.  On the other hand, the imaginary part of the self-energy describes the damping of the quasiparticle, which is the focus of this section.
It is generated by the interaction in the present model.  Unlike the real part of the self-energy, the imaginary part can be calculated without a cutoff; we do not employ an RG method in this section, but integrate over the entire frequency and momentum space at once.

We calculate the quasiparticle decay rate $\Gamma(\bm{k},\omega)$, obtained from the retarded self-energy as
\begin{equation}
\Gamma(\bm{k},\omega) = -\operatorname{Im}\Sigma^R(\bm{k},\omega).
\end{equation}
The retarded self-energy $\Sigma^R(\bm{k},\omega)$ is calculated from the self-energy $\Sigma(\bm{k},\omega_n)$, with the analytic continuation of the Matsubara frequency to the real frequency: $i\omega_n = \omega + i\delta$ ($\delta$: infinitesimal positive quantity).
In the presence of the contact interaction, a finite imaginary part of the self-energy $\Sigma^R$ emerges at two-loop order and higher.  The one-loop correction, or the Hartree term $\Sigma_\text{H}$, does not yield a finite imaginary component.  Here we consider the two-loop diagram (the sunrise diagram) [Fig.~\ref{fig:loop}(d)] to calculate the quasiparticle decay rate $\Gamma$.  Like Eq.~\eqref{eq:two-loop_function}, it is given by
\begin{align}
\label{eq:two-loop_damping}
&\quad \Sigma^{(2)}(\bm{k},\omega_n) \nonumber\\
&= -T\sum_{\omega_p} T\sum_{\omega_q} T\sum_{\omega_l}
\int_{\bm{p}\bm{q}\bm{l}} G_0(\bm{p},\omega_p) G_0(\bm{q},\omega_q) G_0(\bm{l},\omega_l) \nonumber\\
&\quad\times \frac{(2\pi)^d}{T} \delta(\omega_p+\omega_q-\omega_l-\omega_n)  \delta(\bm{p}+\bm{q}-\bm{l}-\bm{k}),
\end{align}
but we do not need a cutoff for the imaginary part.

The calculation of $\Sigma^{(2)}$ is standard and can be found in e.g. Ref.~\cite{AGD}; we also show the derivation in Appendix~\ref{sec:two-loop_lifetime} and just present the result here.  The quasiparticle decay rate to two-loop order is given by $\Sigma^{(2)}$ after the analytic continuation:
\begin{align}
&\quad \Gamma(\bm{k},\omega) \nonumber\\
&= -g^2 \operatorname{Im}\Sigma^{(2)R}(\bm{k},\omega) \nonumber\\
&= \frac{\pi}{4} g^2 \cosh\left(\frac{\omega}{2T}\right) \nonumber\\
&\quad \times \int_{\bm{p}\bm{q}} \frac{\delta(\omega-E_{\bm{p}}-E_{\bm{q}}+E_{\bm{p}+\bm{q}-\bm{k}})}{\cosh\left(\dfrac{E_{\bm{p}}}{2T}\right) \cosh\left(\dfrac{E_{\bm{q}}}{2T}\right) \cosh\left(\dfrac{E_{\bm{p}+\bm{q}-\bm{k}}}{2T}\right)} .
\end{align}
This relation holds for an arbitrary energy dispersion $E_{\bm{k}}$.

We extract the temperature dependence by introducing dimensionless quantities in terms of temperature $T$: we define $\tilde{p}_\pm = p_\pm/T^{1/n_\pm}$, so that the energy dispersion satisfies $E_{\tilde{\bm{p}}}=E_{\bm{p}}/T$.
Here we are interested in the low-frequency limit with $\omega \ll T$.
By substituting $\bm{k}=\bm{0}$ and $\omega=0$, we obtain the temperature dependence \cite{multicritical1}
\begin{align}
\label{eq:decay_rate_0}
\Gamma &= \frac{\pi}{4} g^2 T^{1-2\epsilon} \int_{\tilde{\bm{p}}\tilde{\bm{q}}} \frac{\delta(E_{\tilde{\bm{p}}}+E_{\tilde{\bm{q}}}-E_{\tilde{\bm{p}}+\tilde{\bm{q}}})}{\cosh\left(\dfrac{E_{\tilde{\bm{p}}}}{2}\right) \cosh\left(\dfrac{E_{\tilde{\bm{q}}}}{2}\right) \cosh\left(\dfrac{E_{\tilde{\bm{p}}+\tilde{\bm{q}}}}{2}\right)} \nonumber\\
&\propto T^{1-2\epsilon}.
\end{align}
The integral gives a finite constant without a cutoff.

In the Fermi liquid theory, when the temperature is much smaller than the Fermi energy $\epsilon_F \gg T$, the decay rate is proportional to $T^2$.  This result relies on the existence of the Fermi surface with finite DOS.  On the other hand, the decay rate Eq.~\eqref{eq:decay_rate_0} is distinct from the Fermi liquid results, reflecting the divergent DOS at $\mu=0$.
The behavior is different also from the case for a conventional VHS with a logarithmic DOS, which shows a (marginal) Fermi liquid behavior \cite{Pattnaik,Gopalan,Dzyaloshinskii2,Menashe}.
We note that the result does not depend on whether the power-law divergent DOS is located at a saddle point or a band edge of the energy dispersion.  This is in contrast to the anomalous dimension, which can only be found at a saddle point as we have discussed in Sec.~\ref{sec:anomalous_dimension}.

\section{Summary and discussions}
\label{sec:discussions}

We now summarize our main results, compare supermetal with normal metal and other non-Fermi liquid systems, and discuss experimental signatures of supermetal.

\subsection{Summary}

We have analyzed electron interaction effects near a high-order VHS with a scale-invariant Fermi surface with a power-law divergent DOS.
Scale invariance of the system allows an RG analysis to search for fixed points and a scaling analysis of thermodynamic quantities and correlation functions around the fixed points.

The one-loop RG analysis finds that electron interaction around high-order VHS offers a fermionic analog of the $\phi^4$ theory.  We have identified the two RG fixed points: the noninteracting and interacting fixed points.  The latter is an analog of the Wilson--Fisher fixed point in the $\phi^4$ theory.  Like the $\phi^4$ theory, the noninteracting fixed point is unstable and the interacting fixed point is stable in terms of the RG flow of the coupling constant.

We performed a controlled RG analysis up to two-loop order about the interacting fixed point owing to the smallness of the DOS singularity exponent $\epsilon$.  We reveal that the quantum critical metal at the interacting fixed point is a non-Fermi liquid that exhibits a finite anomalous dimension of electrons, and power-law divergent charge and spin susceptibilities.
We term such a metallic state with various divergent susceptibilities but yet without a long-range order as a supermetal.
In this regard, the noninteracting fixed point can be viewed as a noninteracting supermetal and the interacting fixed point as an interacting supermetal.  

A supermetal appears at the topological transition between electron and hole Fermi liquids. An interacting supermetal is a multicritical state reached by tuning two parameters---chemical potential and detuning of energy dispersion from high-order saddle point. Combining the RG and mean-field analyses, we conjecture a global phase diagram where a supermetal is at the border between electron/hole Fermi liquids and on the verge of becoming ferromagnetic.

\subsection{Comparison with normal metal and other non-Fermi liquids}

It is worth drawing a comparison between a supermetal and a normal metal.  Being at finite density, a normal metal is characterized by a Fermi surface with a characteristic momentum scale.
The RG theory of metals with a closed Fermi surface, commonly referred to as Shankar's RG \cite{Shankar}, requires a judicious RG procedure that only consider electrons within a \textit{small} energy shell around the Fermi surface.
Then, Fermi liquid appears as the RG fixed point in the limit that the energy range is taken to zero. It is characterized by an infinite number of marginal coupling constants, i.e., Landau forward scattering parameters in all angular momentum channels. Moreover, this Fermi liquid fixed point is only stable when BCS interactions in all angular momentum channels are repulsive \cite{Kohn-Luttinger}.

These behaviors of a normal metal should be contrasted with the case of a supermetal. Our theory is formulated with a large UV energy cutoff on the order of bandwidth. The supermetal fixed point is characterized by a single coupling constant---the contact interaction, with all other interactions being irrelevant and without suffering from the Kohn--Luttinger instability to superconductivity.

We have shown that a high-order saddle point with repulsive interaction exhibits the non-Fermi liquid behavior.  Non-Fermi liquids are realized also in e.g., one-dimensional systems, other kinds of quantum critical metals, and doped Mott insulators.
In a one-dimensional electronic system, there is no quasiparticle, but instead, collective charge and spin waves are the elementary excitations \cite{Tomonaga,Luttinger,Haldane}.  Electron interaction as a forward scattering renormalizes the velocities of the charge and spin modes separately, thus leading to a non-Fermi liquid.

Great efforts have been devoted to search for generalized Luttinger liquids  in dimensions greater than one. While significant progress has been made in quasi-one dimensional systems \cite{Kane,Vishwanath}, to our knowledge results are limited on Luttinger liquid type behavior in metals with truly two-dimensional Fermi surface.

Another situation for a non-Fermi liquid arises around a quantum critical point where strong electron interaction drives a phase transition from a metallic state to a symmetry-breaking ordered state at $T=0$ \cite{NFL1,NFL1a,review2,NFL2}.  Seminal works by Hertz \cite{Hertz}, Moriya \cite{Moriya}, and Millis \cite{Millis} deal with the quantum critical phenomenon in itinerant magnets, which describe the coupling between electrons with a finite Fermi surface and bosonic fluctuations of an order parameter near a magnetic transition.  In their theories, low-energy modes of electrons are integrated out to yield a nonlocal singular effective action for bosonic modes.  This challenging problem has invoked intense work and considerable progress \cite{Pankov,Rech,Senthil2,Mross,Lee1,Fitzpatrick1,Fitzpatrick2,NFL3,Berg}.

Non-Fermi liquids have also been proposed in doped Mott insulators close to near superconductivity \cite{review1,review_LNW}, electronic liquid-crystal phases, \cite{Kivelson1,Fradkin,Emery,Oganesyan,Kivelson_review}, fractionalized electron systems \cite{Senthil1,orthogonal,Fisher}, and near superconductor-insulator transition \cite{Feigelman,Das2,Dalidovich,Galitski,Motrunich}.

Unlike these non-Fermi liquids, the supermetal we found near a high-order VHS is obtained under \textit{weak} electron interaction. It relies on the singular DOS instead of singular interaction. This feature enabled us to develop an analytically controlled theory of supermetal with local interaction, using a small parameter---the DOS exponent.

Quantum criticality in nodal semimetals also hosts a non-Fermi liquid \cite{Abrikosov1,Abrikosov2}.  The vanishing DOS allows unscreened long-range Coulomb interaction, which is dressed by the gapless electron spectrum.  Possible platforms include graphene \cite{Gonzalez2,DasSarma,Son}, pyrochlore iridates \cite{Moon,Savary}, and topological phase transition in two dimensions \cite{Isobe2,Cho}.
The RG procedure for semimetals has a similar spirit as ours for supermetal, both of which are Wilson--Fisher type instead Shankar type.
Unlike semimetals though, a supermetal has divergent DOS at Fermi level and an \textit{extended} Fermi surface.

We also note that a saddle point of an energy dispersion in two dimensions $E_{\bm{k}}=k_x^2-k_y^2$ gives a conventional VHS with a logarithmic divergence mentioned above.
This is also scale invariant; however, it requires an additional care in an RG analysis \cite{Kallin,Kapustin}.  As we mentioned in Sec.~\ref{sec:energy-shell}, momentum integrals in a perturbative RG calculation suffer from a singularity at $k\to\infty$.  This cannot be regularized by the UV energy cutoff $\Lambda$ since every energy contour extends to $k\to\infty$, so that a UV momentum cutoff is needed in addition.  This is also related to the non-analyticity of the DOS.  As a result, the UV  cutoff is not eliminated from RG equations.  It occurs as a sequel that the low-energy physics is affected by the UV scale $\Lambda$.

We have listed several other non-Fermi liquid systems.  Supermetal fixed points are regarded as multicritical points in the phase space spanned by the coupling constant, the chemical potential, and a parameter for the energy dispersion (Fig.~\ref{fig:phase_diagram-conj}).  Importantly, our present analysis does not suffer from the difficulties related to a closed Fermi surface, the presence of a length scale, a logarithmic DOS, and a singular bosonic fluctuation.
A small DOS singularity exponent guarantees that only the short-range interaction is relevant.

\subsection{Discussion}

Our analysis is for the case of a single high-order VHS in the Brillouin zone at the energy range in focus.
In reality, materials may have multiple high-order VHS points at the same energy in the Brillouin zone, related by symmetry.  In that case, additional interactions involving different VHS should be included and their presence may lead to symmetry-breaking instabilities \cite{multicritical1}, as opposed to quantum criticality.
However, for a certain parameter range before an ordering instability takes place, there could exist a scaling region where thermodynamic or transport quantities follow scaling properties.
For example, when temperature $T$ and the carrier density $n$ are control parameters, a physical quantity $Q$ follows the scaling relation
\begin{equation}
Q(T,n) = T^a \hat{\mathcal{F}}\left( n T^{-(1-\epsilon)} \right)
\end{equation}
around the noninteracting fixed point, where the exponent $a$ is determined by a dimensional analysis of $Q$ and $\hat{\mathcal{F}}$ is a scaling function.

\begin{acknowledgments}

We acknowledge helpful discussions with A. V. Chubukov, E. Fradkin, S.-S. Lee, and especially E. Berg.  This work was supported by DOE Office of Basic Energy Sciences, Division of Materials Sciences and Engineering under Award DE-SC0018945.  LF was supported in part by a Simons Investigator Award from the Simons Foundation.

\end{acknowledgments}

\appendix

\section{One-loop energy-shell RG analysis at finite temperature}
\label{sec:finite-temperature}

We work on the RG equations at finite temperature $T \neq 0$ with the energy-shell RG analysis to one-loop level.  Temperature has a dimension of energy, so that it is one of relevant perturbations.  The one-loop corrections $\Sigma_\text{H}$, $\Pi_\text{pp}$, and $\Pi_\text{ph}$, shown in Eq.~\eqref{eq:one-loop_T0} at $T=0$, should be calculated at $T\neq 0$.  To order $l$, we obtain
\begin{subequations}
\begin{gather}
\Sigma_\text{H} = T\sum_{\omega_n} \int_{\bm{k}}^> G_0(\bm{k},\omega_n) \simeq -l c_\text{H}(\bar{T}) \Lambda D(\Lambda) , \\
\Pi_{\text{pp}} = T\sum_{\omega_n} \int_{\bm{k}}^> G_0(\bm{k},\omega_n) G_0(-\bm{k},-\omega_n) \simeq l c_\text{pp}(\bar{T}) D(\Lambda) , \\
\Pi_\text{ph} = T\sum_{\omega_n} \int_{\bm{k}}^> G_0(\bm{k},\omega_n) G_0(\bm{k},\omega_n) \simeq -l c_\text{ph}(\bar{T}) D(\Lambda) ,
\end{gather}
\end{subequations}
where we introduce the dimensionless temperature $\bar{T}=T/\Lambda$.
Again, all quantities are evaluated at zero external frequency and momentum, so that the results depend only on the DOS.
The temperature-dependent dimensionless coefficients $c_{\mu}$, $c_\text{pp}$, and $c_\text{ph}$ are
\begin{subequations}
\begin{gather}
\label{eq:c_mu}
c_\text{H}(\bar{T}) =  \frac{1}{2} \left( 1-\frac{D_-}{D_+} \right) \tanh\left(\frac{1}{2\bar{T}}\right), \\
\label{eq:c_pp}
c_\text{pp}(\bar{T}) = \frac{1}{2} \left( 1+\frac{D_-}{D_+} \right) \tanh\left(\frac{1}{2\bar{T}}\right), \\
\label{eq:c_ph}
c_\text{ph}(\bar{T}) = \frac{1}{2\bar{T}} \left( 1+\frac{D_-}{D_+} \right) \frac{1}{2\cosh^2\left(\dfrac{1}{2\bar{T}}\right)}.
\end{gather}
\end{subequations}
Unlike the calculation at $T=0$, $\Pi_\text{ph}$ becomes finite for $T\neq 0$, while the correction to the field or the energy dispersion remains absent to one-loop order.

In the analysis at zero temperature, we rescale the frequency as Eq.~\eqref{eq:omega_tree}.  At finite temperature, rescaling of the Matsubara frequency leads to rescaling of temperature \cite{Millis}.  Temperature obeys the same scaling relation as that for the frequency: $T' = bT$.

Including the temperature-dependent factors $c_{\mu}(\bar{T})$, $c_\text{pp}(\bar{T})$, and $c_\text{ph}(\bar{T})$, we obtain the changes of the parameters at an RG step as
\begin{subequations}
\begin{gather}
T' = b T, \\
\mu' \simeq b [\mu + l c_\text{H}(\bar{T}) g \Lambda D(\Lambda) - l c_\text{ph}(\bar{T}) g \mu D(\Lambda) ], \\
h' = b [h - l c_\text{ph}(\bar{T})g h D(\Lambda)], \\
\Delta' \simeq b [\Delta - l c_\text{pp}(\bar{T}) g \Delta D(\Lambda)], \\
g' \simeq b^{\epsilon} \{ g - l g^2 [c_\text{pp}(\bar{T}) - c_\text{ph}(\bar{T})] D(\Lambda) \}.
\end{gather}
\end{subequations}
Then, we reach the RG equations
\begin{subequations}
\begin{gather}
\frac{d\bar{T}}{dl} = \bar{T}, \\
\frac{d\bar{\mu}}{dl} = [ 1- c_\text{ph} (\bar{T})\bar{g} ] \bar{\mu} + c_\mu(\bar{T}) \bar{g}, \\
\frac{d\bar{h}}{dl} = [ 1- c_\text{ph}(\bar{T})\bar{g} ] \bar{h}, \\
\frac{d\bar{\Delta}}{dl} = [ 1- c_\text{pp}(\bar{T})\bar{g} ] \bar{\Delta}, \\
\frac{d\bar{g}}{dl} = \epsilon \bar{g} - \left[ c_\text{pp}(\bar{T})-c_\text{ph}(\bar{T}) \right] \bar{g}^2.
\end{gather}
\end{subequations}
Since temperature $T$ is relevant and its fixed point is located at $T=0$, the fixed points of the parameters are found at $T=0$, as we discussed in the main part.
The RG equations are consistent with the beta functions derived by the field theory approach in Sec.~\ref{sec:field_RG_one-loop} and \ref{sec:field_RG_h-Delta}.

\section{Ward--Takahashi identity}
\label{sec:ward-takahashi}

\subsection{Derivation}

We show the derivation of the Ward--Takahashi identity, based on a diagrammatic discussion by Peskin and Schroeder \cite{Peskin}.
It consists of two parts: equalities for a through-going line from an initial state to a final state and an internal loop.
Suppose that we calculate a diagram with the same numbers of incoming and outgoing electron lines.  Then we can decompose the diagram into lines that connect an incoming and an outgoing line, and internal loops.
For the present analysis with the contact interaction, we choose each decomposed diagram to include only one spin species.  We set $T=0$ in the following discussion.

To derive the Ward--Takahashi identity, we insert an external line for the bosonic field $\varphi(k)$ with $k=(\bm{k},\omega_k)$.
The coupling between an electron and the bosonic field is given by
\begin{equation}
\alpha_\sigma \bar{\psi}_\sigma(p+k) \varphi(k) \psi_\sigma (p),
\end{equation}
where we introduce the spin-dependent coupling constant $\alpha_\sigma$.

\begin{figure}
\centering
\includegraphics[width=\hsize]{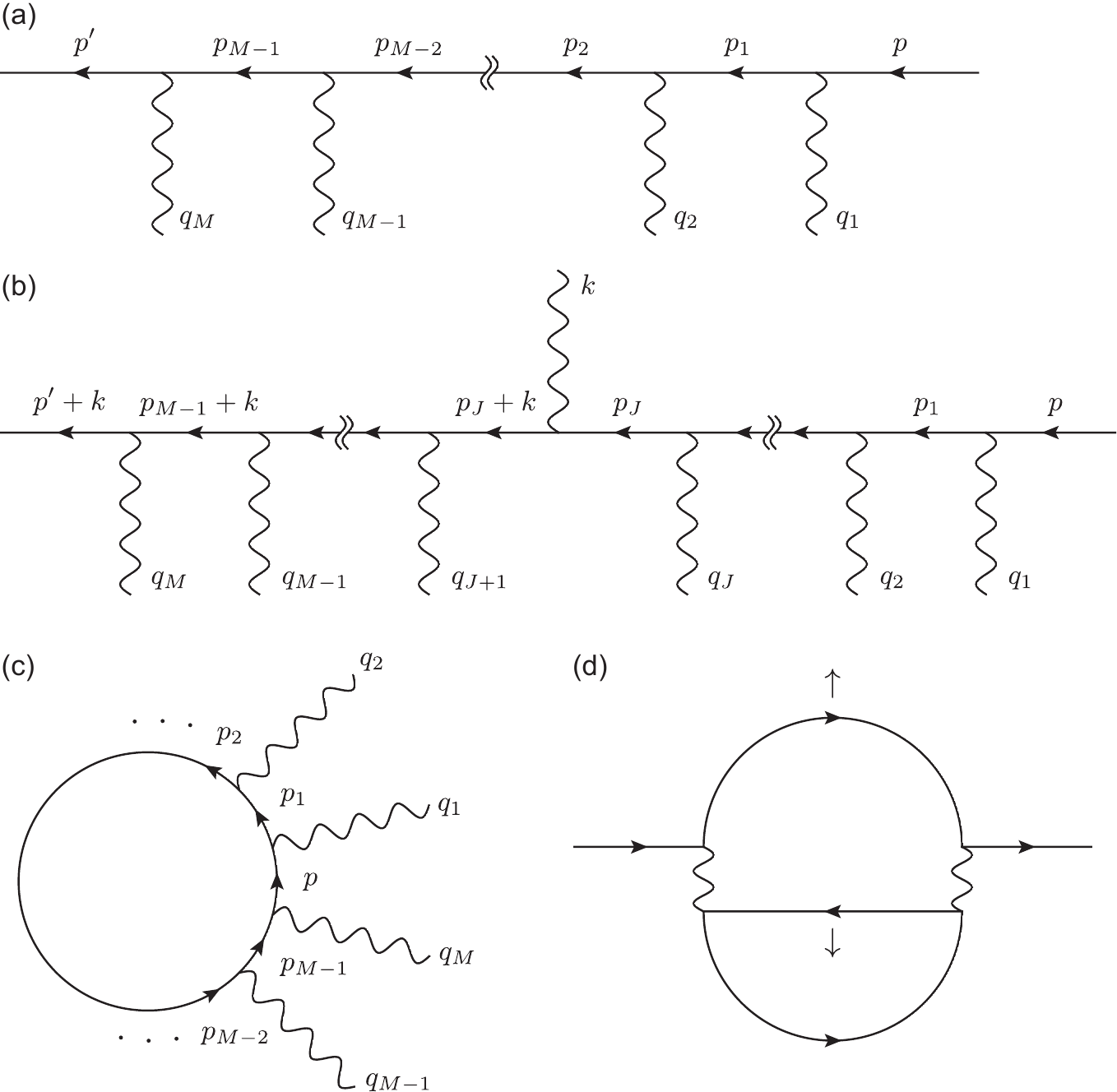}
\caption{
Diagrammatic representations for
(a) $L_M^\sigma(p',p;\{q\})$,
(b) $L_{M,J}^\sigma(p',p;\{q\};k)$, and
(c) $R_M^\sigma(\{q\})$.
(d) Decomposition of a sunrise diagram for $\Gamma_\uparrow^{(2)}$.  It consists of a line $L^\uparrow$ and a loop $R^\downarrow$.
The wavy lines represent a bosonic field; in the present case, internal wavy lines can be regarded as a Hubbard--Stratonovich field for the contact interaction.
}
\label{fig:WT_2}
\end{figure}

\begin{widetext}
\textit{Step 1}.
For a line connecting an initial state and a final state, the corresponding equation contains a product of the noninteracting Green's functions.  When we consider an electron line with $(M+1)$ electron line segments, the product is written as
\begin{equation}
L_M^\sigma(p',p;\{q\}) = G_0(p') G_0(p_{M-1}) G_0(p_{M-2}) \cdots G_0(p_2) G_0(p_1) G_0(p),
\end{equation}
with $p_J = p_{J-1} + q_J$, $p_0=p$, $p_M=p'$, and $\{q\}=(q_1,q_2,\cdots,q_M)$ [Fig.~\ref{fig:WT_2}(a)].
Remember that all connected line segments has the spin index $\sigma$.

Now we insert a vertex to the $J$-th line, connecting a bosonic line carrying momentum and frequency $k$ [Fig.~\ref{fig:WT_2}(b)].  We denote it as $L_{M,J}^\sigma(p',p;\{q\};k)$, given by
\begin{equation}
L_{M,J}^\sigma(p',p;\{q\};k) =
G_0(p'+k) G_0(p_{M-1}+k) \cdots G_0(p_J+k) G_0(p_J) G_0(p_{J-1}) \cdots G_0(p_1) G_0(p).
\end{equation}
For $k=(\bm{0},\omega_k)$, the following equality holds:
\begin{equation}
\label{eq:Ward_Green_0}
G_{0}^{-1}(p+k) - G_{0}^{-1}(p) = i\omega_k,
\end{equation}
or equivalently
\begin{equation}
\label{eq:Ward_Green}
G_0(p+k) (i\omega_k) G_0(p)
= G_0(p) - G_0(p+k).
\end{equation}
Using the equality, we obtain
\begin{equation}
\label{eq:Lnj}
(i\omega_k) L_{M,J}^\sigma(p',p;\{q\};k) =
G_0(p'+k) G_0(p_{M-1}+k) \cdots \left[ G_0(p_J) - G_0(p_J+k) \right] G_0(p_{J-1}) \cdots G_0(p_1) G_0(p).
\end{equation}
If we insert a vertex on the $(J-1)$-st line, we then have
\begin{equation}
\label{eq:Lnj-1}
(i\omega_k) L_{M,J-1}^\sigma(p',p;\{q\};k) =
G_0(p'+k) G_0(p_{M-1}+k) \cdots G_0(p_J+k) \left[ G_0(p_{J-1}) - G_0(p_{J-1}+k) \right] \cdots G_0(p_1) G_0(p).
\end{equation}
We find a cancellation of the second term in the brackets in Eq.~\eqref{eq:Lnj} by the first term in the brackets in Eq.~\eqref{eq:Lnj-1} when we sum the two.
Summing all possible insertions of vertices, we find
\begin{align}
\label{eq:Ward_line}
(i\omega_k) \sum_{J=0}^{M} L_{M,J}^\sigma(p',p;\{q\};k) = L_M^\sigma(p',p;\{q\}) - L_M^\sigma(p'+k,p+k;\{q\}).
\end{align}

\end{widetext}

\textit{Step 2}.
We apply a similar argument for an internal loop.  The difference from the discussion for a line is that internal frequency and momentum that run through the loop have to be integrated out.  Thus, the equation corresponding to the loop diagram is given by
\begin{equation}
R_M^\sigma(\{q\}) = \int_{p} \tilde{L}_M^\sigma(p;\{q\}).
\end{equation}
It is diagrammatically depicted in Fig.~\ref{fig:WT_2}(c).
Here, the loop consists of $M$ electron lines and all lines have the same spin index $\sigma$.  $\tilde{L}^\sigma_M(p;\{q\})$ is defined from $L^\sigma_M(p',p;\{q\})$, by imposing $p=p'$ and removing one $G_0(p)$ to close the loop.

We then insert a vertex on the $J$-th electron line and write it as
\begin{equation}
R_{M,J}^\sigma(\{q\};k) = \int_{p} \tilde{L}_{M,J}^\sigma(p,\{q\};k),
\end{equation}
where $\tilde{L}^\sigma_{M,J}$ is defined from $L^\sigma_{M,J}$ in the same way as $\tilde{L}_M^\sigma$ from $L_M^\sigma$.
The summation of all possible insertions of vertices yields a similar equation as Eq.~\eqref{eq:Ward_line}, but it vanishes as the frequency and momentum are conserved along the loop:
\begin{align}
\label{eq:Ward_loop}
&\quad (i\omega_k) \sum_{J=1}^M R_{M,J}^\sigma(\{q\};k) \nonumber\\
&= \int_p \left[ \tilde{L}^\sigma_{M,J}(p;\{q\}) - \tilde{L}^\sigma_{M,J}(p+k;\{q\}) \right] \nonumber\\
&=0.
\end{align}

\textit{Step 3}.
By combining the results of \textit{Step 1} and \textit{Step 2}, we deduce the Ward--Takahashi identity.  We introduce the $N$-point function $\Gamma^{(N)}(\{p'\},\{p\})$ and the associated vertex function $\Gamma^{(N,\alpha_\sigma)}(\{p'\},\{p\};k)$, which is obtained by attaching an external bosonic line with frequency and momentum $k$ to the $N$-point function $\Gamma^{(N)}$.  $\{p\}$ stands for the set of frequencies and momenta $(p_1,\cdots,p_N)$.
Frequency and momentum should be conserved before and after the process: $\sum p'_J = \sum p_J$ for $\Gamma^{(N)}$ and $\sum p'_J = \sum p_J + k$ for $\Gamma^{(N,g_\sigma)}$.

The $N$-point function can be decomposed into $N/2$ lines $L^\sigma$ and loops $R^\sigma$.  [For example, Fig.~\ref{fig:WT_2}(d) shows a decomposition of a sunrise diagram for a two-point function $(N=2)$ into $L^\uparrow$ and $R^\downarrow$.]
Internal frequencies and momenta $\{q\}$ are to be integrated out.  Therefore, from Eqs.~\eqref{eq:Ward_line} and \eqref{eq:Ward_loop}, we obtain
\begin{align}
\label{eq:Ward_N-point}
&\quad -(i\omega_k) \Gamma^{(N,\alpha_\sigma)}(\{p'\},\{p\};k) \nonumber\\
&= \sum_{J=1}^{N} \left[ \Gamma^{(N)}(\{p'\}_{J,-k}^\sigma,\{p\}) - \Gamma^{(N)}(\{p'\},\{p\}_{J,+k}^\sigma) \right].
\end{align}
We use the notation $\{p\}_{J,+k}^\sigma = (p_1,\cdots, p_J+k, \cdots, p_N)$, where the addition of $k$ occurs only when $p_J$ is associated with an electron with spin $\sigma$.  $\{p\}_{J,+k}^\sigma$ reflects the spin-dependent coupling $\alpha_\sigma$.  The sign $(-1)$ on the left-hand side is solely up to the definition of the vertex function.
Equation~\eqref{eq:Ward_N-point} is the Ward--Takahashi identity for the $N$-point function.
For the present model, internal bosonic lines mediate the contact interaction and external lines represent an external field as a perturbation, such as the chemical potential $(\alpha_\uparrow = \alpha_\downarrow = -\mu)$ and the magnetic field $(\alpha_\uparrow = -\alpha_\downarrow = h)$.

Here, we considered the case where the external bosonic field carries finite frequency only, so that the right-hand side of Eq.~\eqref{eq:Ward_Green_0} is simply proportional to the frequency.  The extension to a case with a finite momentum is straightforward, but the right-hand side becomes not as simple as that for frequency, which depends on the energy dispersion.

\subsection{Ward identity for two-point functions}

The Ward--Takahashi identity originates from conservation laws: the expression Eq.~\eqref{eq:Ward_N-point} obeys due to the charge conservation for each spin.  The identity yields a relation between a self-energy and a vertex function, thus leading to a relation between the field renormalization and an exponent for a thermodynamic quantity.
We can see this from a two-point function $(N=2)$.  The Ward--Takahashi identity for $N=2$ is written as
\begin{equation}
(i\omega_k) \Gamma^{(2,\alpha_\sigma)}_\sigma(p+k,p;k) = \Gamma^{(2)}_\sigma(p+k) - \Gamma^{(2)}_\sigma(p),
\end{equation}
where the subscript $\sigma$ is added to show the external lines correspond to electrons with spin $\sigma$.
The two-point function is written with the self-energy $\Sigma_\sigma$ as
\begin{equation}
\Gamma^{(2)}_\sigma(p) = -i\omega_p + E_{\bm{p}} + \Sigma_\sigma(p).
\end{equation}
Therefore, in the limit $\omega_k\to0$, we obtain the Ward identity, which relates the vertex function $\Gamma^{(2,\alpha_\sigma)}_\sigma$ and the self-energy $\Sigma_\sigma$:
\begin{equation}
\label{eq:Ward_phi-g}
\Gamma^{(2,\alpha_\sigma)}_\sigma(p,p;0) = 1 - \frac{\partial \Sigma_\sigma(p)}{\partial (i\omega_p)}.
\end{equation}

\section{Brief review of the field theory approach to RG analyses}
\label{sec:field-theory_review}

Here we describe a field theory approach to RG equations, in light of the Wilsonian approach.  We derive the Callan--Symanzik equation and the beta functions, which show how scale-dependent parameters affect physical quantities.  We partly owe the following descriptions to the references \cite{Zinn-Justin,Peskin,Amit,Nair}.

For the sake of clarity, we consider a theory with a scalar field $\phi$ and a set of  dimensionless parameters $\{\bar{g}_\rho\}$, where we write the action as $S[\phi;\bar{g}]$.  The partition function is given by
\begin{equation}
\mathcal{Z} = \int D\phi e^{-S[\phi;\bar{g}]}.
\end{equation}
When the model suffers from UV divergences, i.e., perturbative loop corrections have UV divergences, we need to cure them to extract meaningful information.  Those UV divergences can be regularized by removing UV modes from the model.  To this end, we decompose the field $\phi$ depending on the energy range to which they contribute: $\phi_{\Lambda'}^{\Lambda}$ accounts for the energy between $\Lambda'$ and $\Lambda$.
Then, we redefine the partition function as
\begin{equation}
\mathcal{Z}_{\Lambda_0}[\bar{g}_{0}] = \int D\phi_0^{\Lambda_0} e^{-S_{\Lambda_0}[\phi_0^{\Lambda_0};\bar{g}_{0}]}.
\end{equation}
It does not obviously have a UV divergence because no UV modes are included.  Here the energy scale $\Lambda_0$ works as a UV energy cutoff.  (This is equivalent to impose the effective action to be finite at $\Lambda_0$ without a cutoff.  To this end, divergent counterterms should be introduced to cure UV divergences.)

The scale $\Lambda_0$ is an arbitrary energy scale to regularize UV divergences.  The next thing we should check is how a change of the characteristic energy scale affects the theory.
To see it, we define the effective action at an energy scale $\Lambda(<\Lambda_0)$ as
\begin{align}
S_\Lambda^\text{eff}[Z_\Lambda^{1/2} \phi_0^\Lambda;\bar{g}(\Lambda)] = -\ln \left[ \int D\phi_\Lambda^{\Lambda_0} e^{-S_{\Lambda_0}[\phi_0^\Lambda + \phi_\Lambda^{\Lambda_0};\bar{g}_{0}]} \right].
\end{align}
We require that the effective action $S^\text{eff}$ have the same form as the action $S$.
Then, the partition function can be written as
\begin{align}
\mathcal{Z}_{\Lambda_0} [\bar{g}_{0}]
&= \int D\phi_0^\Lambda D\phi_\Lambda^{\Lambda_0} e^{-S_{\Lambda_0}[\phi_0^\Lambda + \phi_\Lambda^{\Lambda_0};\bar{g}_{0}]} \nonumber\\
&= \int D\phi_0^{\Lambda} e^{-S^\text{eff}_\Lambda [Z_\Lambda^{1/2} \phi_0^\Lambda;\bar{g}(\Lambda)]} \nonumber\\
&\equiv \mathcal{Z}_\Lambda [Z_\Lambda^{1/2} \phi_0^\Lambda;\bar{g}(\Lambda)].
\end{align}
This is simply rewriting of the partition function with the effective action at the scale $\Lambda$.  We relate the partition functions at different scales to find
\begin{align}
\mathcal{Z}_\Lambda [Z_\Lambda^{1/2} \phi_0^\Lambda;\bar{g}(\Lambda)]
= \mathcal{Z}_{\Lambda'} [Z_{\Lambda'}^{1/2} \phi_0^{\Lambda'};\bar{g}(\Lambda')].
\end{align}
This equality tells us that we have the same partition function defined at different energy scales $\Lambda$ and $\Lambda'$, together with the changes of the weight $Z$ and the parameters $\bar{g}_\rho$.

We then aim to calculate the $N$-point correlation function with the cutoff $\Lambda$ and the parameters $\bar{g}_\rho(\Lambda)$:
\begin{align}
&\quad \langle \phi_0^{\Lambda}(k_1) \cdots \phi_0^{\Lambda}(k_N) \rangle_{\Lambda;\bar{g}(\Lambda)} \nonumber\\
&= \frac{1}{\mathcal{Z}_{\Lambda}[\phi_0^\Lambda;\bar{g}(\Lambda)]} \int D\phi_0^{\Lambda} \phi_0^{\Lambda}(k_1) \cdots \phi_0^{\Lambda}(k_N) e^{-S^\text{eff}_{\Lambda}[\phi_0^{\Lambda};\bar{g}(\Lambda)]}.
\end{align}
When all momenta $k_a$ correspond to energies below $\Lambda$ and $\Lambda'$, we find $\phi_0^{\Lambda}(k_a) = \phi_0^{\Lambda'}(k_a)$, which enables us to relate the $N$-point correlation functions at different scales as
\begin{align}
&\quad Z_{\Lambda}^{-N/2} \langle \phi_0^{\Lambda}(k_1) \cdots \phi_0^{\Lambda}(k_N) \rangle_{\Lambda;\bar{g}(\Lambda)} \nonumber\\
&= Z_{\Lambda'}^{-N/2} \langle \phi_0^{\Lambda'}(k_1) \cdots \phi_0^{\Lambda'}(k_N) \rangle_{\Lambda';\bar{g}(\Lambda')}.
\end{align}
We now write this relation using the connected $N$-point correlation function $G^{(N)}$:
\begin{align}
Z_{\Lambda}^{-N/2} G_{\Lambda;\bar{g}(\Lambda)}^{(N)}(\{k_a\})
= Z_{\Lambda'}^{-N/2} G_{\Lambda';\bar{g}(\Lambda')}^{(N)}(\{k_a\}).
\end{align}
The scale dependence of this equality can be written in the form of a differential equation:
\begin{align}
\left[ \Lambda\frac{\partial}{\partial\Lambda} -\beta_\rho(\bar{g})\frac{\partial}{\partial\bar{g}_\rho} + \frac{N}{2}\gamma(\bar{g}) \right] G_{\Lambda;\bar{g}(\Lambda)}^{(N)}(\{k_a\}) = 0.
\end{align}
Note that the repeated index is summed over.
This equation is called the Callan--Symanzik equation \cite{Callan,Symanzik1,Symanzik2} for the connected $N$-point correlation function $G^{(N)}$ with the beta functions $\beta_\rho$ and the field renormalization $\gamma$ defined by
\begin{subequations}
\begin{gather}
\beta_\rho(\bar{g}) = -\left(\Lambda\frac{\partial\bar{g}_\rho}{\partial\Lambda}\right)_{\bar{g}_{\rho,0}}, \\
\gamma(\bar{g}) = -\left(\Lambda\frac{\partial}{\partial\Lambda} \ln Z_{\Lambda}\right)_{\bar{g}_{\rho,0}}.
\end{gather}
\end{subequations}

The correlation functions are obtained from perturbative calculations.  Actually, it is rather straightforward to calculate the one-particle irreducible $N$-point function $\Gamma^{(N)}$ instead of the $N$-point correlation function $G^{(N)}$.
When a model involves a quartic interaction $\phi^4$ without a cubic term $\phi^3$, $\Gamma^{(2)}$ and $\Gamma^{(4)}$ are given by
\begin{gather}
\Gamma^{(2)}(k) = [G^{(2)}(k)]^{-1}, \\
\Gamma^{(4)}(k_1,k_2,k_3,k_4) = \frac{G^{(4)}(k_1,k_2,k_3,k_4)}{G^{(2)}(k_1) G^{(2)}(k_2) G^{(2)}(k_3) G^{(2)}(k_4)}.
\end{gather}
For the definition of $\Gamma^{(N)}$ from the effective action, see the references \cite{Zinn-Justin,Peskin,Amit,Nair}.
Roughly speaking, $\Gamma^{(N)}$ corresponds to the coefficient of the $\phi^N$ term in the effective action.  Again, using the fact that $\phi_0^{\Lambda}(k_a) = \phi_0^{\Lambda'}(k_a)$ holds when the energy corresponding to the momentum $k_a$ is smaller than $\Lambda$ and $\Lambda'$, we find the relation
\begin{align}
\label{eq:Gamma_scale}
Z_{\Lambda}^{N/2} \Gamma_{\Lambda;\bar{g}(\Lambda)}^{(N)}(\{k_a\})
= Z_{\Lambda'}^{N/2} \Gamma_{\Lambda';\bar{g}(\Lambda')}^{(N)}(\{k_a\}).
\end{align}
It results in the Callan--Symanzik equation for $\Gamma^{(N)}$:
\begin{align}
\left[ \Lambda\frac{\partial}{\partial\Lambda} -\beta_\rho(\bar{g})\frac{\partial}{\partial\bar{g}_\rho} - \frac{N}{2}\gamma(\bar{g}) \right] \Gamma_{\Lambda;\bar{g}(\Lambda)}^{(N)}(\{k_a\}) = 0.
\end{align}

So far, we have compared $G^{(N)}$ or $\Gamma^{(N)}$ at different cutoffs $\Lambda$ and $\Lambda'$, so that the dependence on $\Lambda$ is explicit.  However, this comparison is still theoretical; i.e., this is a comparison of different systems.  Our aim is to compare the two theories with the same cutoff $\Lambda$.  For this sake, we rescale the coordinate to change the cutoff.  Suppose we have the scaling relations
\begin{subequations}
\begin{gather}
\Lambda' = b \Lambda, \\
k'_j = b^{d_{k_j}} k_j, \\
\phi_0^{\Lambda'}(k') = b^{d_\phi} \phi_0^{\Lambda}(k),
\end{gather}
\end{subequations}
where $k_j$ is a component of $k=(\bm{k},\omega)$ and $d_{\mathcal{O}}$ denotes the scaling (energy) dimension of $\mathcal{O}$.
Those relations lead to
\begin{align}
\label{eq:Gamma_scale_2}
\Gamma_{\Lambda';\bar{g}(\Lambda')}^{(N)}(\{k_a'\})
= b^{d_{\Gamma^{(N)}}} \Gamma_{\Lambda;\bar{g}(\Lambda')}^{(N)}(\{k_a\}),
\end{align}
Rescaling the momentum forces the cutoff $\Lambda'$ back to $\Lambda$ with the overall factor $b^{d_{\Gamma^{(N)}}}$, but this process does not alter the dimensionless parameters $\bar{g}_\rho$.
From Eqs.~\eqref{eq:Gamma_scale} and \eqref{eq:Gamma_scale_2}, we find the relation
\begin{align}
\label{eq:scale_Wilsonian}
\Gamma_{\Lambda;\bar{g}(\Lambda)}^{(N)}(\{ b^{-d_{k_j}} k_{a,j} \})
= \frac{Z_{\Lambda/b;\bar{g}(\Lambda/b)}^{N/2}}{Z_{\Lambda;\bar{g}(\Lambda)}^{N/2}} b^{-d_{\Gamma^{(N)}}} \Gamma_{\Lambda;\bar{g}(\Lambda/b)}^{(N)}(\{ k_{a,j} \}).
\end{align}
Importantly, this equation compares the $N$-point function $\Gamma^{(N)}$ with the same cutoff $\Lambda$ but at different momenta and parameters.
The interpretation of the scale-dependent parameters can be found from this equation: The parameters $\bar{g}_\rho(\Lambda/b)$ describe the physics at scale $k_j/b^{d_{k_j}}$.

This procedure actually illustrates rescaling in the Wilsonian RG scheme.  We rewrite Eq.~\eqref{eq:scale_Wilsonian} as
\begin{align}
\Gamma_{\Lambda;\bar{g}(\Lambda/b)}^{(N)}(\{ k_{a,j} \})
= \frac{Z_{\Lambda;\bar{g}(\Lambda)}^{N/2}}{Z_{\Lambda/b;\bar{g}(\Lambda/b)}^{N/2}} b^{d_{\Gamma^{(N)}}} \Gamma_{\Lambda;\bar{g}(\Lambda)}^{(N)}(\{ b^{-d_{k_j}} k_{a,j} \}),
\end{align}
which can be interpreted in the following way.
We integrate out fluctuations between the range of $(\Lambda/b,\Lambda]$ (corresponding to $Z_{\Lambda;\bar{g}(\Lambda)}^{N/2}/Z_{\Lambda/b;\bar{g}(\Lambda/b)}^{N/2}$) and rescale the field and parameters (multiplying the factor $b^{d_{\Gamma^{(N)}}}$) to obtain the new action with a different coupling constant but with the same cutoff $\Lambda$.

\begin{widetext}
We can also write down the Callan--Symanzik equation to describe the momentum dependence, instead of the cutoff $\Lambda$.  We differentiate Eq.~\eqref{eq:scale_Wilsonian} with respect to $b$ and then set $b=1$ to obtain
\begin{align}
\left[ d_{k_j} k_{a,j} \frac{\partial}{\partial k_{a,j}} + \beta_\rho(\bar{g})\frac{\partial}{\partial\bar{g}_\rho} - d_{\Gamma^{(N)}} + \frac{N}{2}\gamma(\bar{g}) \right] \Gamma_{\Lambda;\bar{g}}^{(N)}(\{k_{a,j}\}) = 0.
\end{align}
Equivalently, we can introduce a factor $b$ to scale all momenta $b^{d_{k_j}} k_{a,j}$ at once, so the Callan--Symanzik equation becomes
\begin{align}
\left[ b\frac{\partial}{\partial b} + \beta_\rho(\bar{g})\frac{\partial}{\partial\bar{g}_\rho} - d_{\Gamma^{(N)}} + \frac{N}{2}\gamma(\bar{g}) \right] \Gamma_{\Lambda;\bar{g}}^{(N)}(\{ b^{d_{k_j}} k_{a,j} \}) = 0.
\end{align}
\end{widetext}

The Callan--Symanzik equation can be solved by the method of characteristics.
With a parameter $l$, we obtain the differential equations
\begin{subequations}
\begin{gather}
\frac{db}{dl} = b, \\
\frac{d\bar{g}_\rho}{dl} = \beta_\rho(\bar{g}), \\
\frac{d\Gamma_{\Lambda;\bar{g}}^{(N)}(\{ b^{d_{k_j}}k_{a,j} \})}{dl} = \left[ d_{\Gamma^{(N)}} -\frac{N}{2}\gamma(\bar{g}) \right] \Gamma_{\Lambda;\bar{g}}^{(N)}(\{ b^{d_{k_j}}k_{a,j} \}).
\end{gather}
\end{subequations}
The solution to the first equation is straightforward:
\begin{subequations}
\begin{equation}
b(l) = e^l,
\end{equation}
with the initial condition $b(0)=1$.  We write the solution to the second formally as
\begin{equation}
\bar{g}_\rho(l) = \int_0^l dl' \beta_\rho(\bar{g}(l')).
\end{equation}
Also, the formal solution to the third equation is
\begin{align}
&\quad \Gamma_{\Lambda;\bar{g}(l)}^{(N)}(\{ e^{d_{k_j} l} k_{a,j} \}) \nonumber\\
&= e^{d_{\Gamma^{(N)}}l} \Gamma_{\Lambda;\bar{g}(0)}^{(N)}(\{ k_{a,j} \}) \exp \left[ -\frac{N}{2}\int_{0}^{l} dl' \gamma(\bar{g}(l')) \right].
\end{align}
\end{subequations}
Changing $k_{a,j}$ to $e^{-d_{k_j} l} k_{a,j}$, we can express the $N$-point function as
\begin{align}
&\quad \Gamma_{\Lambda;\bar{g}(0)}^{(N)}(\{ e^{-d_{k_j} l} k_{a,j} \}) \nonumber\\
&= e^{-d_{\Gamma^{(N)}}l} \Gamma_{\Lambda;\bar{g}(l)}^{(N)}(\{ k_{a,j} \}) \exp \left[ \frac{N}{2}\int_{0}^{l} dl' \gamma(\bar{g}(l')) \right].
\end{align}
This equation describes the parameters $\bar{g}_\rho$ effectively behaves as if they are $\bar{g}_\rho(l)$ with small momenta $e^{-d_{k_j} l} k_{a,j}$ $(l>0)$.

Before concluding the section, we note that $\gamma$ corresponds to the anomalous dimension.  Let us consider the two-point function $\Gamma^{(2)}$, which corresponds to the inverse of the two-point correlation function.  At a fixed point, the beta function vanishes, and hence both $\bar{g}$ and $\gamma$ are constant; we write them as $\bar{g}^*$ and $\eta$, respectively.  Here we consider the frequency dependence and we hence choose $k_j = \omega_0$.  Since $d_\omega=1$, we obtain
\begin{align}
\Gamma_{\Lambda;\bar{g}^*}^{(2)}(e^{-l} \omega_0) = e^{-d_{\Gamma^{(2)}} l} \Gamma_{\Lambda;\bar{g}^*}^{(2)}(\omega_0) e^{\eta l}.
\end{align}
As $l$ is arbitrary, we set $l=\ln(\omega_0/\omega)$ to find
\begin{align}
\Gamma_{\Lambda;\bar{g}^*}^{(2)}(\omega) = \omega^{d_{\Gamma^{(2)}}-\eta} \omega_0^\eta \Gamma_{\Lambda;\bar{g}^*}^{(2)}(\omega_0)
\propto \omega^{d_{\Gamma^{(2)}}-\eta}.
\end{align}
A naive power counting predicts $\Gamma^{(2)}\propto \omega^{d_{\Gamma^{(2)}}}$, but actually it behaves differently with the exponent $d_{\Gamma^{(2)}}-\eta$.
The deviation $\eta$ corresponds to the anomalous dimension.

\begin{widetext}

\section{Two-loop self-energy for the quasiparticle lifetime}
\label{sec:two-loop_lifetime}

The quasiparticle damping is captured by a finite imaginary part of the self-energy $\Sigma$.  In a series of perturbative expansions, the lowest order correction appears at second order, which is diagrammatically shown in Fig.~\ref{fig:loop}(d).  The result is shown in Eq.~\eqref{eq:two-loop_damping} and here we calculate it explicitly.

We first perform the Matsubara summations.  With the standard procedure, we can convert the summations into the contour integrals on the complex plane, to obtain
\begin{align}
&\quad \Sigma^{(2)}(\bm{k},\omega_n) \nonumber\\
&= -\frac{1}{(2\pi)^2} \int_{\bm{p}\bm{q}\bm{l}} (2\pi)^d \delta(\bm{p}+\bm{q}-\bm{l}-\bm{k}) \int_{-\infty}^{\infty} d\omega_p d\omega_l \nonumber\\
&\quad \times \bigg\{ G_0(\bm{p},\omega_n-i\omega_p) [ \operatorname{Im}G_0^R(\bm{q},\omega_l) \operatorname{Im}G_0^R(\bm{l},\omega_p+\omega_l) -\operatorname{Im}G_0^R(\bm{q},\omega_l-\omega_p) \operatorname{Im}G_0^R(\bm{l},\omega_l) ]  \coth\left(\frac{\omega_p}{2T}\right) \tanh\left(\frac{\omega_l}{2T}\right) \nonumber\\
&\qquad + \operatorname{Im}G_0^R(\bm{p},\omega_p) [ G_0(\bm{q},\omega_n+i\omega_p-i\omega_l) \operatorname{Im}G_0^R(\bm{l},\omega_l) + \operatorname{Im}G_0^R(\bm{q},\omega_l) \cdot G_0(\bm{l},-\omega_n-i\omega_p-i\omega_l) ] \tanh\left(\frac{\omega_p}{2T}\right) \tanh\left(\frac{\omega_l}{2T}\right) \bigg\},
\end{align}
We define the noninteracting retarded (advanced) Green's function $G_0^R$ $(G_0^A)$ as
\begin{equation}
G_0^{R/A}(\bm{k},\omega) = \frac{1}{\omega - E_{\bm{k}} \pm i\delta}.
\end{equation}

The retarded function is obtained by the analytic continuation $i\omega_n = \omega + i\delta$.  We insert $\int d\omega_q \delta(\omega_p+\omega_q-\omega_l-\omega)$ to write the self-energy $\Sigma^{(2)R}$ in a symmetric form:
\begin{align}
&\quad \Sigma^{(2)R}(\bm{k},\omega) \nonumber\\
&= -\frac{1}{(2\pi)^2} \int_{\bm{p}\bm{q}\bm{l}} (2\pi)^d \delta(\bm{p}+\bm{q}-\bm{l}-\bm{k}) \int_{-\infty}^{\infty} d\omega_p d\omega_q d\omega_l \delta(\omega_p+\omega_q-\omega_l-\omega) \nonumber\\
&\quad\times \bigg\{ G_0^R(\bm{p},\omega_p) \operatorname{Im}G_0^R(\bm{q},\omega_q) \operatorname{Im}G_0^R(\bm{l},\omega_l)
\coth\left(\frac{\omega_l-\omega_q}{2T}\right) \left[ \tanh\left(\frac{\omega_q}{2T}\right) - \tanh\left(\frac{\omega_l}{2T}\right) \right] \nonumber\\
&\qquad + \operatorname{Im}G_0^R(\bm{p},\omega_p) \cdot G_0^R(\bm{q},\omega_q) \operatorname{Im}G_0^R(\bm{l},\omega_l) \tanh\left(\frac{\omega_p}{2T}\right) \tanh\left(\frac{\omega_l}{2T}\right) \nonumber\\
&\qquad + \operatorname{Im}G_0^R(\bm{p},\omega_p) \operatorname{Im}G_0^R(\bm{q},\omega_q) \cdot G_0^A(\bm{l},\omega_l) \tanh\left(\frac{\omega_p}{2T}\right) \tanh\left(\frac{\omega_q}{2T}\right) \bigg\}.
\end{align}
Now we take the imaginary part to obtain
\begin{align}
&\quad \operatorname{Im}\Sigma^{(2)R}(\bm{k},\omega) \nonumber\\
&= \frac{1}{(2\pi)^2} \cosh\left(\frac{\omega}{2T}\right) \int_{\bm{p}\bm{q}\bm{l}} (2\pi)^d \delta(\bm{p}+\bm{q}-\bm{l}-\bm{k}) \int_{-\infty}^{\infty} d\omega_p d\omega_q d\omega_l \delta(\omega_p+\omega_q-\omega_l-\omega) \nonumber\\
&\quad\times \operatorname{Im}G_0^R(\bm{p},\omega_p) \operatorname{Im}G_0^R(\bm{q},\omega_q) \operatorname{Im}G_0^R(\bm{l},\omega_l)
\frac{1}{\cosh\left(\dfrac{\omega_p}{2T}\right) \cosh\left(\dfrac{\omega_q}{2T}\right) \cosh\left(\dfrac{\omega_l}{2T}\right)} \nonumber\\
&= -\frac{\pi}{4} \cosh\left(\frac{\omega}{2T}\right) \int_{\bm{p}\bm{q}} \frac{\delta(\omega-E_{\bm{p}}-E_{\bm{q}}+E_{\bm{p}+\bm{q}-\bm{k}})}{\cosh\left(\dfrac{E_{\bm{p}}}{2T}\right) \cosh\left(\dfrac{E_{\bm{q}}}{2T}\right) \cosh\left(\dfrac{E_{\bm{p}+\bm{q}-\bm{k}}}{2T}\right)} ,
\end{align}
where we use the relation $\operatorname{Im}G_0^R(\bm{k},\omega) = -\operatorname{Im}G_0^A(\bm{k},\omega) = -\pi \delta(\omega-E_{\bm{k}})$.

\end{widetext}

\section{Susceptibilities at a high-order VHS}
\label{sec:susceptibilities}

A divergent DOS $D(E)$ accompanies divergent susceptibilities.  Here we consider the noninteracting susceptibilities in the particle-hole and particle-particle channels, $\chi_\text{ph}$ and $\chi_\text{pp}$, respectively:
\begin{gather}
\chi_\text{ph} (\bm{q},\omega;T) = \int_{\bm{p}} \frac{f(\xi_{\bm{p}+\bm{q}})-f(\xi_{\bm{p}})}{\omega+i\delta-\xi_{\bm{p}+\bm{q}}+\xi_{\bm{p}}}, \\
\chi_\text{pp} (\bm{q},\omega;T) = \int_{\bm{p}} \frac{f(\xi_{\bm{p}+\bm{q}})-f(-\xi_{-\bm{p}})}{\omega+i\delta-\xi_{\bm{p}+\bm{q}}-\xi_{-\bm{p}}},
\end{gather}
where $f(\xi)=(e^{\xi/T}+1)^{-1}$ is the Fermi--Dirac distribution and $\xi_{\bm{p}} = E_{\bm{p}}-\mu$.  In the following, we focus on the static susceptibilities $(\omega=0)$.  At $\mu = 0$, the noninteracting susceptibilities follow the scaling relations for momentum $\bm{q}$ and temperature $T$, described by
\begin{gather}
\chi_\text{ph}(\bm{q},\omega=0;T) = \nu^{-\epsilon} \hat{\chi}_\text{ph} \left( \frac{q_+^{n_+}}{\nu}, \frac{q_-^{n_-}}{\nu}, \frac{T}{\nu} \right), \\
\chi_\text{pp}(\bm{q},\omega=0;T) = \nu^{-\epsilon} \hat{\chi}_\text{pp} \left( \frac{q_+^{n_+}}{\nu}, \frac{q_-^{n_-}}{\nu}, \frac{T}{\nu} \right),
\end{gather}
where $\hat{\chi}_\text{ph}$ and $\hat{\chi}_\text{pp}$ are the scaling functions.  Those scaling behaviors of the noninteracting susceptibilities lead to the approximate relations
\begin{gather}
\chi_\text{ph}(\bm{q},\omega=0;T) \sim \max(T^{-\epsilon}, q_+^{-\epsilon n_+}, q_-^{-\epsilon n_-}), \\
\chi_\text{pp}(\bm{q},\omega=0;T) \sim \max(T^{-\epsilon}, q_+^{-\epsilon n_+}, q_-^{-\epsilon n_-}).
\end{gather}

Although the explicit forms depend on the specific form of the energy dispersion, the temperature dependence reflects only the DOS Eq.~\eqref{eq:DOS}:
\begin{gather}
\label{eq:diff_ph}
\chi_\text{ph}(T) = T^{-\epsilon} (D_++D_-) (2^{1+\epsilon}-1) \Gamma(1-\epsilon) [-\zeta(-\epsilon)],  \\
\label{eq:diff_pp}
\chi_\text{pp}(T) = \frac{1}{\epsilon} T^{-\epsilon} (D_++D_-) (2^{1+\epsilon}-1) \Gamma(1-\epsilon) [-\zeta(-\epsilon)],
\end{gather}
where $\zeta(s)$ is the Riemann zeta function.  We see that $\chi_\text{pp}$ is larger than $\chi_\text{ph}$ by the numerical factor $1/\epsilon$.

For the case of a high-order VHS with a power-law divergent DOS, differences between $\chi_\text{ph}$ and $\chi_\text{pp}$ appear as the prefactors of the terms $T^{-\epsilon}$, $q_+^{-\epsilon n_+}$, $q_-^{-\epsilon n_-}$ as we have shown in Eqs.~\eqref{eq:diff_ph} and \eqref{eq:diff_pp}.
This is in contrast to the conventional VHS with a logarithmically divergent DOS; while $\chi_\text{ph}$ has a logarithmic divergence reflecting the DOS, $\chi_\text{pp}$ exhibits a double logarithmic divergence.  The additional logarithm in the BCS channel appears in the presence of time-reversal symmetry, which can be regarded as the Fermi surface nesting with itself.

\end{document}